\pdfoutput=1
 \documentclass{src/DOEproposal}

\usepackage{amsmath, amsthm, amssymb, ulem}
\usepackage{latexsym}
\usepackage{url}
\usepackage{epsfig}
\usepackage{ragged2e}
\usepackage{tabularx}
\usepackage{epstopdf}
\usepackage[range-phrase=\mbox{--},range-units=single,per-mode=symbol]{siunitx}
\usepackage{subcaption}
\usepackage[x11names,dvipsnames,table]{xcolor} %
\usepackage{colortbl}
\usepackage{array}
\usepackage{multirow}
\usepackage[version=4]{mhchem}
\usepackage{graphicx}
\usepackage{xargs}
\usepackage{xfrac}
\usepackage{etoolbox}
\usepackage[total={6.5in,9in},top=1in,headsep=0.1in,headheight=1in]{geometry}
\usepackage{graphbox}
\usepackage[final]{pdfpages}
\usepackage{kotex}
\usepackage{tabularx}
\usepackage{xltabular}
\usepackage{caption}
\usepackage{booktabs}
\usepackage{tikz}
\usepackage{hyperref}
\usetikzlibrary{backgrounds}
\usetikzlibrary{arrows.meta}
\usetikzlibrary{trees}
\usetikzlibrary{shadings}
\usetikzlibrary{automata}

\pgfdeclarelayer{background}
\pgfdeclarelayer{foreground}
\pgfsetlayers{background,main,foreground}  

\makeatletter
\newif\ifAPS@firstauthor
\makeatother

\usepackage{hyperref}
\hypersetup{
    colorlinks=true,        %
    linkcolor=blue,         %
    citecolor=blue,        %
    filecolor=magenta,      %
    urlcolor=blue,       %
    pdfstartview=           %
}                           %
\usepackage{enumitem}
\setlist{nolistsep}
\setlist[enumerate]{itemsep=0mm}
\newcolumntype{P}[1]{>{\raggedright\arraybackslash}p{#1}}
\newcolumntype{C}[1]{>{\centering\arraybackslash}p{#1}}
\def\bit{\begin{itemize}
  }
  \def\eit{\end{itemize}
  }
\usepackage[dvipsnames]{xcolor}

\DeclareSIUnit\c{\mbox{$c$}}
\DeclareSIUnit\magn{\mbox{$\times$}}
\DeclareSIUnit\min{min}
\DeclareSIUnit\week{week}
\DeclareSIUnit\month{mo}
\DeclareSIUnit\months{mos}
\DeclareSIUnit\year{yr}
\DeclareSIUnit\years{years}
\DeclareSIUnit\yr{yr}
\DeclareSIUnit\standard{std}
\DeclareSIUnit\str{sr}
\DeclareSIUnit\ppm{ppm}
\DeclareSIUnit\ppb{ppb}
\DeclareSIUnit\ppt{ppt}
\DeclareSIUnit\pe{PE}
\DeclareSIUnit\spe{SPE}
\DeclareSIUnit\pdm{PDM}
\DeclareSIUnit\ev{events}
\DeclareSIUnit\ct{counts}
\DeclareSIUnit\neutron{\mbox{$n$}}
\DeclareSIUnit\smp{samples}
\DeclareSIUnit\Sample{S}
\DeclareSIUnit\ch{ch}
\DeclareSIUnit\hit{hit}
\DeclareSIUnit\hits{hits}
\DeclareSIUnit\bin{(\mbox{5-PE}~bin)}
\DeclareSIUnit\sgm{\mbox{$\sigma$}}
\DeclareSIUnit\rms{RMS}
\DeclareSIUnit\keVee{\mbox{keV$_{e{\rm e}}$}}
\DeclareSIUnit\keVr{\mbox{keV$_{\rm nr}$}}
\DeclareSIUnit\eVee{\mbox{eV$_{\rm ee}$}}
\DeclareSIUnit\eVr{\mbox{eV$_{\rm nr}$}}
\DeclareSIUnit\ph{photon}
\DeclareSIUnit\el{\mbox{$e^-$}}
\DeclareSIUnit\pm{\mbox{PMT}}
\DeclareSIUnit\pixel{\mbox{pixel}}
\DeclareSIUnit\inch{''}
\DeclareSIUnit\foot{'}
\DeclareSIUnit\ft{\mbox{ft}}
\DeclareSIUnit\bit{bit}
\DeclareSIUnit\sample{samples}
\DeclareSIUnit\barn{barn}
\DeclareSIUnit\bara{bar}
\DeclareSIUnit\barg{barg}
\DeclareSIUnit\mlardepth{\mbox(meter~of~\LAr~depth)}
\DeclareSIUnit\Curie{Ci}
\DeclareSIUnit\psf{psf}
\DeclareSIUnit\pcf{pcf}
\DeclareSIUnit\parsec{pc}
\DeclareSIUnit\liveday{\mbox{live-days}}
\DeclareSIUnit\days{\mbox{days}}
\DeclareSIUnit\miles{\mbox{miles}}
\DeclareSIUnit\lumens{\mbox{lm}}
\DeclareSIUnit\degreeC{\mbox{$^{\circ}$C}}
\DeclareSIUnit\degreeF{\mbox{$^{\circ}$F}}
\DeclareSIUnit\electron{\mbox{$e^-$}}
\DeclareSIUnit\Euro{\mbox{\euro}}
\DeclareSIUnit\cph{cph}
\DeclareSIUnit\neq{neq}
\DeclareSIUnit\normal{\mbox{N}}

\newcommand{\gr}{$\gamma$-ray}

\newcommand{\alphan}{$(\alpha,n)$}

\newcommand{\ngamma}{$(n,\gamma)$}

\let\OLDthebibliography\thebibliography
\renewcommand\thebibliography[1]{%
  \OLDthebibliography{#1}%
  \raggedright       %
  \sloppy            %
}

\usepackage{orcidlink}
\usepackage{fontawesome}

\usepackage{titlesec}
\titlespacing*{\subsection}
  {0pt}               %
  {12pt plus 4pt minus 2pt}   %
  {6pt plus 2pt minus 1pt}    %
\titlespacing*{\section}
  {0pt}               %
  {12pt plus 4pt minus 2pt}   %
  {6pt plus 2pt minus 1pt}    %

\investigator{Jaehoon Yu}
\institute{University of Texas, Arlington}
\phone{(817) 272-2814}
\email{jaehoon@uta.edu}

\title{The DAMSA Experiment - Conceptual Design Report}

\affiliation{uta}{University of Texas at Arlington, Arlington,~TX~76019,~USA}
\affiliation{fnal}{Fermi National Accelerator Laboratory,~Batavia,~Illinois,~USA}
\affiliation{wau}{Washington University, St. Louis,~MO~63130,~USA}
\affiliation{tamu}{Texas~A\&M~University,~College~Station,~TX~77843,~USA}
\affiliation{ucr}{Department of Physics \& Astronomy, University of California, Riverside,~CA~92521,~USA}
\affiliation{bnl}{Brookhaven National Laboratory, Upton, NY 11973,~USA}
\affiliation{itpnn}{Department of Physics and Institute of Theoretical Physics Nanjing Normal University, Nanjing, 210023, China}
\affiliation{ibs}{Particle Theory and Cosmology Group, Center for Theoretical Physics of the Universe, \\ Institute for Basic Science (IBS), Daejeon 34126,~Republic~of~Korea}
\affiliation{usd}{Department of Physics, University of South Dakota, Vermillion,~SD~57069,~USA
}
\affiliation{snu}{Department of Physics \& Astronomy, Seoul National University, 1~Gwanak-ro,~Gwanak-gu,~Seoul~08826,~Republic~of~Korea}
\affiliation{knu}{Department of Physics, Kyungpook~National~University,~Daegu 41566,~Republic~of~Korea}
\affiliation{uchicago}{Department of Physics, University of Chicago, Chicago,~IL~60637,~USA}
\affiliation{pitt}{University of Pittsburgh,~Pittsburgh,~PA~15260,~USA}
\affiliation{kus}{Department of Accelerator Science, Korea University Sejong Campus, 2511~Sejong-ro,~Sejong~30019,~Republic~of~Korea}
\affiliation{chungnam}{Department of Physics and Institute for Sciences of the Universe, Chungnam National University, Daejeon~34134,~Republic~of~Korea}
\affiliation{sdm}{South Dakota School of Mines and Technology, Rapid City,~SD~57701,~USA}
\affiliation{umd}{Department of Physics, University of Maryland, College Park,~MD~20742,~USA}
\affiliation{jnu}{Laboratory for Symmetry and Structure of the Universe, Department of Physics, Jeonbuk~National~University,~Jeonju,~Jeonbuk~54896,~Republic~of~Korea}
\affiliation{nu}{Northwestern~University,~Evanston,~IL~60208,~USA}

\author[uta]{Prithak~Bhattarai}
\author[uta]{Andrew~Brandt}
\author[fnal]{Alan~Bross}
\author[uta]{Brad~Brown}
\author[uta]{Samriddha~Chakraborty}
\author[wau]{Haohui~Che}
\author[wau]{Bhupal~Dev}
\author[tamu]{Bhaskar~Dutta}
\author[fnal]{Juan~V.~Estrada}
\author[uta]{Eric~Garcia}
\author[ucr]{Anthony~Gomez}
\author[uta]{Gajendra~Gurung}
\author[uta]{Brian~Joshua~Gomez~Hernandez}
\author[uta]{Wooyoung~Jang}
\author[bnl]{Jay~Hyun~Jo}
\author[itpnn,ibs]{Krzysztof~Jod\l{}owski}
\author[usd]{Doojin~Kim}
\author[snu]{Eunsu~Kim}
\author[snu]{Hyunyong~Kim}
\author[knu]{Shin~Hyung~Kim}
\author[uchicago]{Young-Kee~Kim}
\author[usd]{Jing~Liu}
\author[knu]{Chang-Seong Moon}
\author[pitt]{Donna~Naples}
\author[uta]{David~Nygren}
\author[snu]{Minseok~Oh}
\author[pitt]{Vittorio~Paolone}
\author[kus]{Hyangkyu~Park}
\author[chungnam]{Jong-Chul~Park}
\author[fnal]{Nathaniel~J.~Pastika}
\author[uta]{Rohit~Raut}
\author[sdm]{Juergen~Reichenbacher}
\author[fnal]{Paul~Rubinov}
\author[umd]{Eunsuk~Seo}
\author[ucr]{Veronika~Shalamova}
\author[jnu]{Seodong~Shin}
\author[uchicago]{Melvin~Shochet}
\author[nu]{Adrian~Thompson}
\author[uchicago]{Yau~Wah}
\author[ucr]{Shawn~Westerdale}
\author[bnl]{Guang~Yang}
\author[snu]{Un-Ki~Yang}
\author[snu]{Inseok~Yoon}
\author[uta]{Jaehoon~Yu}

\begin{abstract}
    We present the conceptual design of the DAMSA (DArk Messenger Searches at an Accelerator) experiment. 
    DAMSA is a novel ultra-short-baseline accelerator/beam-dump experiment aimed at probing short-lived physics processes, including searches for evidence of a dark sector of particle physics and well-motivated rare Standard Model signals. Motivated by open questions in neutrino physics and the absence of conclusive evidence for conventional weakly interacting massive particles, DAMSA targets MeV-to-sub-GeV dark-sector messengers with feeble couplings that can be produced in abundance at a beam dump/target. By employing an ultra-short-baseline configuration, DAMSA is uniquely positioned to overcome the beam-dump “ceiling” that limits sensitivity to fast-decaying particles in longer-baseline experiments. The conceptual design emphasizes a beam-dump production scheme combined with a compact detector, optimized for rare decays while mitigating intense neutron-induced backgrounds, using detector capabilities.
To validate the experimental strategy and detector technologies even before the precision tracking system is ready, DAMSA Phase I focuses on axion-like particles decaying into two photons, as the benchmark physics case, operating at the 8 GeV electron beams at SLAC Linac-to-ESA (LESA) facility. 
A successful realization of DAMSA Phase I establishes the feasibility of the DAMSA approach, enabling a broad and powerful program to explore short-lived new physics and precision Standard Model processes in a previously inaccessible regime.  
The subsequent DAMSA Phase II expands physics to include the $e^{+}e^{-}$ final state, taking advantage of its full configuration, incorporating the precision silicon tracker under a magnetic field.  
This paper outlines the details of DAMSA's physics goals, key experimental challenges, and how to overcome them with the baseline detector capabilities, as well as a few options under consideration for further improvements.
\end{abstract}

\begin{document}

    \maketitle

    \phantomsection
    \section{Introduction}
\label{sec:Introduction}

A compelling observational motivation for new physics comes from the existence of dark matter that makes up about 25\% of the Universe's mass-energy budget and is strongly supported by numerous astrophysical and cosmological observations through its gravitational effects. Its particle properties, including the mass scale, however, remain unknown. Scenarios involving GeV-scale weakly interacting massive particles (WIMPs) have garnered significant attention, as these candidates are thermally produced, independent of specific model details. Moreover, since WIMPs predict non-gravitational interactions between dark matter and the Standard Model (SM) particles, extensive experimental and theoretical efforts have been devoted to their study over the past few decades (see, e.g., Ref.~\cite{Bertone:2004pz}). However, no conclusive evidence has been found, prompting the exploration of alternative ideas which are now being actively investigated.

Among these alternatives, MeV-scale (light) dark matter is a promising candidate, as it can be thermally produced and its parameter space remains largely unexplored. Additionally, portal scenarios suggest that other dark-sector particles--such as messenger particles mediating the interactions between dark matter and SM particles--of similar mass should exist, with feeble interactions with SM particles (see, e.g., Refs.~\cite{Holdom:1985ag,Patt:2006fw,Pospelov:2007mp,Pospelov:2008jd,Pospelov:2008zw,Falkowski:2009yz,Arcadi:2019lka,Batell:2022xau}). The predicted mass scale is within reach of existing beam facilities, which can produce these dark-sector particles, including MeV-scale dark matter, while the very weak coupling strengths motivate the need for experiments with high intensity beams to accommodate the rare production of such particles. In addition to scenarios involving light dark matter, visibly-decaying messenger particles are often proposed to explain various experimental anomalies, such as the MiniBooNE low-energy excess~\cite{MiniBooNE:2008yuf,MiniBooNE:2018esg,MiniBooNE:2020pnu} and the LSND anomaly~\cite{LSND:2001aii}.

Beam dump facilities are expected to produce the aforementioned feebly-interacting messenger particles in abundance, shedding light on dark-sector physics. The discovery of dark-sector particles at an accelerator would pave the way for a deeper understanding of the nature of dark matter. In particular, high-intensity beam experiments have provided leading constraints on messenger particles in the MeV-to-sub-GeV range. Schematically, messenger particles produced at the target or the dump must survive sufficiently long to reach the detector and decay within its fiducial volume. The sensitivity of an experiment depends on factors such as baseline distance, effective decay volume, beam intensity, and beam energy. A key challenge is probing messenger particles that decay relatively early, as longer baselines reduce detection sensitivity, which is referred to as beam-dump ``ceiling''~\cite{Dutta:2023abe,Kim:2024vxg}. To address this, experiments with very short baselines are better suited for exploring these regions of parameter space~\cite{Dutta:2023abe,Kim:2024vxg}.

In this regard, a novel experiment for searching for dark-sector particles, DAMSA ({\bf{\underline {DA}}}rk {\bf{\underline {M}}}essenger {\bf{\underline {S}}}earches at an {\bf{\underline {A}}}ccelerator - pronounced ``da-m-sa'' (담사) and means rumination) has been proposed, with a baseline scale less than a meter, inspired by the experimental scheme proposed in Ref.~\cite{Jang:2022tsp}. An essential element of DAMSA is the beam-dump facility, which can produce dark-sector particles in the dump through the beam interactions with it. Given the potential of DAMSA to explore the dark sector, it is crucial to place this effort within the broader scope of dark matter research. 

The use of a sufficiently thick beam dump and the proximity of the detector inevitably lead to a high flux of low-energy neutrons, which may induce unwanted backgrounds. In this proposal, we aim to overcome this challenge by selecting specific final states, optimizing beam configurations, and employing a detector system designed to mitigate backgrounds from neutron interactions. 
To this end, we will first construct the {\bf DAMSA Phase I}, which will focus on the two-photon final state of the axion-like particles (ALP) and operate using \SI{8}{\GeV} electron beams---potentially offering a more controlled environment regarding neutron-induced backgrounds---at SLAC's LESA facility~\cite{markiewicz2022slaclinacesalesa}
with the detector system placed $\sim10$~cm away from the average ALP production point in the target. 
A successful completion of the proposed Phase I will allow us not only to possibly discover the messenger particle but also to build confidence in the experimental techniques, paving the way for many more additional discoveries through the next phase DAMSA experiment and other future efforts in the field.
The subsequent DAMSA Phase II will have its full detector configuration with the precision silicon tracking system under a 1 Tesla magnetic field, enabling the expansion of the signal to $e^{+}e^{-}$ final states, at SLAC's LESA facility.
Since DAMSA detector with full configuration will be completed after Phase II, the discussions of this conceptual design focuses up to Phase II.

DAMSA aims to eventually take data at CERN's beam-dump facility (BDF)~\cite{Albanese:2878604} to take advantage of higher energy proton beams which will enable exploration of different production mechanisms, as can be seen in Table~\ref{tab:detector_phases}, a summary of the phases of the DAMSA experiment.
While the collaboration has been working on securing funds necessary to make further progress towards the technical design of the experiment, we expect this paper to provide the concept for an innovative experimental proposal.

An outline of the paper is as follows; section 2 describes the physics goals, section 3 details the experiment and detector, including a few options under consideration for the final detector configurations to further improve physics performance, and section 4 presents the background mitigation plans, using the baseline detector configurations presented in this conceptual design report.

\begin{table}[h]
\caption{Summary of DAMSA phases at different facilities, including their potential physics targets in searches for long-lived particle species $X$. The last column characterizes the maximum energy endpoint of the $X$ species produced from electromagnetic cascades, which can be up to the full beam energy in electron beam dumps but limited by the secondary cascade endpoint in proton beam dumps.}
\label{tab:detector_phases}
\begin{center}
\begin{tabular}{l|l|l|l|l}
\hline
\textbf{Phases} & \textbf{Goals}                   & \textbf{Facility} & \textbf{Beam/Energy} & \textbf{$X$ Energy Scale}\\ \hline
Phase 0           & Detector Development             & FAST@FNAL         & 300 MeV electrons     & $\lesssim 300$~MeV  \\
                  & \& Background Validation         &                   &              &         \\
Phase I           & $X$ —\textgreater $\gamma\gamma$      & LESA@SLAC         & 8 GeV Electrons   &  $\lesssim 8$ GeV   \\
Phase II          & $X$ —\textgreater $\gamma\gamma$, $e^+e^-$       & LESA@SLAC         & 8 GeV Electrons  & $\lesssim 8$ GeV     \\
Phase III         & $X$ —\textgreater{}$\gamma\gamma$, $e^+e^-$, $\mu^+\mu^-$ & BDF@CERN          & 400 GeV Protons & $\lesssim 320$~GeV
\end{tabular}
\end{center}
\end{table}

\vspace{0.5em}
\section{Physics Goals}  
\label{sec:PhysicsGoals}
This section summarizes the potential physics topics that DAMSA can explore. Full details can be found in Ref.~\cite{DAMSA:2026fiw}.
By exploiting the short baseline between the target and detector, DAMSA can search for short-lived particles in a region of phase space inaccessible to present and future experiments. 
\paragraph{Axion-like Particles Coupling to Photons:} 
Axion-like particles provide a well-motivated and economical extension of the SM in which a new pseudoscalar field \(a\) interacts through the leading electromagnetic portal. The QCD axion~\cite{Peccei:1977hh,Wilczek:1977pj,Weinberg:1977ma}, introduced to resolve the strong-CP problem, provides a canonical benchmark within this broader framework; more general ALPs motivate an experimentally driven exploration of $m_a$ (ALP mass), $g_{a\gamma\gamma}$~(parameterizes the ALP-photon coupling) parameter space. From a phenomenological standpoint, the $a\gamma\gamma$ portal is especially attractive because it yields clean, versatile experimental signatures that map directly onto our detector and infrastructure choices. These channels provide multiple, complementary discovery handles with controllable systematics, enabling DAMSA to define a clear and broadly interpretable physics program in the standard ($m_a, g_{a\gamma\gamma}$) plane while retaining sensitivity across wide ranges of lifetime and mass.

\begin{figure}[t]
    \centering
    \includegraphics[width=0.75\linewidth]{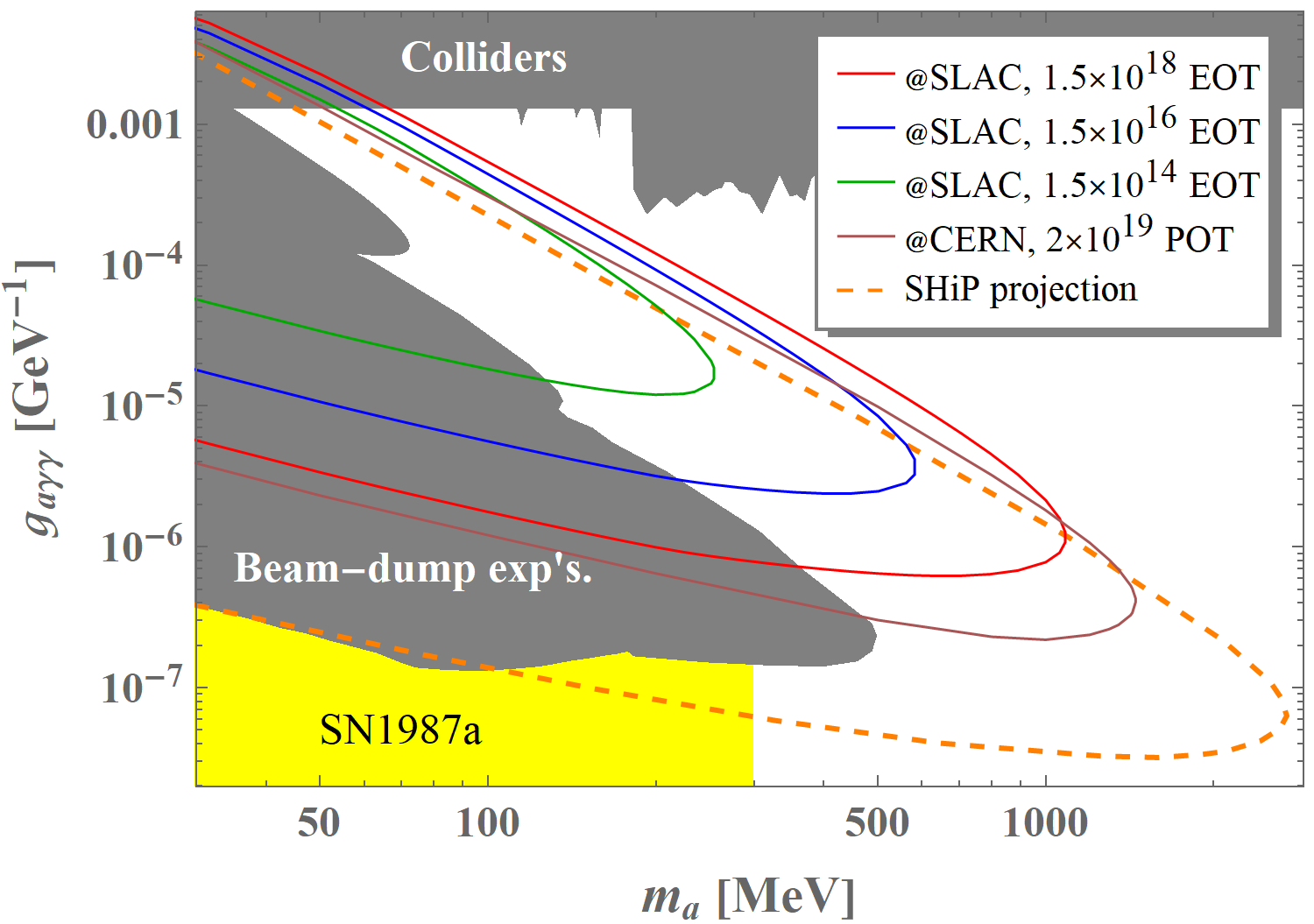}
    \caption{Sensitivity prospects for DAMSA to ALPs interacting with the SM photon at SLAC's 8 GeV LESA electron-beam facility for different amounts of data (solid green, blue and red).  This plot also presents physics reach DAMSA at CERN's beam dump facility (solid brown) and that of the SHiP experiment, clearly demonstrating the complementarity of the two experiments.}
    \label{fig:alpsensitivity}
\end{figure}

The sensitivity projection in Fig.~\ref{fig:alpsensitivity} from our preliminary study shows that the 8 GeV electron beam at SLAC’s LESA facility enables us to explore a broad range of previously unprobed parameter space using the beam target, decay chamber, and calorimeter described in Section~\ref{sec:Detector} and assuming several representative cumulative beam intensities. Existing laboratory-based (e.g., beam-dump-type and collider experiments) and astrophysics-based constraints are shown as the gray-shaded and yellow-shaded regions, respectively. 
The signal sensitivity projection is based on the dedicated simulation framework described in Ref.~\cite{DAMSA:2026fiw}. The photon flux and its energy-angular distribution inside the tungsten target are obtained with GEANT4~\cite{GEANT4:2002zbu}, and the resulting photons are converted into ALPs through the Primakoff process~\cite{Primakoff:1951iae}. The calculation subsequently includes the ALP propagation and decay probability within the fiducial decay volume, together with the geometric acceptance, as described in Ref.~\cite{DAMSA:2026fiw}.

For the LESA electron-beam configuration, an important prompt background originates from electromagnetic shower particles, primarily photons, electrons, and positrons, leaking from the tungsten target toward the detector. We estimate this component again with a dedicated GEANT4 simulation by investigating the energy distributions of particles entering the detector volume and, in particular, by identifying high-energy photons that could form signal-like di-photon events.
To reduce such accidental di-photon backgrounds, we impose the requirements that each decay photon carry at least 500 MeV of energy and that the angular separation between the two photons be greater than $1^\circ$. We expect this selection to yield a nearly background-free search. To highlight the Phase I DAMSA's physics potential at SLAC's LESA facility for three different total number of electrons on target (EOT) in green ($1.5\times 10^{14}$), blue ($1.5\times 10^{16}$) and red ($1.5\times 10^{18}$) solid lines. 
We also present, for comparison, the projected sensitivity of SHiP~\cite{Albanese:2878604} and the expected sensitivity reach of DAMSA at CERN’s BDF facility~\cite{Kim:2024vxg} with $2\times 10^{19}$ total number of protons on target (POT) in solid brown. The representative experimental configuration, expected signal characteristics, beam-related background environment and mitigation considerations, and physics feasibility of such a compact ultra-short-baseline implementation are discussed in detail in Ref.~\cite{Kim:2024vxg}.

We emphasize that DAMSA Phase I at SLAC LESA and the prospective fully-equipped DAMSA implementation at a high-energy proton-beam facility are distinct experimental phases, producing physics results at each phase.
DAMSA Phase I is a compact electron-beam experiment designed primarily to establish the ultra-short-baseline detection strategy, validate the detector technologies and background-rejection methods, and search for ALPs through the $a\to \gamma \gamma$ channel in a comparatively controlled electromagnetic-shower environment. 
The subsequent DAMSA Phase II at SLAC LESA facility will complete the detector system by adding the precision silicon tracking system under magnetic field and expand the signal into $a\to e^{+}e^{-}$ channel. 

A future proton-beam implementation, such as that considered for the CERN BDF, would retain the basic target-decay-volume-detector concept but would require a facility-specific redesign of the target, shielding, detector dimensions, and readout to accommodate the substantially higher beam energy and hadronic radiation environment. It would also extend the physics program to additional production mechanisms and final states, including $e^+e^-$, $\mu^+\mu^-$, and potentially hadronic channels. Correspondingly, its dominant backgrounds would include hadronic-shower particles, neutral-meson decays, and beam-related neutrons at rates and energies different from those at LESA. The sensitivity curve shown here for the CERN BDF is therefore a prospective physics benchmark based on the portable compact-detector framework of Ref.~\cite{Kim:2024vxg}, rather than a claim that the Phase I apparatus could be installed there without substantial modification.

\paragraph{Axion-like Particles Coupling to Electrons}
We consider ALPs with dominant couplings to electrons, which can arise by the mediation of an extended Higgs sector that lifts the fermion mass terms to pseudoscalar currents, as in the DFSZ (Dine-Fischler-Srednicki-Zhitnitsky) variants of the QCD axion~\cite{Dine:1982ah,Zhitnitsky:1980tq,Sun:2020iim,DiLuzio:2020wdo}, and others~\cite{Han:2020dwo,Ganguly:2022imo}. 
This phenomenological coupling allows for the production of ALPs in the DAMSA targets in several ways. In the case of an electron beam impinging on the target, ALPs can be emitted in the electron bremsstrahlung process. In addition, secondary electrons, positrons, and photons produced in the beam target can support ALP emission through resonant and non-resonant annihilation of positrons with target electrons, $e^+ e^- \to a$ and $e^+ e^- \to a \gamma$, through Compton scattering ($e^- \gamma \to a e^-$) and through further bremsstrahlung emission from secondary positrons and electrons. 
The projected reach at DAMSA expands upon existing beam dump limits and gains access to heavier ALPs approximately in the $g_{ae} \in [10^{-7}, 10^{-5}]$ coupling range. Pushing the sensitivity envelope in this direction complements collider probes~\cite{Bauer:2017ris, Eberhart:2025lyu, Alimena:2025kjv}, which test regions of parameter space at higher couplings and larger masses.

\paragraph{Dark Photons}
We consider a model in which a dark photon interacts via a kinetic mixing term.
Any interactions involving the photon are also allowed for the dark photon, but will be scaled by a coupling factor, as well as kinematic factors due to the mass of the dark photon $m_{A^\prime}$. 
The dominant production processes for photons above 10 MeV are pion decay and bremsstrahlung. In addition, vector meson mixing with $A^\prime$ could also lead to enhanced $A^\prime$ production in the target, see for example ref.~\cite{Schuster:2021mlr}.
The dark photon can decay into $e^{+} e^{-}$, $\mu^{+} \mu^{-}$, or hadrons.
Like the sensitivity to electron-coupled ALPs, we expect di-lepton final states to drive the sensitivity to this dark photon search, with the possibility of hadronic decay modes opening up for higher energy beams. 

\paragraph{Large Extra Dimensions}
Massive spin-2 fields are a common feature of many extensions of the Standard Model of particle physics. They naturally appear in extra-dimensional models, such as the Randall-Sundrum model~\cite{Randall:1999ee}, Linear Dilaton (LD)~\cite{Antoniadis:2001sw}, and the ADD model \cite{Arkani-Hamed:1998jmv}, among others, which aim to solve the electroweak hierarchy problem. In these models, the SM fields are confined to a 3-brane, while gravity propagates in the bulk. As a result of propagating in the extra dimension, the 4D Lagrangian contains an infinite tower of massive spin-2 fields, the Kaluza-Klein (KK) modes of the metric. The lightest of these modes, the graviton, is massless and mediates the gravitational interaction between the SM fields with interactions suppressed by the Planck mass. The heavier KK modes, on the other hand, are massive and can mediate new interactions between the SM fields, whose strength can be much larger than for the massless graviton, leading to rich phenomenology.

On the other hand, massive spin-2 field $G$ can also appear in the context of thermal dark matter (DM) models, where $G$ mediates the interaction between the DM and the SM~\cite{Lee:2013bua,Kang:2020huh}.
In such a case, the origin of the mass of the spin-2 field $G$ is typically left unspecified, but it can be generated by the St\"ukelberg formalism, among other possibilities~\cite{Hinterbichler:2011tt}.
The leading decay channel of a light massive spin-2 field $G$ is to two photons~\cite{Lee:2013bua,Kang:2020huh}, which allows for the possibility of observing $G$ decays in DAMSA.
The leading production mechanism of producing sub-GeV spin-2 field $G$ is through the Primakoff-like process of conversion of on-shell photons into $G$: $\gamma N \to G N$~\cite{Jodlowski:2023yne}, while bremsstrahlung is typically subdominant.
DAMSA can probe previously unexplored parameter space, such as E137, NuCal, and colliders. We note that the default DAMSA setup, characterized by beam $E_p=0.8\,\mathrm{GeV}$, can probe most of the gap between the LHC and supernovae limits. This is due to DAMSA's main characteristics, which allow us to probe the intermediate long-lived regime - the one between collider-stable species and species decaying on astrophysical length scales.

\paragraph{Light Dark Matter}
Identification of dark matter remains a long-standing challenge in the particle physics community.
DAMSA can contribute to this effort by probing light dark matter (LDM) in the sub-GeV mass range, as motivated by theories with light mediators such as dark photons, dark scalars, axion-like particles, or neutrino-philic mediators.
At low-energy accelerator facilities, LDM can be produced through processes such as meson decays and electromagnetic radiation, depending on the underlying dark matter scenario~\cite{Bjorken:2009mm,Batell:2014mga,Batell:2022xau,Choi:2025wbw}. From the experimental perspective, two broad categories of signals can be considered: elastic and inelastic dark matter scattering. The first category corresponds to the elastic scattering of dark matter off electrons, nucleons, or nuclei (coherently), producing visible recoil signals in the detector. The second category involves inelastic scattering of dark matter, in which either an excited dark sector state~\cite{Tucker-Smith:2001myb, Izaguirre:2014dua, Kim:2016zjx} or an excited target nucleus~\cite{Dutta:2024kuj, Choi:2024ism} is produced.

Depending on the model parameters and the amount of energy transferred, these excited states can give rise to secondary signatures, providing distinctive experimental features compared to elastic scattering processes.
Light dark matter particles produced at the target may scatter elastically or inelastically within the detector volume, generating low-energy electromagnetic or hadronic activity. 
The extremely short source–detector distance of DAMSA enhances the expected event rates compared to conventional beam-dump or long-baseline configurations, allowing competitive sensitivity even for very small couplings. 
Moreover, DAMSA’s proximity to the target opens opportunities to probe regions of parameter space where the dark sector particle is moderately long-lived, but would decay before reaching far detectors.
For a facility with a high intensity but pulse-type beam, the utilization of timing information can be important~\cite{Dutta:2019nbn,Dutta:2020vop}.
DAMSA can provide complementary coverage to existing beam-dump and collider experiments by probing LDM scenarios in which production is dominated by meson decays or photon-induced processes at low-momentum transfer.

\paragraph{The Bread-and-Butter Standard Model Physics: Mesons and Physics Validation}

The intense electron beam source impinging on the tungsten target also leads to meson production. Depending on the mass, lifetime, boost factor, and closeness of the production site to the decay pipe of these mesons when they decay, their final states can act as backgrounds to new physics signatures described in the previous sections. Quality vertex resolution for decay signatures (e.g. $\gamma\gamma$, $e^+ e^-$) will be very important for discriminating \textit{short-lived} particles decaying within the target from those decaying within the decay pipe. This also presents an opportunity to reconstruct mass resonances of promptly decaying mesonic states, both for the purposes of physics validation as well as studies of rarer decay channels predicted within the SM.

An example is the production of neutral $\pi^0$ from the beam, in addition to possible production of $\eta$, and $J/\psi$ in the beam target, %
depending on the beam energy. Their decays to $\gamma\gamma$ and $e^+ e^-$ final states will serve as benchmark decays to test the function and calibration of the detectors and selection efficiency to demonstrate the sensitivity to BSM particle searches with similar final states, and can complement searches for invisible final states~\cite{Schuster:2021mlr}. The lifetimes of these particles are incredibly short, so they need to be produced close to the end of the target geometry so that their decay products can escape into the decay pipe without absorption or significant energy loss before detection. %

\paragraph{Physics Summary}

{
Table~\ref{tab:signals} for Phases I and II, using the 8\,GeV electron beam at
SLAC's LESA facility.
At LESA dark sector particles are produced primarily through electromagnetic
processes (Primakoff, bremsstrahlung, and photon-induced reactions in the EM shower).
}

\definecolor{promptcol}{RGB}{70,130,210}
\definecolor{beamcol}{RGB}{220,120,0}
\definecolor{delaycol}{RGB}{150,80,150}
\definecolor{neglcol}{RGB}{140,140,140}
\definecolor{signalcol}{RGB}{200,20,50}
\definecolor{promptbg}{RGB}{230,240,255}
\definecolor{beambg}{RGB}{255,240,220}
\definecolor{delaybg}{RGB}{245,230,245}
\definecolor{neglbg}{RGB}{240,240,240}
\definecolor{signalbg}{RGB}{255,235,235}

\renewcommand{\arraystretch}{1.25}
\setlength{\tabcolsep}{3pt}
\begin{xltabular}{\textwidth}{P{1.4cm} C{1.0cm} P{3.0cm} C{1.4cm} C{2.0cm} X}
\caption{%
{Summary of dark sector signal channels for the DAMSA Phases I and II at SLAC's LESA (8\,GeV $e^-$ beam on tungsten). The travel distance column gives the characteristic distance the dark sector particle must travel from its production point to its decay or interaction point. The vacuum decay chamber is 30\,cm long; the total target-to-ECal baseline is ${\sim}\,1$\,m.}}
\label{tab:signals} \\
\toprule
\textbf{Physics Channel}
  & \textbf{Dark Sector Particle}
  & \textbf{Production Mechanism}
  & \textbf{Final State}
  & \textbf{Travel Distance}
  & \textbf{Appearance Mechanism} \\
\midrule
\endfirsthead
\toprule
\textbf{Physics Channel}
  & \textbf{Dark Sector Particle}
  & \textbf{Production Mechanism}
  & \textbf{Final State}
  & \textbf{Travel Distance}
  & \textbf{Appearance Mechanism} \\
\midrule
\endhead
\bottomrule
\multicolumn{6}{r}{\scriptsize\textit{Continued on next page}} \\
\endfoot
\bottomrule
\endlastfoot

ALP $\to\gamma\gamma$ (photon coupling)
  & $a$
  & Primakoff effect: $\gamma + Z \to a + Z$,
  including secondary photons throughout the EM cascade in the thick target. 
  & $\gamma\gamma$
  & $\mathcal{O}(\text{cm})$--1\ m (target exit to ECal back face)
  & ALP decays to two photons in vacuum decay chamber; displaced vertex reconstructed using ECal \\
\midrule

ALP $\to e^+e^-$ (electron coupling)
  & $a$
  & $e^-$ bremsstrahlung ($e^-Z\!\to\! e^-Za$); resonant and non-resonant $e^+e^-$ annihilation ($e^+e^-\!\to\! a$, $e^+e^-\!\to\! a\gamma$); Compton scattering ($e^-\gamma\!\to\! a\,e^-$)
  & $e^+e^-$
  & $\mathcal{O}(\text{cm})$--1\ m (target exit to ECal back face)
  & ALP decays to $e^+e^-$ pair in vacuum decay chamber; pair reconstructed in tracker \\
\midrule

Dark Photon
  & $A'$
  & Dark bremsstrahlung via kinetic mixing ($e^-Z\!\to\! e^-ZA'$); $\gamma$--$A'$ mixing;
  & $e^+e^-$, $\mu^+\mu^-$, hadrons
  & $\mathcal{O}(\text{cm})$--1\ m (target exit to ECal back face)
  & $A'$ decays visibly in vacuum chamber; di-lepton final states dominate; hadronic modes at higher mass \\
\midrule

Large Extra Dimensions (KK graviton)
  & $G$
  & Primakoff-like: on-shell $\gamma N \!\to\! GN$
  & $\gamma\gamma$, $e^+e^-$
  & $\mathcal{O}(\text{cm})$--1\ m (target exit to ECal back face)
  & Spin-2 KK graviton decays in vacuum chamber; $\gamma\gamma$ dominant, $e^+e^-$ subdominant \\
\midrule

LDM elastic scattering
  & $\chi$
  & $A'\!\to\!\chi\bar{\chi}$ (from dark brems.\ or 
  any mechanisms producing $A'$ in target)
  & recoil $e^-$, $N$, or nucleus
  & $\sim$\,30--100\,cm (target to detector)
  & $\chi$ scatters elastically off electrons, nucleons, or nuclei in detector, producing visible recoil \\
\midrule

LDM inelastic scattering
  & $\chi,\,\chi^*$
  & $A'\!\to\!\chi\chi^*$ or $\chi$ up-scatters to $\chi^*$ in detector
  & $e^+e^-$ or $\gamma$ from $\chi^*$ decay
  & $\sim$\,30--100\,cm (target to detector)
  & Excited state $\chi^*$ de-excites: $\chi^*\!\to\!\chi\,e^+e^-$ or $\chi^*\!\to\!\chi\,\gamma$; secondary EM/hadronic activity in detector \\

\end{xltabular}
\normalsize

{Figures~\ref{fig:backgrounds} provides overview
of the dominant Standard Model background processes for the DAMSA Phases I and II at LESA.
For the Phases I and II DAMSA detector at electron-beam (Fig.~\ref{fig:backgrounds}), the primary challenge is suppressing beam-related neutrons (BRN) produced by photo-nuclear reactions in the EM shower. 
The rates are shown relative to the EM shower photon flux escaping the target.
The normalizations reflects the dominant backgrounds in absolute rates per electron-on-target (EOT)}

\begin{figure}[p]
\centering
\begin{tikzpicture}[
    >=Stealth,
    processname/.style={font=\footnotesize\bfseries},
    processdesc/.style={font=\scriptsize},
    ratelabel/.style={font=\scriptsize\bfseries, anchor=east},
    plabel/.style={font=\scriptsize},
    catlabel/.style={font=\scriptsize\itshape, text width=1.6cm, align=center},
  ]
  \def\xrate{-0.2}  \def\xinc{0.7}  \def\xaL{1.7}  \def\xaR{4.8}
  \def\xout{5.0}  \def\xproc{7.5}  \def\xcat{13.0}
  \def\yA{0}  \def\yB{-1.8}  \def\yC{-3.4}  \def\yD{-5.0}
  \def\yE{-6.8}  \def\yF{-8.4}  \def\yG{-10.0}
  \def\ySA{-11.8}  \def\ySB{-13.2}
  \node[font=\scriptsize\bfseries] at (\xrate-0.3, 1.0) {Relative};
  \node[font=\scriptsize\bfseries] at (\xrate-0.3, 0.7) {Rate$^*$};
  \node[font=\scriptsize\bfseries] at (\xinc, 0.85)  {In};
  \node[font=\scriptsize\bfseries] at (\xout+0.8, 0.85) {Out};
  \node[font=\scriptsize\bfseries] at (\xproc+1.5, 0.85) {Process};
  \draw[thick] (-1.2, 0.5) -- (12.2, 0.5);
  \begin{scope}[on background layer] \fill[promptbg, rounded corners=2pt] (-1.2, 0.45) rectangle (12.2, -0.85); \end{scope}
  \node[ratelabel] at (\xrate, \yA) {$10^{0}$}; \node[plabel] at (\xinc, \yA) {$e^-$};
  \draw[->, thick, promptcol] (\xaL, \yA) -- (\xaR, \yA);
  \node[plabel, anchor=west] at (\xout, \yA) {$\gamma,\; e^\pm$};
  \node[processname, anchor=west, promptcol] at (\xproc, \yA) {EM shower leakage};
  \node[processdesc, anchor=west] at (\xproc, \yA-0.35) {prompt $\gamma/e^\pm$ escaping target};
  \node[catlabel, promptcol, anchor=west] at (\xcat, \yA-0.2) {prompt BG\\(geometry,\\timing)};
  \begin{scope}[on background layer] \fill[beambg, rounded corners=2pt] (-1.2, -1.1) rectangle (12.2, -5.9); \end{scope}
  \node[ratelabel] at (\xrate, \yB) {$10^{-1}$}; \node[plabel] at (\xinc, \yB) {$e^-$};
  \draw[->, thick, beamcol] (\xaL, \yB) -- (\xaR, \yB);
  \node[plabel, anchor=west] at (\xout, \yB) {$n$ (fast)};
  \node[processname, anchor=west, beamcol] at (\xproc, \yB) {Fast beam-related neutrons};
  \node[processdesc, anchor=west] at (\xproc, \yB-0.35) {photo-nuclear: $\gamma\, W \to n + X$};
  \node[ratelabel] at (\xrate, \yC) {$10^{-2}$}; \node[plabel] at (\xinc, \yC) {$e^-$};
  \draw[->, thick, beamcol] (\xaL, \yC) -- (\xaR, \yC);
  \node[plabel, anchor=west] at (\xout, \yC) {$\pi^\pm,\; p$};
  \node[processname, anchor=west, beamcol] at (\xproc, \yC) {Charged hadron escape};
  \node[processdesc, anchor=west] at (\xproc, \yC-0.35) {hadronic shower products from target};
  \node[ratelabel] at (\xrate, \yD) {$10^{-3}$}; \node[plabel] at (\xinc, \yD) {$e^-$};
  \draw[->, thick, beamcol] (\xaL, \yD) -- (\xaR, \yD);
  \node[plabel, anchor=west] at (\xout, \yD) {$\gamma\gamma$};
  \node[processname, anchor=west, beamcol] at (\xproc, \yD) {Neutral meson $\boldsymbol{\to\gamma\gamma}$};
  \node[processdesc, anchor=west] at (\xproc, \yD-0.4) {$\pi^0\!/\eta$ at target exit};
  \node[processdesc, anchor=west] at (\xproc, \yD-0.7) {decay $\gamma$'s escape into decay pipe};
  \node[catlabel, beamcol, anchor=west] at (\xcat, -3.5) {beam-induced\\BG (primary\\challenge)};
  \begin{scope}[on background layer] \fill[delaybg, rounded corners=2pt] (-1.2, -6.1) rectangle (12.2, -9.25); \end{scope}
  \node[ratelabel] at (\xrate, \yE) {$10^{-5}$}; \node[plabel] at (\xinc, \yE) {$n$};
  \draw[->, thick, delaycol] (\xaL, \yE) -- (\xaR, \yE);
  \node[plabel, anchor=west] at (\xout, \yE) {$\gamma$ cascade};
  \node[processname, anchor=west, delaycol] at (\xproc, \yE) {$n$-capture $\gamma$ cascades};
  \node[processdesc, anchor=west] at (\xproc, \yE-0.4) {$n{+}\text{material}{\to}\gamma$ (2--11\,MeV)};
  \node[processdesc, anchor=west] at (\xproc, \yE-0.7) {delayed several 100\,ns};
  \node[ratelabel] at (\xrate, \yF) {$10^{-7}$}; \node[plabel] at (\xinc, \yF) {cosmic};
  \draw[->, thick, delaycol] (\xaL, \yF) -- (\xaR, \yF);
  \node[plabel, anchor=west] at (\xout, \yF) {$n$};
  \node[processname, anchor=west, delaycol] at (\xproc, \yF) {Environmental neutrons};
  \node[processdesc, anchor=west] at (\xproc, \yF-0.4) {cosmic-ray induced};
  \node[processdesc, anchor=west] at (\xproc, \yF-0.7) {${\sim}10^{-4}$--$10^{-2}$\,n/cm$^2$/s};
  \node[catlabel, delaycol, anchor=west] at (\xcat, -7.6) {delayed /\\environ.\ BG\\(shielding,\\beam timing)};
  \begin{scope}[on background layer] \fill[neglbg, rounded corners=2pt] (-1.2, -9.35) rectangle (12.2, -10.85); \end{scope}
  \node[ratelabel] at (\xrate, \yG) {$<\!10^{-10}$}; \node[plabel] at (\xinc, \yG) {$\pi/K$};
  \draw[->, thick, neglcol, dashed] (\xaL, \yG) -- (\xaR, \yG);
  \node[plabel, anchor=west] at (\xout, \yG) {$\nu$ (no vis.)};
  \node[processname, anchor=west, neglcol] at (\xproc, \yG) {Neutrinos};
  \node[processdesc, anchor=west] at (\xproc, \yG-0.35) {effectively invisible; negligible};
  \node[catlabel, neglcol, anchor=west] at (\xcat, \yG-0.15) {negligible};
  \draw[thick, dashed, gray!70] (-1.2, -11.1) -- (12.2, -11.1);
  \node[font=\scriptsize\bfseries, signalcol] at (5.7, -11.35) {--- Signal processes ---};
  \begin{scope}[on background layer] \fill[signalbg, rounded corners=2pt] (-1.2, -11.55) rectangle (12.2, -13.85); \end{scope}
  \node[ratelabel, signalcol] at (\xrate, \ySA) {${\sim}10^{-15}$}; \node[plabel, signalcol] at (\xinc, \ySA) {$e^-$};
  \draw[->, thick, signalcol, densely dotted] (\xaL, \ySA) -- node[above, font=\tiny, signalcol] {$A'\!/a/G$} (\xaR, \ySA);
  \node[plabel, anchor=west, signalcol] at (\xout, \ySA) {$\gamma\gamma$ / $e^+e^-$};
  \node[processname, anchor=west, signalcol] at (\xproc, \ySA) {Dark sector decay};
  \node[processdesc, anchor=west] at (\xproc, \ySA-0.4) {$A'/a/G$ produced in target};
  \node[processdesc, anchor=west] at (\xproc, \ySA-0.7) {decays in vacuum chamber};
  \node[plabel, signalcol] at (\xinc, \ySB) {$e^-$};
  \draw[->, thick, signalcol, densely dotted] (\xaL, \ySB) -- node[above, font=\tiny, signalcol] {$\chi$} (\xaR, \ySB);
  \node[plabel, anchor=west, signalcol] at (\xout, \ySB) {recoil / $e^+e^-$};
  \node[processname, anchor=west, signalcol] at (\xproc, \ySB) {LDM scattering};
  \node[processdesc, anchor=west] at (\xproc, \ySB-0.35) {$\chi$ elastic/inelastic scattering in detector};
\end{tikzpicture}
\caption{%
  Relative rates of Standard Model background processes for DAMSA Phases I and II at SLAC's LESA (8\,GeV $e^-$ beam on tungsten). Backgrounds are grouped into prompt electromagnetic (blue), beam-induced hadronic/neutron (orange), and delayed/environmental (purple). Signal processes are shown below the dashed line.            
    $^*$Relative rates are approximate and normalized to the EM shower photon flux escaping the target (dimensionless). The signal rate is shown only schematically to indicate that dark sector production is many orders of magnitude rarer than SM backgrounds; absolute rates depend strongly on the dark sector particle's mass and coupling, with sensitivity contours for Phase I (ALP $\to\gamma\gamma$) shown in Fig.~\ref{fig:alpsensitivity}.}  
\label{fig:backgrounds}
\end{figure}

\definecolor{promptcol}{RGB}{70,130,210}
\definecolor{beamcol}{RGB}{220,120,0}
\definecolor{delaycol}{RGB}{150,80,150}
\definecolor{neglcol}{RGB}{140,140,140}
\definecolor{signalcol}{RGB}{200,20,50}
\definecolor{promptbg}{RGB}{230,240,255}
\definecolor{beambg}{RGB}{255,240,220}
\definecolor{delaybg}{RGB}{245,230,245}
\definecolor{neglbg}{RGB}{240,240,240}
\definecolor{signalbg}{RGB}{255,235,235}

\vspace{0.5em}
\section{Experiment and The Detector}      
\label{sec:Detector}
\begin{figure}[htb]
    \centering
    \includegraphics[scale=0.119]{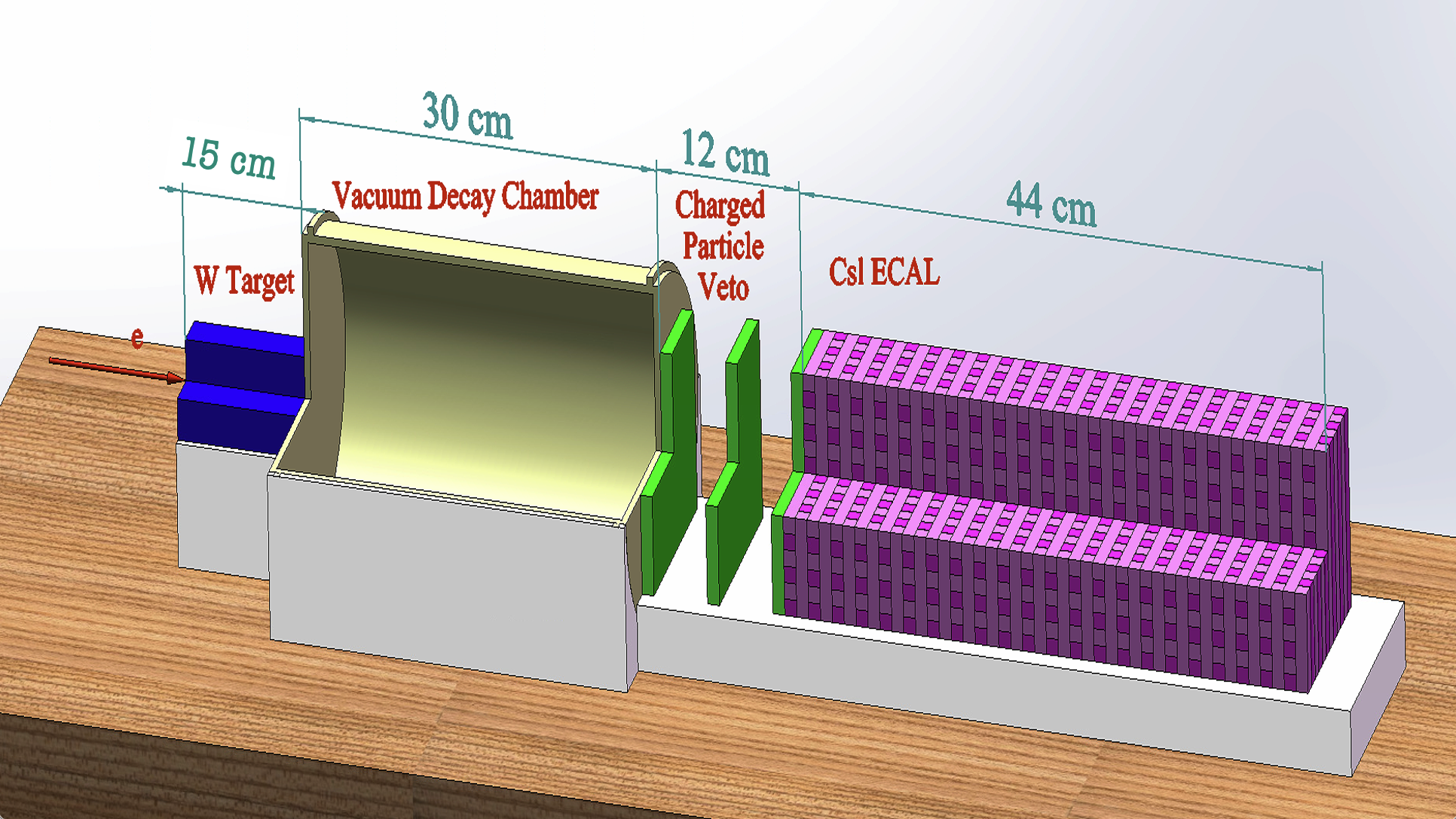}     \includegraphics[width=0.495\linewidth]{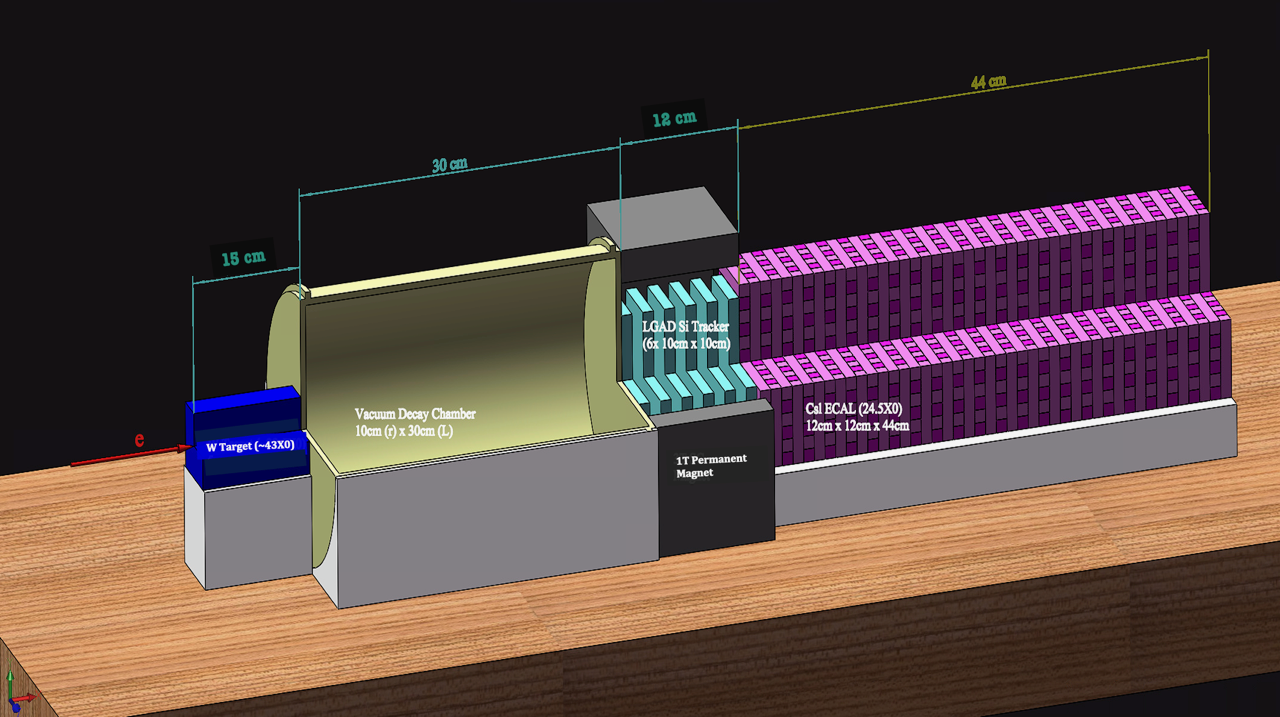}
    \caption{
        (Left) DAMSA Stage I which consists of the 15~cm tungsten target, the vacuum decay chamber, a three scintillator charged particle ID system and the 4D total absorption ECal;
        (Right) Final version of DAMSA with magnet and 6 LGAD Si tracking planes.
    }
    \label{fig:stage1}
\end{figure}

DAMSA~\cite{Jang:2022tsp} is a very short baseline beam dump experiment, aiming to probe the parameter space inaccessible in previous beam dump experiments. Given the proximity to the beam dump/target, the primary background comes from the large number of beam-related neutrons, resulting from the beam interactions in the dump. 
Our extensive simulation-based studies~\cite{Kim:2024vxg} show that DAMSA can access the targeted parameter space with a tabletop scale experiment equipped with a vacuum decay chamber, a precision tracking detector under a magnetic field and a fine granular 4D total absorption electromagnetic calorimeter (ECal), as shown by the 3-D models in Fig.~\ref{fig:stage1}.
This section presents the DAMSA experiment components in detail.

\textcolor{black}{Table~\ref{tab:detector_summary} summarizes the key design parameters and expected performance of each detector subsystem. The total baseline from the upstream face of the tungsten target to the downstream end of the ECal is approximately
1\,m, with the active detector elements spanning from $z=45$\,cm (tracker entrance) to $z=101$\,cm (ECal exit) measured from the target face.}

\begin{table}[htb]
\centering
\footnotesize
\textcolor{black}{\caption{Summary of DAMSA detector subsystem parameters and expected performance.}\label{tab:detector_summary}}
\begin{tabularx}{\textwidth}{P{2.4cm} P{3.5cm} X}
\toprule
\textcolor{black}{\textbf{Subsystem}} & \textcolor{black}{\textbf{Key Parameters}} & \textcolor{black}{\textbf{Expected Performance}} \\
\midrule
\textcolor{black}{W Target} & \textcolor{black}{$5{\times}5{\times}15$\,cm$^3$; $42.9\,X_0$} & \textcolor{black}{Contains EM shower; produces ALP/dark photon via Primakoff and bremsstrahlung} \\
\textcolor{black}{Vacuum Chamber} & \textcolor{black}{$\varnothing$\,20\,cm $\times$ 30\,cm; $P < 10^{-3}$\,mbar} & \textcolor{black}{ALP/$A'$ decay volume; minimizes neutron scattering} \\
\textcolor{black}{Permanent Magnet} & \textcolor{black}{NdFe or SmCo (radiation-hard); $B = 0.65$\,T (with yoke); 12\,cm gap} & \textcolor{black}{$e^+/e^-$ charge separation; $\delta p/p \approx 4$\% at 100\,MeV/$c$ (measurement $\oplus$ Multiple Scattering (MS))} \\
\textcolor{black}{Tracking Detector} & \textcolor{black}{$10{\times}10{\times}12$\,cm$^3$; 4 layers baseline (LGAD or $\mu$RWELL)} & \textcolor{black}{Efficiency $>$95\% ($p>30$\,MeV/$c$); vertex $\sigma < 1$\,mm; timing 50\,ps (LGAD)} \\
\textcolor{black}{CsI ECal} & \textcolor{black}{$12{\times}12{\times}44$\,cm$^3$; ${\sim}24\,X_0$; SiPM readout} & \textcolor{black}{Targeting at $\sigma_E/E < 6\%/\sqrt{E[\text{GeV}]}$; timing ${\sim}1$\,ns; two-photon vertex reconstruction} \\
\bottomrule
\end{tabularx}
\end{table}

\subsection{The Target}
The target of DAMSA detector will be made of a block of tungsten of $\sim$\qtyproduct{5x5}{\cm} in the transverse direction and $15~{\rm cm}$ long in the beam direction, providing $42.9X_{0}$ radiation lengths.
\textcolor{black}{The transverse dimensions are chosen to fully contain the electromagnetic shower core produced by the 8\,GeV electron beam (the Moli\`ere radius of tungsten is $\sim$0.93\,cm), while minimizing the overall target mass and the associated neutron production rate.}

The detailed configuration and shape of the target will be optimized to further reduce the flux of the electromagnetic particles, such as X-rays.
For example, we may add a \SI{1}{\cm}-thick lead sheet at the end of the \SI{15}{\cm} W target to absorb a large fraction of X-rays that could be produced in the target.
We may also explore a rotating, cylindrical shape target that is positioned slightly off center with respect to the incident beam position to help further dissipate heat from the beam interactions.
This is an efficient and cost-effective alternative to rastering the beam to further disperse the heat.  This kind of target geometry may be necessary if the beam pulse intensity at a future experimental configuration increases.

\begin{figure}[htb]
\centering
\includegraphics[width=0.48\linewidth]{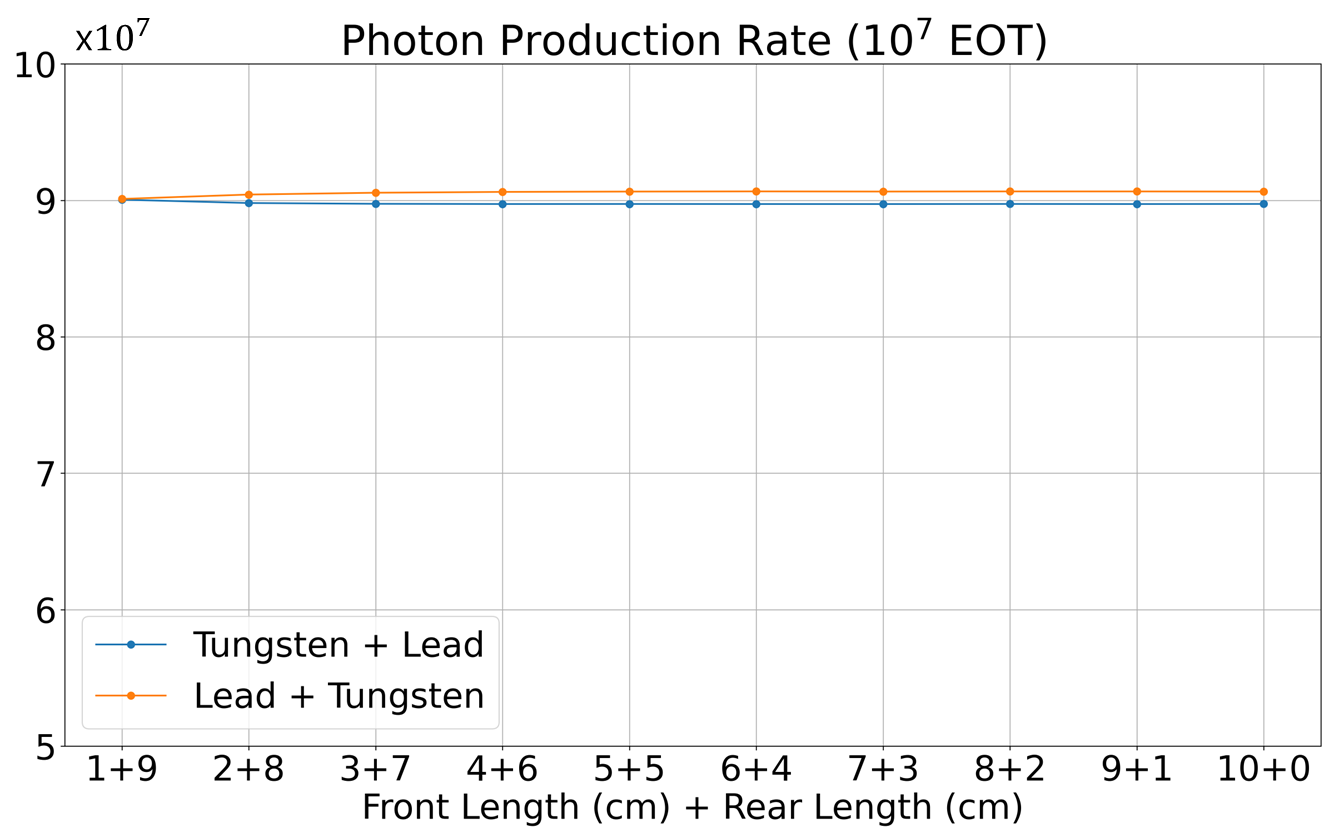}
\includegraphics[width=0.48\linewidth]{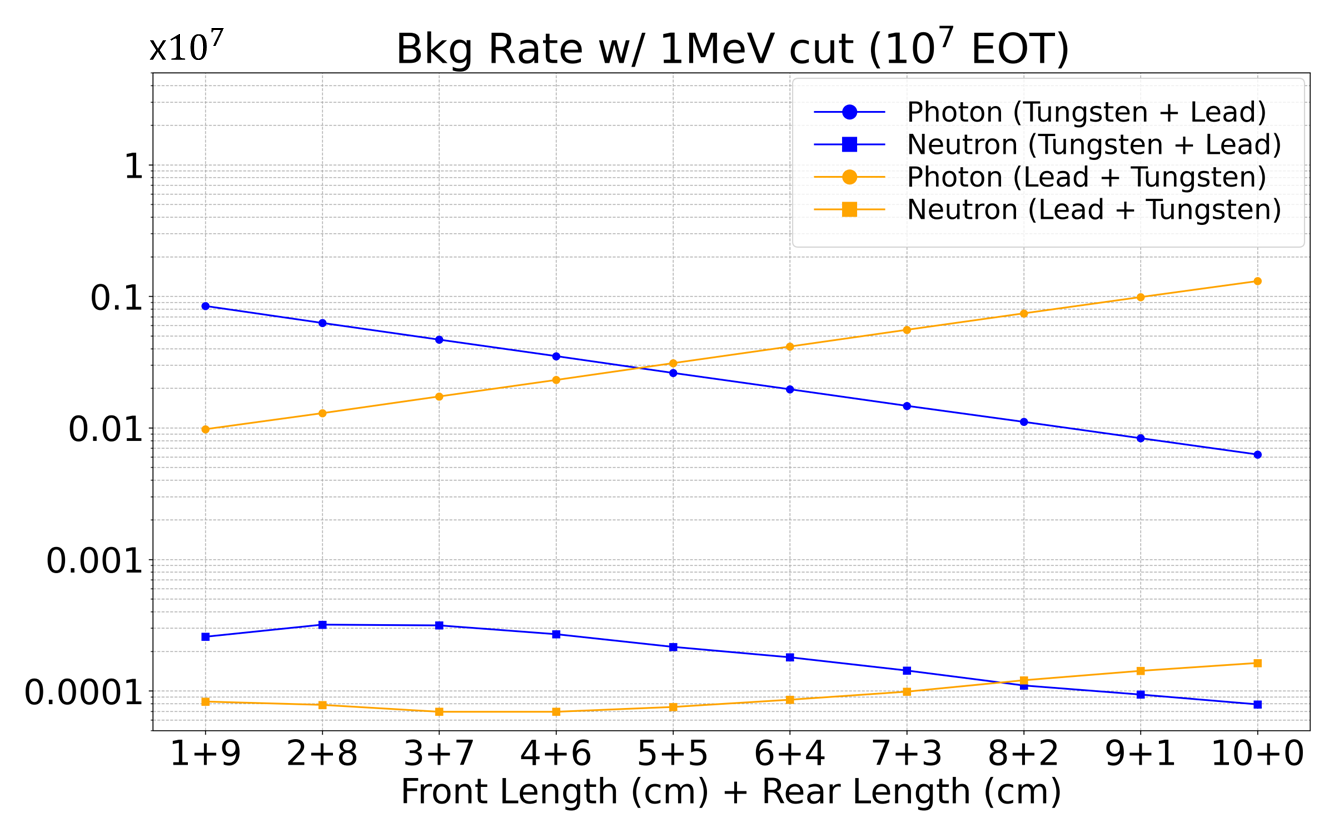}
\caption{Photon production rate (left) and background rate (right)} 
\label{fig:Target}
\end{figure}

We have studied the optimal target composition of tungsten and lead, using \textsc{GEANT4} simulations, considering both photon production and backgrounds. For photon production, we calculated the number of photons produced with energy above 10 MeV and positive P$_z$, which are potential candidates for ALP generation. As shown in the left plot of Fig.~\ref{fig:Target}, the difference is at most 1\%. For background, electrons and positrons are omitted in right plot of Fig.~\ref{fig:Target} because they exhibit the same trend as photons. 
Considering both photon production and background, we have determined that a full tungsten target is optimal when only tungsten and lead are considered for target composition. 

\subsection{The Vacuum Decay Chamber}

The vacuum decay chamber is an essential element of the DAMSA experiment.
It is located immediately downstream of the target, as shown in Fig.~\ref{fig:stage1} and, in particular, allows unstable dark-sector particles like ALPs to decay inside.
The decay vertex will then be reconstructed from the fine granular electromagnetic calorimeter, mitigating the background from random overlap of the photons from neutron interactions which require materials to interact.
The diameter of the vacuum chamber is 20~cm to provide sufficient angular coverage to capture a large fraction of the unstable particles produced in the target.
The length of the chamber is 30~cm to allow sufficient separation between the two decay products in the ECal downstream.
The small size of the vacuum decay chamber makes fabrication easy and cost effective.
We plan to utilize a commercial vendor for the fabrication.
A chamber with convex heads and side pump out is the baseline.
\textcolor{black}{The chamber vessel will be fabricated from stainless steel to minimize out-gassing and provide compatibility with the high-radiation environment.}
Maintaining a high level of vacuum is important in minimizing neutron interactions inside the vacuum volume.
\textcolor{black}{The baseline target pressure is $< 10^{-3}$\,mbar, which ensures a neutron nuclear interaction mean free path far exceeding the 30\,cm chamber length (the nuclear scattering probability in residual gas at this pressure is $< 10^{-8}$ per traversal), thus rendering in-vacuum neutron scattering negligible. Turbo-molecular pumping with a dry roughing stage is foreseen. The small volume ($\sim$9.4\,litres) allows rapid pump-down.}
One of the outcomes of DAMSA Stage 1 is to measure the vacuum level dependence of the backgrounds and determine the optimal level of vacuum for the subsequent data taking.

\subsection{The Magnet}\label{sec:magnet}
The magnet provides the dipole field for the tracking volume located between the vacuum decay chamber and the electromagnetic calorimeter. This field enables the separation of charged particles from photons, determination of the charge sign, and momentum reconstruction of charged tracks. Owing to its compact size and maintenance-free operation, a permanent magnet is preferred over an electromagnet. Among the permanent-magnet options, NdFe magnets provide the highest field strength, while SmCo magnets offer improved radiation tolerance.
The magnetized region is constrained to a \SI{12}{\cm}-long tracking volume. In the present design, two NdFe magnet blocks of dimensions \SI{30}{\cm}\,$\times$\,\SI{25}{\cm}\,$\times$\,\SI{12}{\cm} are separated by a \SI{12}{\cm} gap, giving a maximum central field of \SI{0.55}{\tesla}. With an added yoke structure, the central field increases to \SI{0.65}{\tesla}.

\textcolor{black}{The baseline design adopts the yoke configuration with $B = 0.65$\,T. In this field, a charged particle with momentum $p$ receives a sagitta $s \approx 0.3\,B\,L^2/(8p) \approx 3.5~\text{mm}\times(100~\text{MeV}/c)/p$ over the 12\,cm tracking length $L$.
With four LGAD layers providing $\sim$0.1\,mm position resolution per hit, the Gluckstern formula gives a measurement contribution to the momentum resolution of $(\delta p/p)_{\text{meas}} \approx 3.4\%$ at $p=100$\,MeV/$c$.
Multiple scattering (MS) in four LGAD layers ($\sim$0.3\%\,$X_0$ each) contributes an additional $(\delta p/p)_{\text{MS}} \approx 2\%$ at 100\,MeV/$c$, yielding a combined resolution of
$\delta p/p \approx 4\%$ at $p=100$\,MeV/$c$, scaling approximately linearly with $p$ for the measurement term.
This is sufficient for $e^+/e^-$ charge identification and invariant mass reconstruction of ALP and dark photon decays in the 10--500\,MeV range.}

\subsection{The Tracking Detector}

The tracking detector is placed immediately downstream of the vacuum decay chamber and upstream of the electromagnetic calorimeter. Its main role is to reconstruct charged-particle tracks, in particular $e^{+}e^{-}$ pairs, and to separate them from photon-induced activity in the calorimeter. The baseline tracking technology is the Low Gain Avalanche Detector (LGAD) silicon pixel sensor array developed for the CMS timing detector upgrade, while a micro-resistive well ($\mu$RWELL) detector is also under consideration as an alternative option, especially for the early stage of the experiment, given the anticipated completion timing of the LGAD detectors.

The tracker covers a transverse area of \SI{10}{\cm}\,$\times$\,\SI{10}{\cm} over a length of \SI{12}{\cm} along the beam direction and operates in the dipole field described in Section~\ref{sec:magnet}, with a central field of \SIrange{0.55}{0.65}{\tesla} depending on the magnet-yoke configuration. 
\textcolor{black}{The tracker will have 4 layers along the beam line as the baseline configuration, with the possibility of expanding to 6 layers if space permits.}
The performance goals are a tracking efficiency above 95\% for momenta above $30\,\mathrm{MeV}/c$ and a vertex resolution better than \SI{1}{\mm} for short-lived particles decaying to an $e^{+}e^{-}$ pair. The expected timing resolution is about \SI{50}{\pico\second} for the LGAD option and about \SI{5}{\nano\second} for the $\mu$RWELL option.

\textcolor{black}{Table~\ref{tab:tracker_comparison} compares the two tracking technology options. The baseline plan is to deploy $\mu$RWELL detectors for the initial DAMSA Stage 1 run, taking advantage of their rapid fabrication, low cost, and adequate spatial resolution, while the LGAD option is pursued in parallel for the full DAMSA detector, where
sub-100\,ps timing is critical for rejecting BRN-induced backgrounds.}
\begin{table}[htb]
\centering
\footnotesize
\textcolor{black}{\caption{Comparison of tracking technology options for DAMSA.}\label{tab:tracker_comparison}}
\begin{tabularx}{\textwidth}{P{2.8cm} X X}
\toprule
\textcolor{black}{\textbf{Parameter}} & \textcolor{black}{\textbf{LGAD Si Pixel}} & \textcolor{black}{\textbf{$\mu$RWELL}} \\
\midrule
\textcolor{black}{Position resolution} & \textcolor{black}{$\sim$0.4\,mm (pad pitch/$\sqrt{12}$)} & \textcolor{black}{60--70\,$\mu$m (capacitive sharing)} \\
\textcolor{black}{Timing resolution} & \textcolor{black}{30--50\,ps} & \textcolor{black}{$\sim$5\,ns} \\
\textcolor{black}{Material budget / layer} & \textcolor{black}{$\sim$0.3\% $X_0$} & \textcolor{black}{$<$0.1\% $X_0$} \\
\textcolor{black}{Channels (4 layers)} & \textcolor{black}{$\sim$4$\times$256 = 1024} & \textcolor{black}{$\sim$4$\times$256 = 1024} \\
\textcolor{black}{Readout} & \textcolor{black}{Bump-bonded ASIC} & \textcolor{black}{VMM/SRS} \\
\textcolor{black}{Radiation tolerance} & \textcolor{black}{Moderate (gain degradation)} & \textcolor{black}{High (gas detector)} \\
\textcolor{black}{Maturity for DAMSA} & \textcolor{black}{R\&D ongoing} & \textcolor{black}{Ready for Stage 1} \\
\bottomrule
\end{tabularx}
\end{table}

\subsubsection{LGAD Si Pixel Detector}

The baseline tracking detector DAMSA is the silicon tracker, designed to identify the electron and positron originating from the decay of an ALP or dark photon. The silicon tracker will be installed in \textcolor{black}{4 layers as the baseline configuration. The total amount of space required for two-pixel layers along the z-axis for the detector is about 45 mm.}
An additional \SI{20}{\mm} of space on each side of the silicon tracker is required for the thermal screen.
Each pixel layer will use LGAD sensors being developed for the CMS experiment, which enable precision timing and position reconstruction for charged particles. The sensors, with a \SI{50}{\micro\meter} active region in a standard \SI{300}{\micro\meter}-thick silicon wafer and a thin implanted gain layer, are expected to provide the desired performance. Previous studies have demonstrated a timing resolution of about \SIrange{30}{50}{\pico\second}.
The LGAD sensor is a \numproduct{16x16} pixel array consisting of square pads, with each pixel measuring \qtyproduct[product-units=single]{1.3x1.3}{\square\mm}, resulting in a total sensor size of \qtyproduct[product-units=single]{21x21}{\square\mm}. Each module is read out by an ASIC of approximately \qtyproduct[product-units=single]{20x20}{\square\mm}, which processes signals from a \numproduct{16x16} submatrix. The sensor size is larger than the chip size to allow the ASIC to be bump-bonded to it for electrical connection.
A set of readout chips interfaces with an on-detector board, known as the service hybrid, which supplies DC power, bias voltage, communication for both slow and fast control, and monitoring for the ASICs. The service hybrid consists of two components: a readout board and a power board. The pixel layer will be operated with a cooling system to provide temperatures of at least \SI{-30}{\celsius}.
\newline

\subsubsection {Optional Tracking Detector Technology}

\begin{figure}[htb]
    \centering
    \includegraphics[scale=0.40]{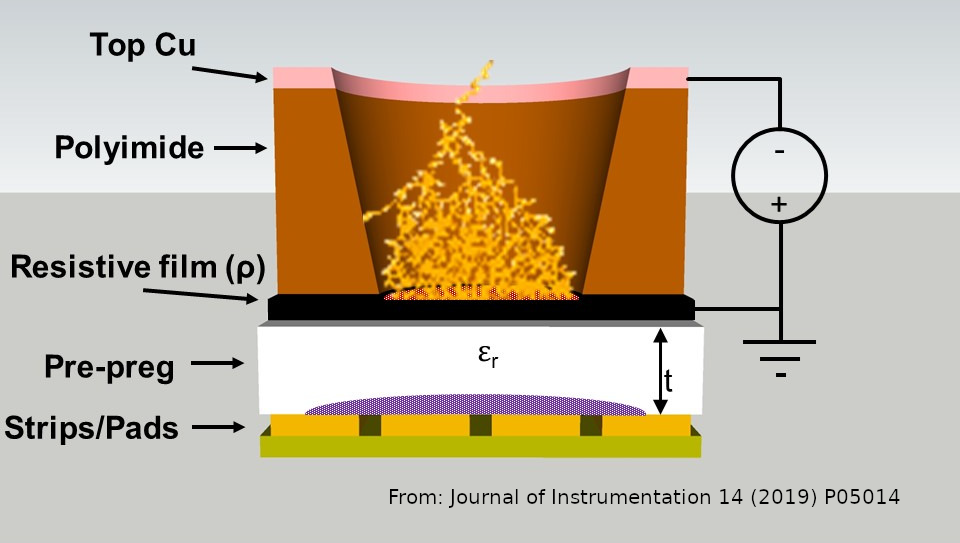}
    \caption{
          Structure of the $\mu$RWELL unit cell.
          {Copied from Ref.~\cite{Bencivenni_2019}. Figure reused under CC BY 3.0 license}
    }
    \label{fig:detector:uRWELL_Unit_Cell}
\end{figure}

Another technology to consider is $\mu$RWELL, the resistive variant of gas electron multiplier (GEM) \cite{POLILENER2016565}.
Figure~\ref{fig:detector:uRWELL_Unit_Cell} shows the unit cell structure of $\mu$RWELL.
The strong electric field provided by the GEM microelectrodes amplifies electrons generated by primary ionization, producing a measurable signal.
The resistive film composed of diamond-like carbon (DLC) suppresses discharges by inducing local voltage drop when  streamers form.
Through this mechanism, $\mu$RWELL suppresses discharge occurrence, and even if a discharge occurs, the discharge current is highly quenched.
Therefore, a higher voltage can be applied to the GEM foil, enabling a single GEM foil to achieve a high gain.

Such discharge suppression offers various advantages.
First, it mitigates channel loss in the readout electronics caused by discharges.
Additionally, it eliminates the need for spatial separation between the avalanche region and the induction region.
The GEM foil can be directly stacked onto the readout PCB.
As a result, the GEM foil obtains self rigidity, eliminating the need for foil stretching.
This enables a simpler detector structure and significantly simplifies the assembly process.
This makes the $\mu$RWELL detector more cost-effective.

Depending on the readout configuration, $\mu$RWELL can achieve an excellent position resolution of \SI{70}{\micro\meter} or slightly better~\cite{POLILENER2016565}.
The time resolution depends on the drift-gap size, gas mixture, and
front-end readout electronics, and values of about \SI{6}{\nano\second}
have been reported for $\mu$RWELL detectors~\cite{MORELLO2019497}.
Other advantages of the $\mu$RWELL detector in this experiment include its thin structure, allowing it to fit within limited space, and its low material budget.

Additionally, capacitive sharing readout is being considered to keep the number of readout channels low.
This method allows signal sharing between adjacent readout channels through capacitive coupling, ensuring high strip multiplicity even with a large strip pitch \cite{GNANVO2023167782}.
As a result, it enables high position resolution while maintaining a reduced number of readout channels. 
By using the capacitive sharing readout with an XY strip structure, a $\SI{10}{\centi\meter}\times\SI{10}{\centi\meter}$ area can be covered with only 2$\times$128 channels, achieving a position resolution of around \SI{60}{\micro\meter}.

For data acquisition (DAQ) for this system, we are considering VMM front-end chips~\cite{6551184} with
the scalable readout system (SRS)~\cite{LUPBERGER201891}, a widely used
architecture for micro-pattern gaseous detector readout.
With four $\mu$RWELL detectors using capacitive-sharing readout (\(2\times128\)
channels per detector), eight VMM hybrids can be operated using a single
adapter card.

One concern is the highly constrained space imposed by the magnet.
Due to this limitation, the commonly used readout PCB for $\SI{10}{\centi\meter}\times\SI{10}{\centi\meter}$ detectors cannot be utilized.
Instead, a compact readout PCB must be designed and used, minimizing space for signal routing and services.
Additionally, it is important to find a balance between minimizing potential damage to the VMM hybrids from the expected high neutron flux and the space constraints that limit their optimal placement for radiation protection.

\subsection{The Total Absorption 4D Electromagnetic Calorimeter}
\begin{figure}[htb]
    \centering
    \begin{minipage}{0.37\linewidth}
    \includegraphics[width=\linewidth]{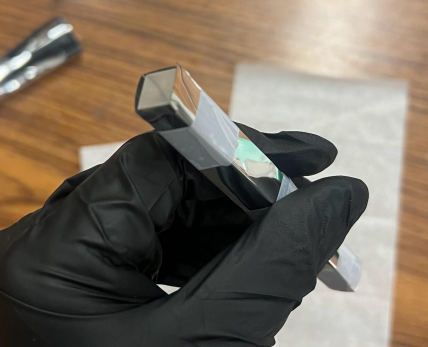}
    \end{minipage}
    \begin{minipage}{0.62\linewidth}
    \includegraphics[width=\linewidth]{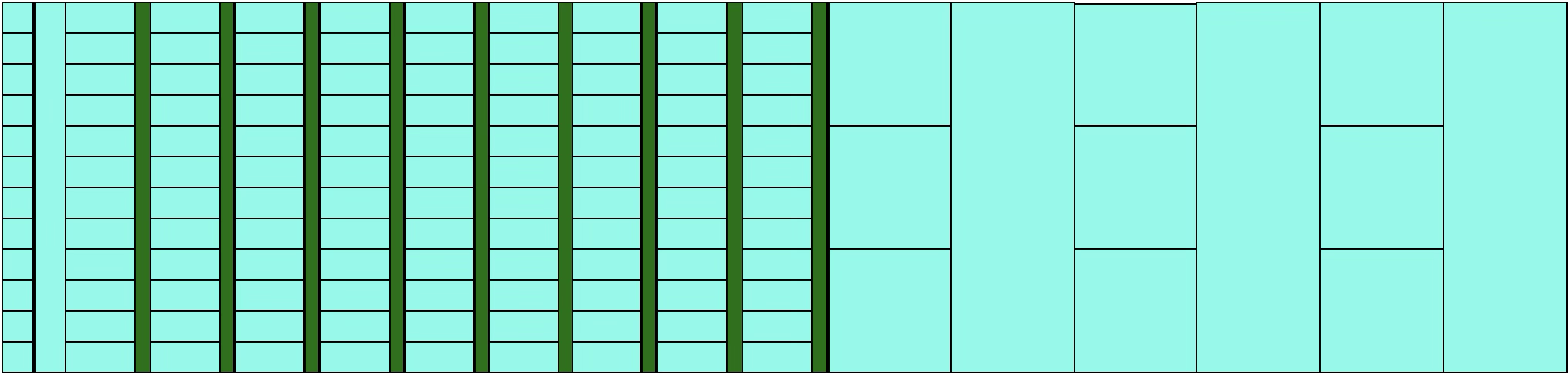}
    \includegraphics[width=\linewidth]{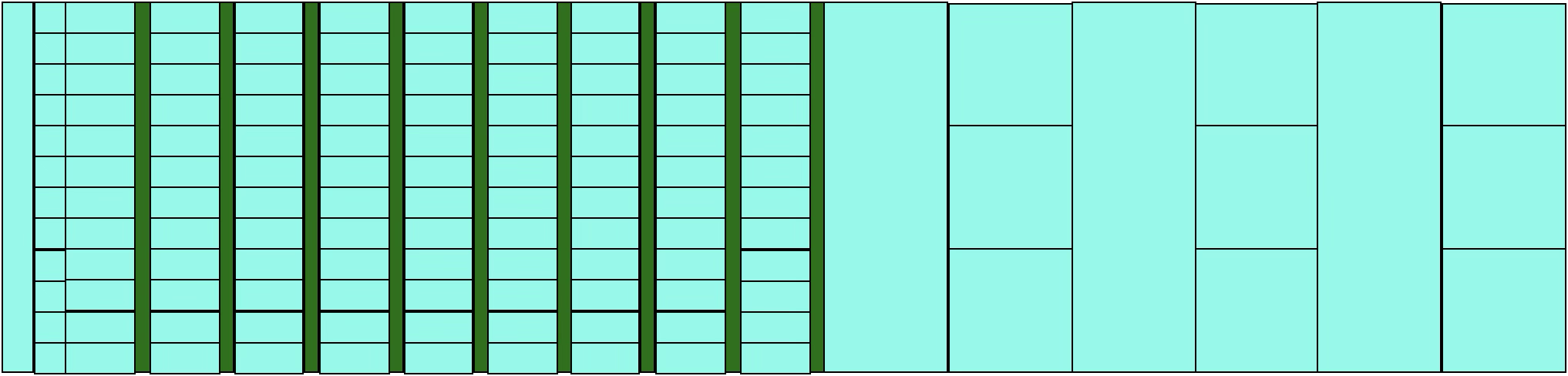}
    \end{minipage}
    \caption{(Left) Photograph of a CsI crystal wrapped in aluminized mylar. (Right) 3D ECal layout. (Top) An $x-z$ side view and (Bottom) $y-z$ side view). The first two layers consist of $1\times1\times12~\rm{cm}^3$ CsI crystal bars stacked along the x and y directions. The next nine layers are cubic \numproduct{12x12} array composed of $1\times1\times2~\rm{cm}^3$ CsI bars. The final six layers are $4\times4\times12~\rm{cm}^3$ CsI bars in the same configuration as the first two layers.}
    \label{fig:ecal_design}
\end{figure}

DAMSA will use a total absorption electromagnetic calorimeter comprised of an array of undoped CsI scintillating crystals, each viewed from both ends by silicon photomultiplier (SiPM) photo-detectors with nanosecond timing resolution mounted on a PCB, as the baseline ECal.
This baseline detector design and its geometry are illustrated in Fig.~\ref{fig:ecal_design}.
Each crystal in the CsI array will be \qtyproduct[product-units=single]{1x1x12}{\cubic\cm} and wrapped in a reflector to increase light collection efficiency.
Undoped CsI scintillates at \SI{310}{\nm} with a \SI{\sim10}{\ns} time constant at room temperature and is radiation hard.
Aluminized mylar, 3M$^\mathrm{TM}$ Enhanced Specular Reflector (ESR), among other materials, are being considered for the reflector.
\textcolor{black}{An important consideration is that ESR reflectivity drops sharply below $\sim$395\,nm and the film becomes absorbing in the deep UV.  Since undoped CsI emits at $\sim$310\,nm, direct use of ESR without wavelength shifting would result in significant light loss.  To address this, wavelength-shifting (WLS) films or coatings that convert the 310\,nm scintillation light to visible wavelengths ($>$400\,nm) before reflection are being actively evaluated.  Candidate solutions include thin-film WLS coatings applied to the crystal surface or to the inner face of the ESR wrapping.  Aluminized mylar, which maintains high reflectivity into the UV, serves as the conservative baseline reflector option pending WLS optimisation.  The ongoing prototype tests at BNL, UC Riverside, Kyungpook National University, and Seoul National University (Section~3.5.1) are specifically designed to quantify the light collection efficiency for each reflector--WLS combination and determine the optimal configuration.}
\textcolor{black}{The total depth of the ECal is 44\,cm, corresponding to approximately
$24\,X_0$ in CsI ($X_0^{\text{CsI}} = 1.86$\,cm). This depth provides $>$99\%
containment of electromagnetic showers initiated by photons up to $\sim$1\,GeV;
at higher photon energies (e.g.\ 5\,GeV), longitudinal leakage increases to
$\sim$2\%, which will be accounted for in the energy calibration.}

\textcolor{black}{
For comparison, the KOTO experiment achieves $1.81\%/\sqrt{E} \oplus 0.66\%$ using large $7\times7\times50$\,cm$^3$ crystals ($27\,X_0$) with PMT readout optimized for the CsI UV emission.
The DAMSA stochastic term is larger than KOTO's, which is
understood from three main contributions: (1)~the narrow $1\times1$\,cm$^2$ crystal cross-section causes substantial position-dependent light-collection non-uniformity along the 12\,cm bar (Section~3.5.2), requiring corrections that introduce residual fluctuations; (2)~SiPM readout has lower photon detection efficiency at 310\,nm compared to
UV-sensitive PMTs, and the SiPM dark noise and gain non-uniformity add stochastic contributions; and (3)~the need for wavelength-shifting to address the ESR reflector UV cutoff (see above) introduces an additional conversion efficiency loss. Full \textsc{GEANT4} shower simulations with the complete ECal geometry are underway to refine this estimate.
The nanosecond-scale timing of the undoped CsI also provides event-by-event discrimination between prompt signal photons (arriving within $\sim$10\,ns of the beam pulse) and delayed neutron-induced backgrounds (arriving several hundred nanoseconds later), as demonstrated in the \textsc{GEANT4} timing studies of Section~4.5.}

The ECal baseline design shown in Fig.~\ref{fig:ecal_design} consists of three types of CsI crystal sections, each optimized for a specific function. 
The first section uses an array of \qtyproduct[product-units=single]{1x1x12}{\cubic\cm} crystal bars to detect pile-up; since most signal photons begin to interact after approximately \SI{2}{\cm}, this section is not critical for signal reconstruction. 
The middle section is designed to measure signal photons with high energy and spatial resolution. The final section is optimized to measure electromagnetic shower tails, providing high energy resolution with reduced spatial resolution.

\subsubsection{Optional Total Absorption Tracking ECal - TATE}

\begin{figure}[t]
    \centering
    \includegraphics[width=0.9\linewidth]{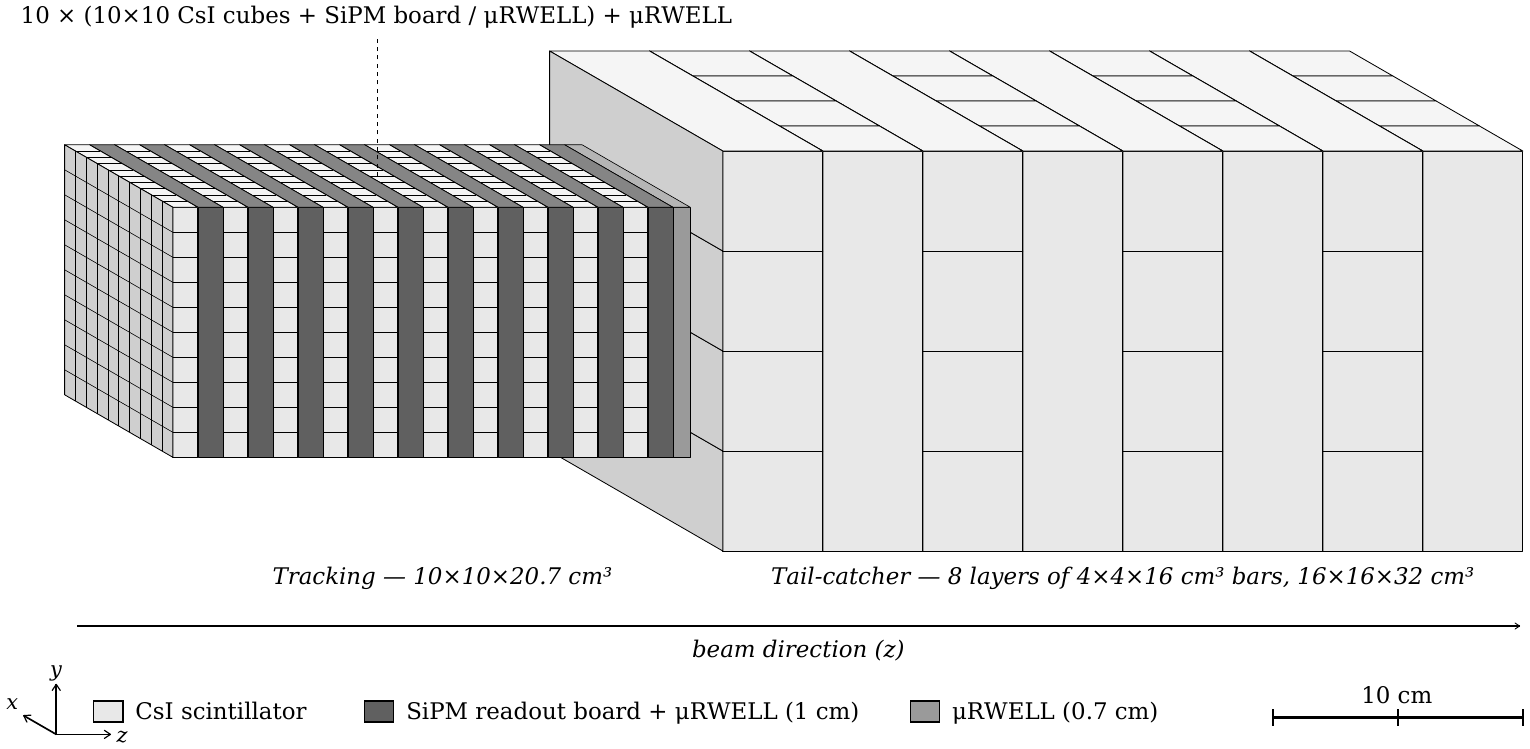}
    \caption{Schematic of the detector, drawn to scale (oblique view from upstream; beam along $z$). 
    The total absorption tracking calorimeter ($10\times10\times20.7$~cm$^3$) consists of ten layers, each a $10\times10$ array of 1~cm$^3$ CsI cubes followed by a 1~cm SiPM readout board and $\mu$RWELL, plus an additional 0.7~cm $\mu$RWELL after the last layer. It is followed directly by a tail-catcher ($16\times16\times32$~cm$^3$) of eight layers of four $4\times4\times16$~cm$^3$ CsI bars, with alternating $x$/$y$ orientation in successive layers. 
    Each cube is read out by one SiPM and each bar by two SiPMs, giving $10\times100 = 1000$ channels in the tracking calorimeter, $8\times4\times2 = 64$ in the tail-catcher, and 1064 SiPM channels in total.}
    \label{fig:TATE}
\end{figure}

DAMSA aims the ALP-to-two electromagnetic particle signal, namely $a\to\gamma\gamma$ and $a\to e^{+}e^{-}$ final states, from ALPs produced in the dump. ALPs that escape the target and decay within the decay volume, and are thus
detectable, are necessarily highly boosted, so the two EM particles are emitted nearly collinear. 
The opening angle scales as $\theta_{\gamma\gamma, e^{+}e^{-}}\gtrsim 2m_a/E_a$, which in the target parameter space corresponds to only a few tens of mrad. To resolve the two EM particles and reconstruct their invariant mass,
$M_{\gamma\gamma, e^{+}e^{-}}^2 = 2E_1E_2(1-\cos\theta_{\gamma\gamma, e^{+}e^{-}})$, with adequate precision, the EM particle directions must be measured with an angular resolution of a few mrad. 
This requirement exceeds what the granularity of the 1~cm CsI cubes alone can provide, so the photon directions must be measured with a dedicated position-sensitive detector~\cite{DAMSeBeam}.

To improve the detector sensitivity, an optional solution under consideration is to intersperse tracking layers equipped with a strip-type readout plane, either
an MPGD or a silicon strip detector (see Fig.~\ref{fig:TATE}) in the fine granular portion of in the upstream end of the ECal, which includes the EM shower maximum. 
In the baseline configuration, each $10\times10$~cm$^2$ plane provides 128 $x$ and 128 $y$ strips (0.78~mm pitch); with charge-sharing readout, the expected position resolution is $\sim$100~$\mu$m, which yields an intrinsic angular resolution of a few mrad when the conversion pair is fitted over three consecutive planes. The tracking section converts more than 98\% of the signal EM particles. The CsI cubes act as both the conversion absorber and the calorimeter, measuring the energy deposited by the showering photons and electrons cube by cube, while the $e^+e^-$ pairs from the photon conversions are tracked through the downstream strip planes to reconstruct the photon directions. 
An additional readout plane after the last CsI layer ensures that photons converting in the final layer are also measured at two points. The overall angular resolution, including multiple scattering of the conversion electrons in the CsI, will be determined with full simulation.

\subsubsection{Validation tests}
\begin{figure}
    \centering
    \begin{minipage}{0.1\linewidth}
    \includegraphics[width=\linewidth]{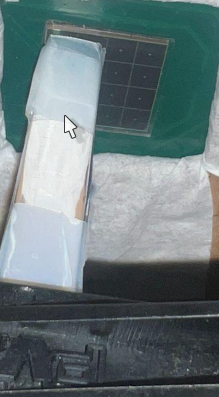}
    \end{minipage}
    \begin{minipage}{0.4\linewidth}
    \includegraphics[width=\linewidth]{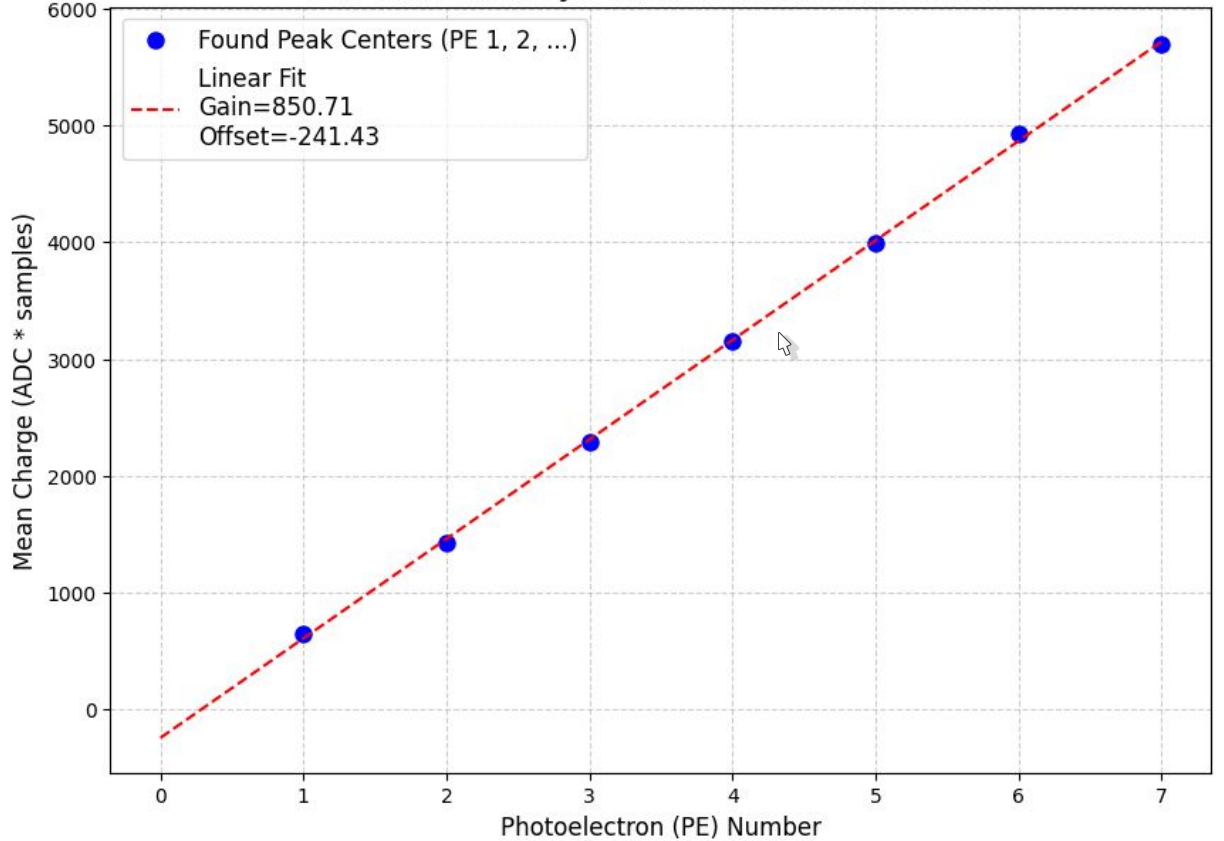}
    \end{minipage}
    \begin{minipage}{0.42\linewidth}
    \includegraphics[width=\linewidth]{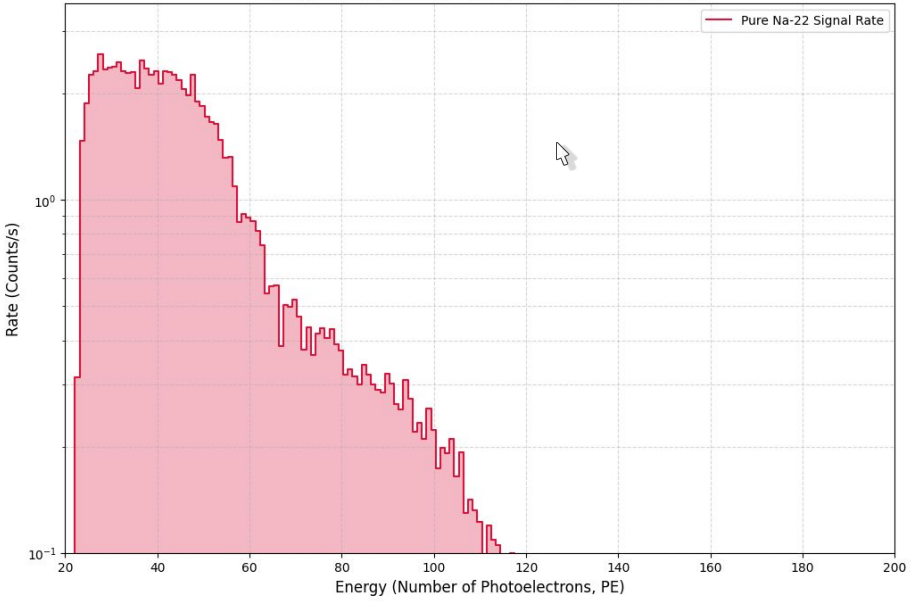}
    \end{minipage}
    \caption{(Left) Illustration of the CsI crystal calibration measurements at UC\,Riverside, using a Hamamatsu S14161 SiPM array coupled to the crystal using a silicone pad. (Middle) Single photoelectron calibration measurement using dark counts at room temperature. (Right) Measured photoelectron spectrum with a \ce{^22Na} calibration source, after background subtraction, showing Compton spectra with edges at \SI{511}{\keV} and \SI{1.27}{\MeV}.}
    \label{fig:ucr_calibration}
\end{figure}

Ongoing studies at Brookhaven National Laboratory (BNL), University of California--Riverside, Kyungpook National University, and Seoul National University are performing prototype tests to optimize and validate the ECal design. These studies aim to test various ECal designs to determine the configuration that gives the best energy and timing resolution and to ensure the spatial uniformity of each crystal.
While tests are still underway, early results using a \ce{^22Na} calibration source, with \SI{511}{\keV} and \SI{1.27}{\MeV} $\gamma$-lines, are shown in Fig.~\ref{fig:ucr_calibration}.
After subtracting backgrounds and correcting for the single photonelectron (PE) gain of the SiPM, the observed spectrum from these $\gamma$-rays Compton scattering in the crystal are estimated to be in the range of \SIrange{100}{150}{PE\per\MeV} with a SiPM only on one side of the crystal. 

Similar studies at BNL used a \ce{^90Sr} $\beta$-source (\SI{546}{\keV} endpoint) at secular equilibrium with a \ce{^90Y} $\beta$-source (\SI{2.278}{\MeV} endpoint) to compare the SiPM performance with visible and VUV-sensitive Hamamatsu SiPMs. 
These studies found a higher light yield with visible SiPMs, consistent with their higher quantum efficiency for \SI{310}{\nano\meter} photons. 

\subsubsection{Optical simulations}
\begin{figure}[b]
    \centering
    \includegraphics[width=0.325\linewidth]{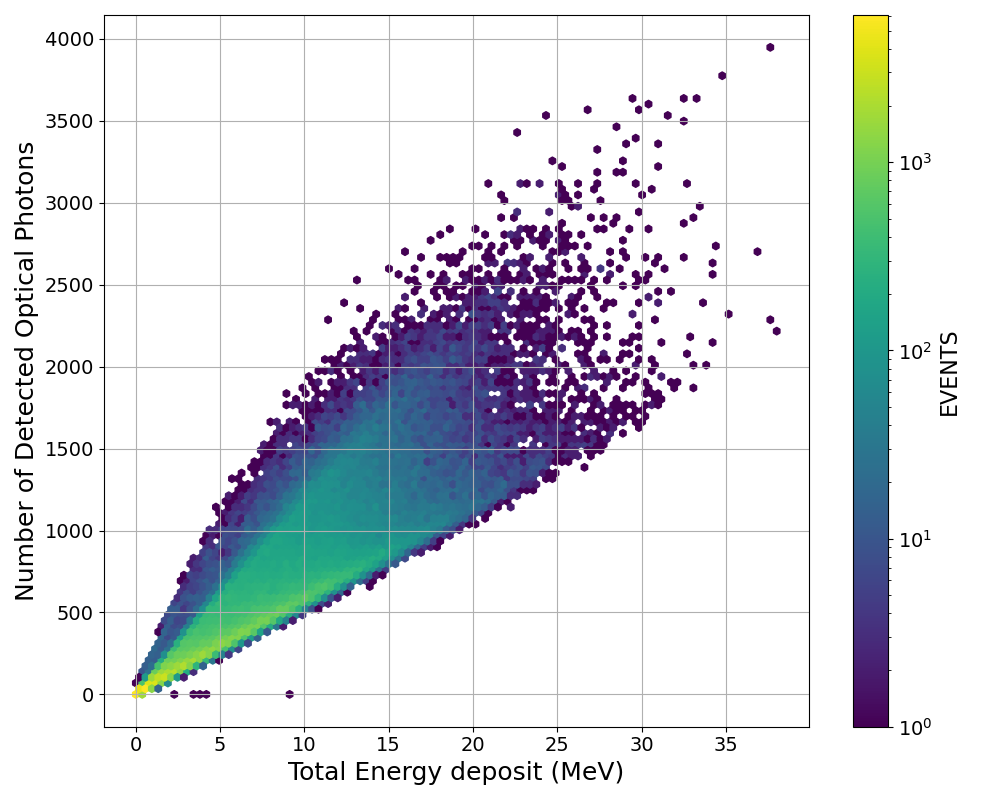}
    \includegraphics[width=0.325\linewidth]{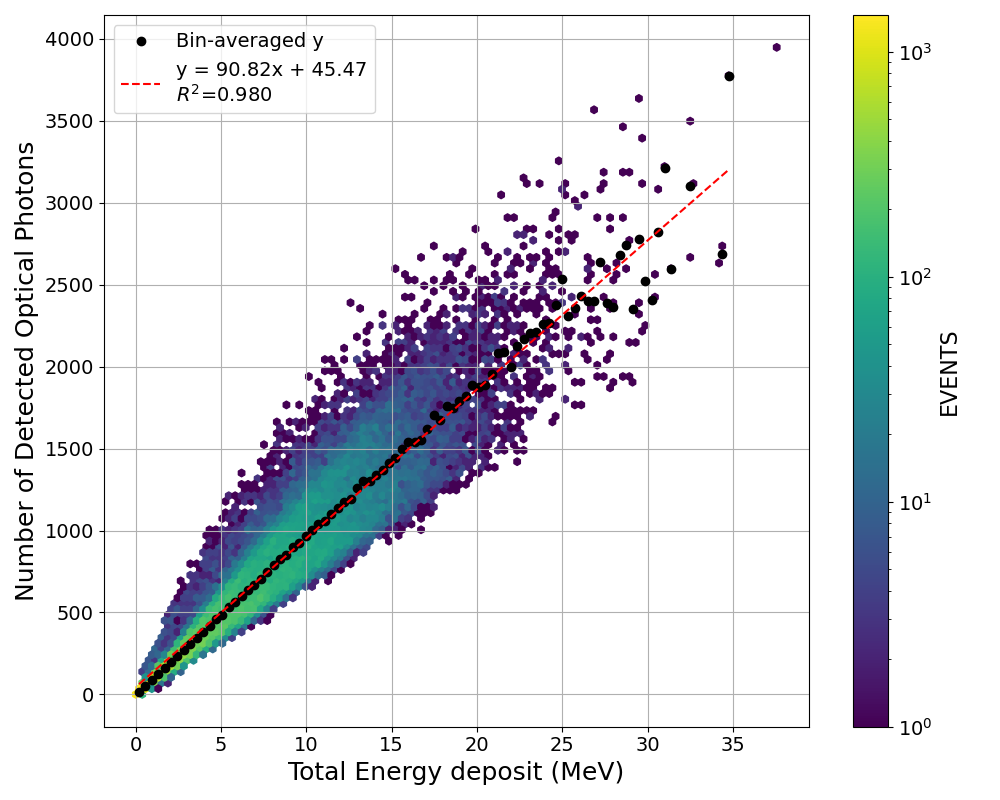}
    \includegraphics[width=0.325\linewidth]{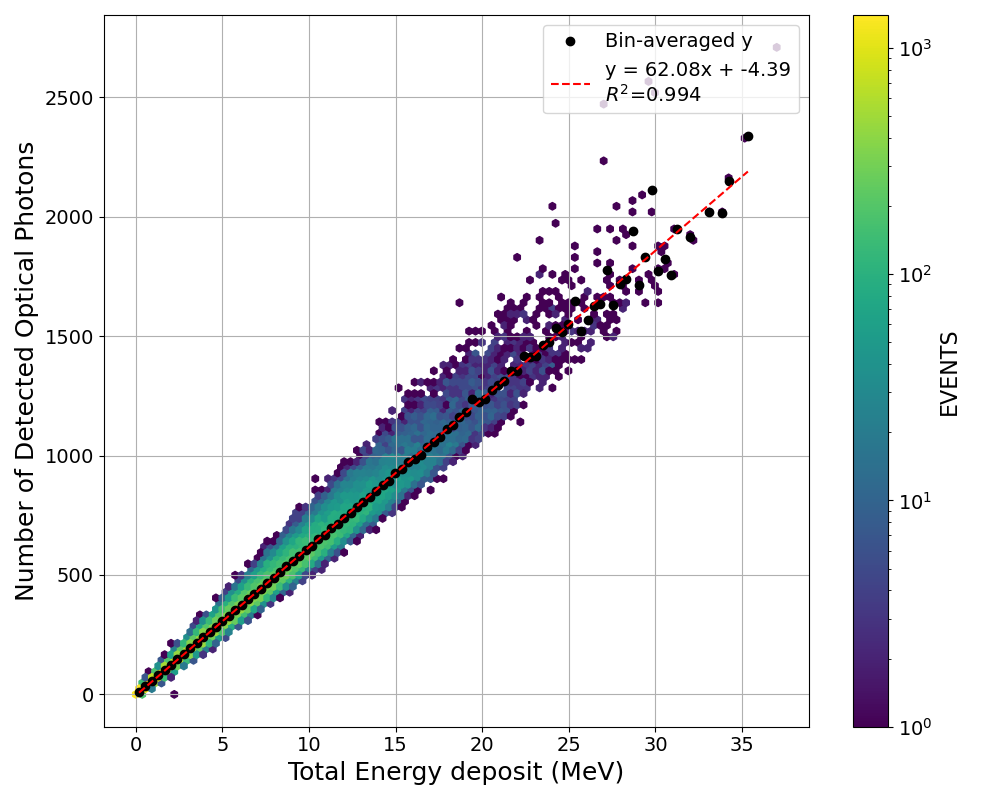}
    \caption{\textsc{Geant4} optical simulations a CsI detector showing the measured scintillation light versus deposited energy for \SIrange{0}{35}{\MeV} depositions in the CsI (left) for all events, (middle) for energy deposited within \SI{3}{\cm} of the SiPM at either end of the CsI bar, and (right) for energy deposited in the central \SI{6}{\cm}.}
    \label{fig:CsI_Geant4_conf}
\end{figure}

A \textsc{Geant4} simulation was performed to estimate the energy resolution of a single CsI crystal bar, in order to calculate the scintillation light yield by propagating optical photons through the crystal. The result is shown in Fig.~\ref{fig:CsI_Geant4_conf}.
Each plot shows the number of photons detected as a function of the energy deposited in the detector.
The leftmost plot shows this distribution for all simulated energy depositions and has a very broad spread: an event for which \num{15000} photons is detected could be anywhere from \SI{10}{\MeV} to \SI{25}{\MeV}.
This spread is driven by the spatial dependence of the light yield: energy deposited near the SiPMs (middle plot of Fig.~\ref{fig:CsI_Geant4_conf}) produces more detected light than energy deposited near the edges (rightmost plot), due to the large number of photons required for photons starting far from the SiPMs.
Better energy resolution is therefore achieved with shorter or wider bars. 
Reconstructing the transverse position may allow corrections to improve the resolution.

\textcolor{black}{When the longitudinal hit position is reconstructed from the SiPM signal ratio at the two ends of each bar, a position-dependent correction can be applied. Preliminary studies indicate that this correction reduces the single-bar energy spread. At the full ECal
level, where shower energy is distributed over many crystals, the stochastic averaging further improves the resolution.}

\begin{figure}[htb]
    \centering
    \begin{minipage}[c]{0.45\linewidth}
    \includegraphics[width=\linewidth]{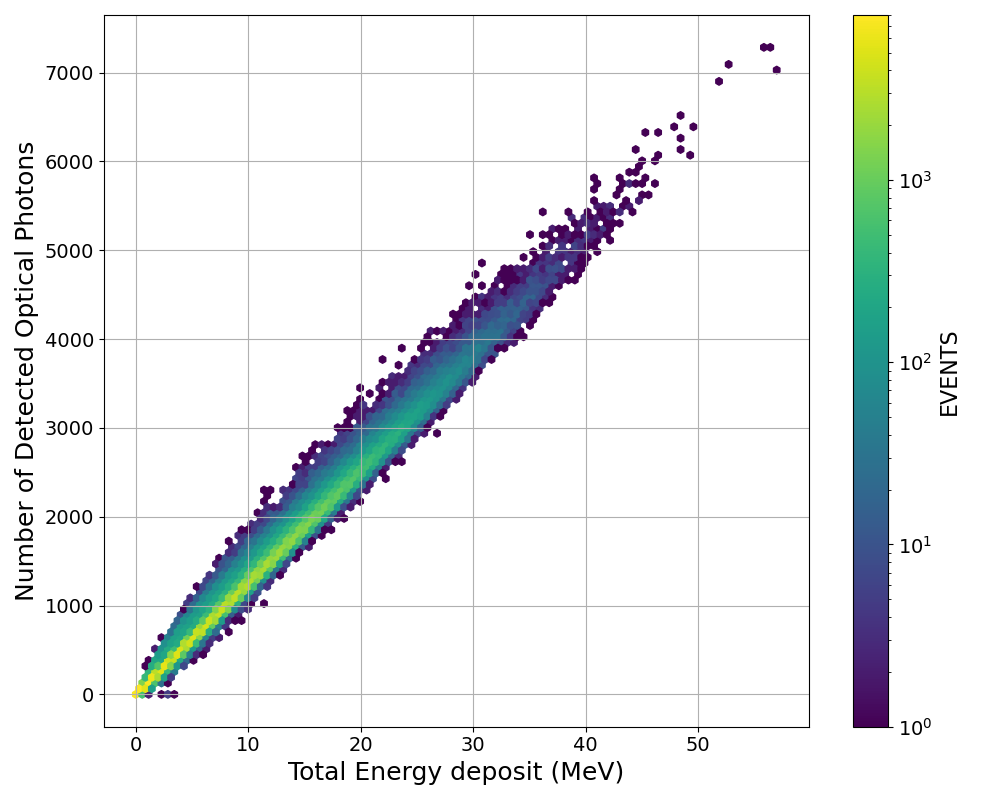}
    \end{minipage}
    \begin{minipage}[c]{0.45\linewidth}
    \includegraphics[width=\linewidth]{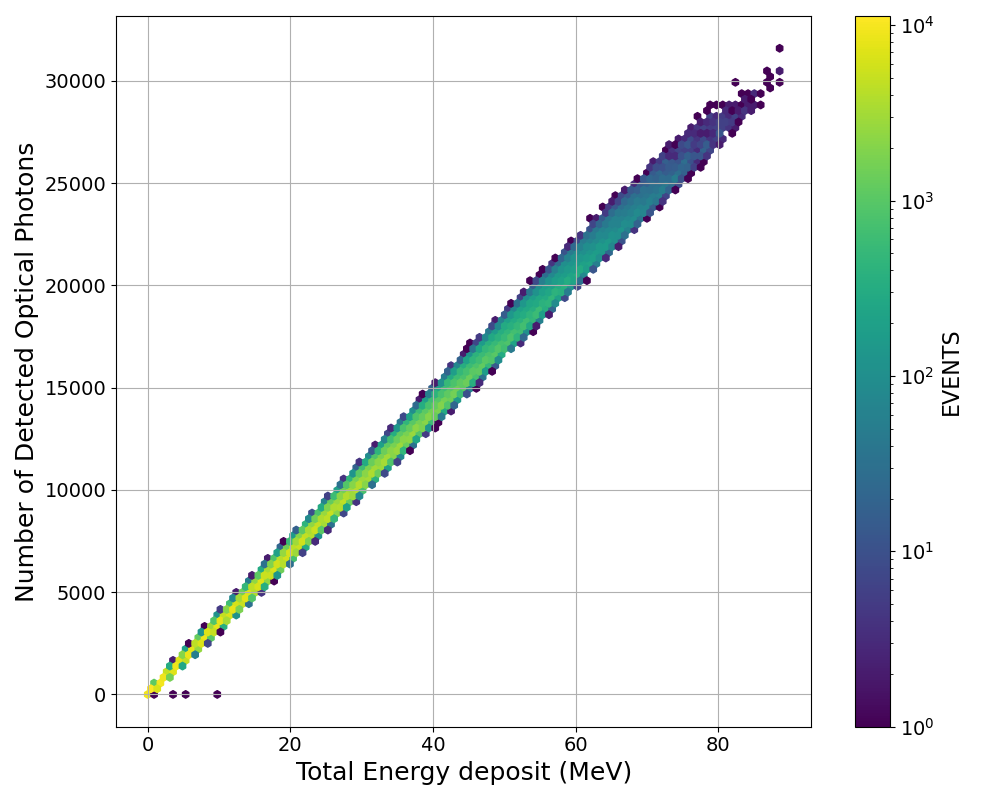}
    \end{minipage}
    \caption{Detection photons as a function of energy deposited in the CsI for (Left) \qtyproduct[product-units=single]{1x1x2}{\cubic\cm} cube-like crystals being considered for an alternative design and (Right) \qtyproduct[product-units=single]{4x4x12}{\cubic\cm} crystals forming the final six layers of the nominal design.}
    \label{fig:cube_like}
\end{figure}

An alternative design using \qtyproduct[product-units=single]{1x1x2}{\cubic\cm} is also being considered.
The improved energy resolution with this design is shown in Fig.~\ref{fig:cube_like} (left).
An ECal composed of such cube-like crystals arranged in a two-dimensional array over multiple layers enhances both the energy resolution and the suppression of ghost hits. In the current design using \qtyproduct[product-units=single]{1x1x12}{\cubic\cm} crystal bars, the bars are stacked along the $x$- and $y$-axes, and these two layers are used to reconstruct the hit position. However, when multiple hits occur simultaneously, the fired $x$- and $y$-axis bars can produce ghost hits at each of their crossing points.

A thicker \qtyproduct[product-units=single]{4x4x12}{\cubic\cm} crystal improves the energy resolution by reducing internal reflections, as shown in Fig.~\ref{fig:cube_like} (right). Although this geometry degrades the spatial resolution, the 3D ECal employs it for energy measurements of the electromagnetic shower produced by signal photons. High spatial resolution is not required at this stage, since after the incident, energetic photons initiate an electromagnetic shower, the transverse spread of secondary particles increases, and the total deposited energy becomes the primary observable. Consequently, the \qtyproduct[product-units=single]{4x4x12}{\cubic\cm} crystal bar reduces the number of DAQ channels while improving the energy measurement.

\subsubsection{Radiation Hardness}
\begin{figure}[htb]
    \centering
    \includegraphics[width=\linewidth]{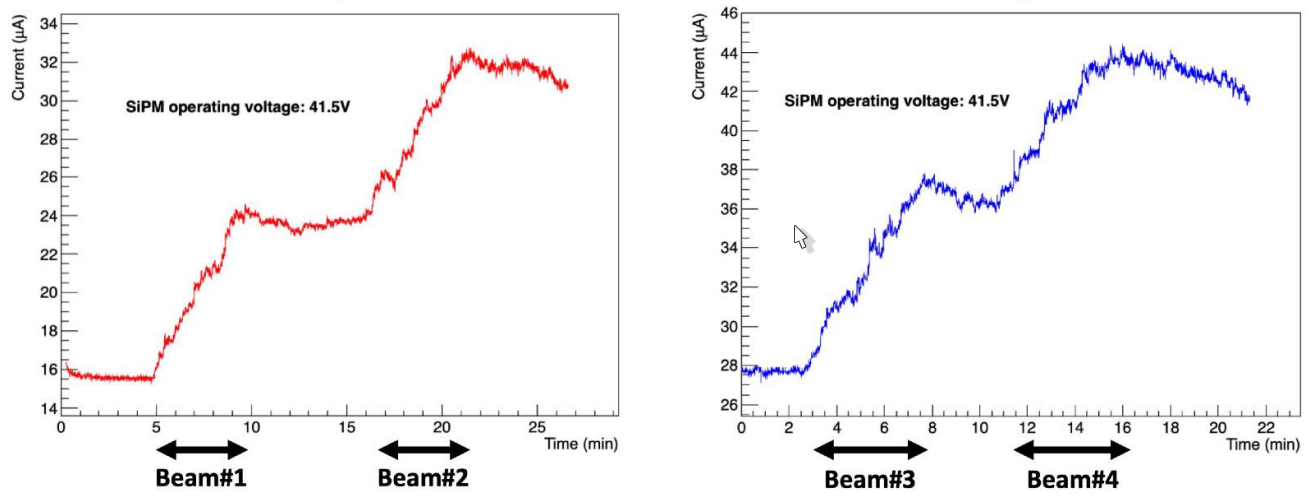}
    \caption{Monitored current of a Hamamatsu S14161 SiPM array during intermittent irradiation at KOMAC. The neutron field is produced by a 100 MeV proton beam on a beam dump, providing a broad spectrum below \SI{90}{\MeV} with a total fluence of approximately \SI{2e7}{\MeV~eq~n\per\square\cm}. The result corresponds to a single readout channel of the array.}
    \label{fig:sipm_hardness}
\end{figure}
Figure~\ref{fig:sipm_hardness} shows the response of a Hamamatsu S14161 \numproduct{4x4} SiPM array exposed to a neutron field at KOMAC. The irradiation was performed using secondary neutrons produced by a 100 MeV proton beam on a beam dump, resulting in a broad spectrum below \SI{90}{\MeV}, with a total fluence of approximately \SI{2e7}{\MeV~eq~n\per\square\cm}. In this measurement, one channel of the array was read out and the current was monitored at the power supply. An increase in the current was observed during irradiation, followed by a gradual recovery after beam-off periods.

\textcolor{black}{While these results demonstrate the SiPM resilience to fast-neutron irradiation at the $10^4$\,cm$^{-2}$ fluence level, operational beam dump environments typically produce integrated fluences of $10^{8}$--$10^{12}$\,n/cm$^2$ over a full run, several orders of magnitude higher than the KOMAC test.  The KOMAC measurement therefore establishes only a lower bound on the SiPM radiation tolerance.
The actual BRN fluence at the ECal location ($\sim$57\,cm from the target face) during extended data taking will be evaluated with full 8\,GeV \textsc{GEANT4} simulations. Additional irradiation campaigns at higher fluences and with mixed $\gamma$/neutron fields are planned, including tests at facilities capable of delivering $10^{10}$\,n/cm$^2$ or above, to establish the SiPM operational lifetime and determine whether periodic annealing or SiPM replacement will be necessary.}

\subsection{Data Acquisition and Trigger}

The data acquisition and trigger are being developed as a single integrated architecture. One candidate solution is a pipelined FADC-based readout in which digitized calorimeter waveforms are used both for event readout and for real-time trigger formation. In this approach, trigger quantities such as the total energy deposited in the electromagnetic calorimeter and the relevant veto information can be formed within the DAQ, reducing the need for separate external trigger hardware. If a commercial FADC solution is adopted instead, DAMSA would likely require a dedicated external trigger system, including hardware to form the calorimeter energy sum and incorporate veto signals. The final choice will be driven by latency, channel count, subsystem integration, and overall system complexity.

\textcolor{black}{The primary trigger for the ALP$\to\gamma\gamma$ search requires two spatially separated energy clusters in the ECal with a combined energy exceeding 500\,MeV and an opening angle $>$1$^\circ$, in coincidence with the beam pulse. For the $e^+e^-$ channel, at least one charged track in the tracker matched to an ECal cluster is additionally required. Assuming our data is $\mathcal{O}(10^3)$\,Hz at $10^4$ electrons per pulse with a 1\,kHz repetition rate, dominated by prompt EM shower leakage,
the total DAQ data rate, with additional assumptions of $\sim$2000 ECal channels and $\sim$1000 tracker channels digitized at 12 bits per sample with $\sim$20 samples per event, is estimated at $\sim$90\,MB/s before zero suppression, reducing to $\sim$30--50\,MB/s after sparsification, well within the bandwidth of a modern Ethernet-based readout system.}

\subsection{Background Validation and future detector upgrade with a 3D-Projection Scintillator Detector}

\begin{figure}[htb]
    \centering
    \includegraphics[scale=0.4]{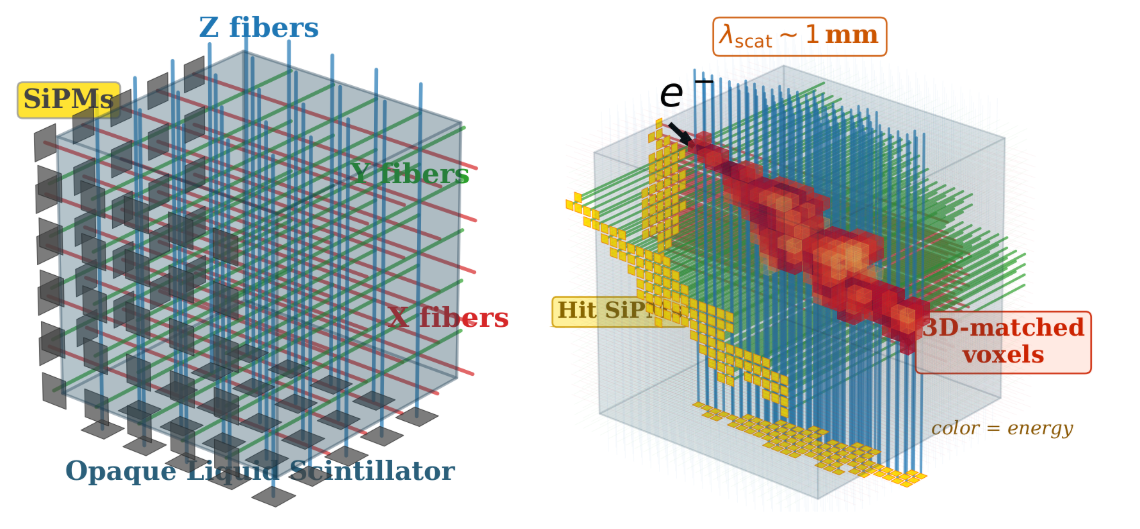}
    \caption{
The concept of the 3D-projection detector. The material and the method of putting a physical boundary for light confinement are flexible. 
    }
    \label{fig:pilot_concept}
\end{figure}

A powerful background validation tool is being considered:
a plastic 3D-projection scintillator detector~\cite{SuperFGD_concept,SuperFGD_full}
that can be deployed in the background
characterization runs at FAST. This detector concept, based on three orthogonal
arrays of wavelength-shifting (WLS) fibers embedded in a scintillator volume, provides
true 3D imaging of particle interactions with sub-centimetre voxel resolution and
sub-nanosecond timing~\cite{Benchmarks3View}.
The concept is shown in Figure~\ref{fig:pilot_concept}.

The 3D-projection readout offers several advantages for BRN
characterizations that are complementary to the borated liquid scintillator detectors
described in Section~\ref{sec:BackgroundMitigation}. First, the three-view fiber geometry enables unambiguous 3D
reconstruction of neutron interaction vertices and secondary particle topologies,
allowing direct visualization of the spatial and angular distribution of BRN-induced
backgrounds. Second, the fine granularity (${\lesssim}\,1$\,cm voxels) permits
discrimination between electromagnetic and hadronic shower profiles on an
event-by-event basis---a capability critical for separating signal-like two-photon
events from neutron-induced $\gamma$-ray cascades. Third, the scintillator provides
both prompt energy deposition and pulse-shape discrimination capabilities, enabling
the same timing-based neutron/photon separation demonstrated with EJ-301 (Section~4.2)
while simultaneously providing full 3D topology information.

A compact 3D-projection prototype, consisting of
$\sim$20\,cm-scale plastic scintillator modules read out by SiPM-coupled WLS fibers,
could be placed at varying distances and angles from the tungsten target during
dedicated background runs. Such measurements would provide a high-granularity,
three-dimensional map of the BRN field that directly benchmarks the \textsc{GEANT4} simulations
of Sections~4.5 and~4.6, particularly in the transition region between prompt and
delayed backgrounds where the current simulations have limited validation data.

Looking beyond the baseline CsI electromagnetic calorimeter design
described in Section~3.5, a 3D-projection scintillator
detector~\cite{SuperFGD_concept,SuperFGD_full}
represents a promising alternative geometry for a finer-granularity calorimeter in DAMSA Stages~2 and beyond. 
A staged approach is envisioned: the plastic 3D-projection detector would first serve
as a background validation tool during initial runs, providing
high-granularity BRN characterisation data. Pending the outcome of ongoing R\&D on
heavy-metal-loaded opaque liquid
scintillator~\cite{WbLS_Yeh,WbLS_30ton,oWbLS_char}, a full 3D-projection calorimeter
module could then be evaluated as a complement to the CsI ECal in a
future DAMSA upgrade, offering the unique combination of fine granularity, full
activity, nanosecond timing, and tunable radiation length in a single detector
volume~\cite{Seg3D_WbLS}.

\vspace{0.5em}
\section{Background Mitigation Strategy} \label{sec:BackgroundMitigation}
As stated in Section~\ref{sec:PhysicsGoals}, DAMSA is an ultra-short baseline table-top beam dump experiment that aims to greatly increase the accessible parameter space by minimizing the distance between the target and the detector. In this process, as the distance between the detector and the target decreases, managing the beam-induced background, as summarized in Fig.~\ref{fig:backgrounds} becomes important.
This section will describe the primary backgrounds that must be discriminated against to realize this potential, and the plan for modeling and mitigating the dominant backgrounds.

\subsection {Environmental Neutrons}
\label{subsec:neutron-flux}
Since neutrons can be captured by materials surrounding the detector to produce \SIrange{2}{11}{\MeV} \gr\ cascades, they can pose a significant background to the expected dark matter signals. Neutrons are produced by natural processes in the surrounding environment as well as by the beam itself.

Near sea-level, the ambient neutron flux is dominated by cosmic-rays and the products of their interactions.
Although cosmogenic neutron fluxes reported in the literature vary between different measurements and studies, typical values range from \SIrange{e-4}{e-2}{n\per\square\cm\per\second}, with an energy spectrum spanning from thermal to fast energies~\cite{cStudyEnvironmentalNeutron2003}
At an altitude of \SI{340}{\meter}, Ref.~\cite{korunMeasurementAmbientNeutron1996} measures a thermal-neutron flux of \SI{1.9\pm0.7e-3}{n\per\square\cm\per\second} and a fast-neutron flux of \SI{16\pm3e-3}{n\per\square\cm\per\second}, consistent with other measurements.
This creates a steady-state background that will need to be mitigated with tight beam-timing cuts.

\subsection {Beam-Related Neutron Flux}
\begin{figure}
    \centering
    \includegraphics[width=0.7\linewidth]{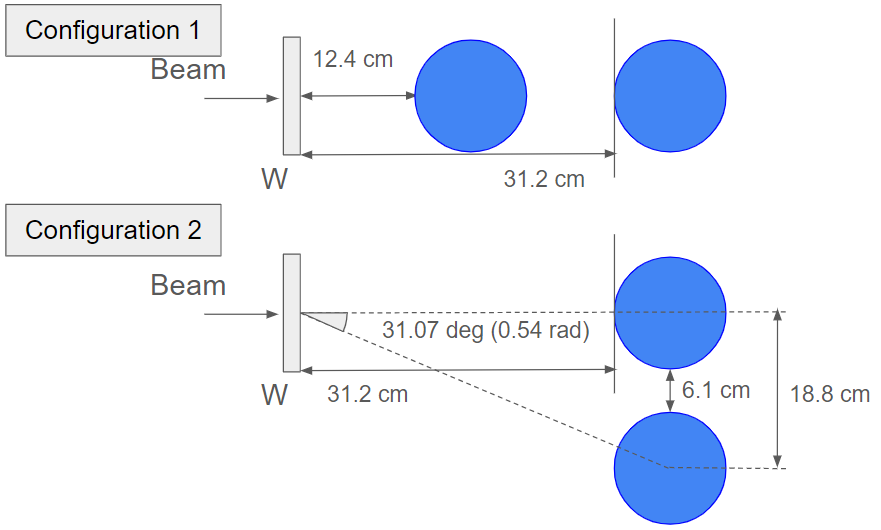}
    \caption{Detector setups for observing variations in background signals based on distance and angle. Detectors are shown in blue, targets in gray, and the beam direction is indicated by arrows. {\it Top} -- A configuration where two detectors are placed at different distances from the target along the beam on-axis to measure the effect of distance. {\it Bottom} -- A configuration where one detector is placed on-axis and the other off-axis to measure the effect of angle.}
    \label{fig:detector-configurations}
\end{figure}

To understand the beam-induced background, we previously measured the neutron and photon production when an electron beam interacts with a tungsten target, using actual tungsten targets and detectors at the Fermilab Test Beam Facility (FTBF)~\cite{FTBF_ref}. FTBF provides a 120 GeV primary proton beam and secondary mixed particle beams, consisting of pions and electrons with energies as low as 1 GeV. 
FTBF is also equipped with differential Cherenkov counters for particle identification. We have already conducted one measurement and plan to carry out several more in the near future. We used two liquid scintillation detectors, each consisting of a cylindrical aluminum-housed cell, \SI{50}{\mm} in diameter and \SI{50}{\mm} in length ($\sim$\SI{0.1}{\liter} active volume), filled with EJ-301 liquid scintillator and closed by a borosilicate-glass optical window viewed by a single photomultiplier tube (PMT) assembly~\cite{Eljen:2021eljen}, and multiple tungsten targets, each with a thickness of 4 mm, varying the tungsten thickness to observe how the beam-induced background changes. Additionally, we set up two different detector configurations to compare particle populations on-axis, where the beam is directed, and off-axis. In the on-axis configuration, the two detectors were placed along the beam axis at distances of 12.4 cm and 31.2 cm from the target, respectively. In the off-axis configuration, one detector was positioned 31.2 cm from the target along the beam axis, while the other was located 18.8 cm perpendicular to the beam axis. Figure~\ref{fig:detector-configurations} shows these two different configurations.

Using EJ-301, we can distinguish between photons and neutrons by employing pulse-shape discrimination (PSD). This method is widely used for identifying neutrons and photons, with numerous studies supporting its application to EJ-301 liquid scintillators~\cite{DAS2022167405, LANG201726, doi:10.14407/jrpr.2019.44.2.53}. The PSD technique relies on differences in the fall times of pulse signals generated by photons and neutrons. Typically, neutron signals have a longer fall time compared to photon signals, resulting in a longer tail in the pulse~\cite{Roy_2022}.

By applying PSD to beam data, we will compare the results with \textsc{GEANT4} simulations to validate the simulations' accuracy. This validation will then be used to inform the experimental design. See Sections~\ref{subsec:300-MeV-geant4-simulation} and \ref{sec:2-GeV-electron-beam-geant4-simulation} for the details of the simulations.

\subsection {Electron and Photon Flux Out of Target}
ALPs can be generated through Primakoff reactions from energetic photons, making it crucial to understand the characteristics of photons produced in the interaction between the accelerator's beam and the target material. As discussed in Section~\ref{subsec:neutron-flux}, PSD enables the identification of neutrons and photons entering the liquid scintillator detector. We plan to use this method to compare the number of photons generated by accelerator beams of known energy interacting with the target material to simulated results, which will assist in planning the experiments.

\subsection {Mitigation Strategy}
\begin{figure}
    \centering
    \includegraphics[width=0.7\linewidth]{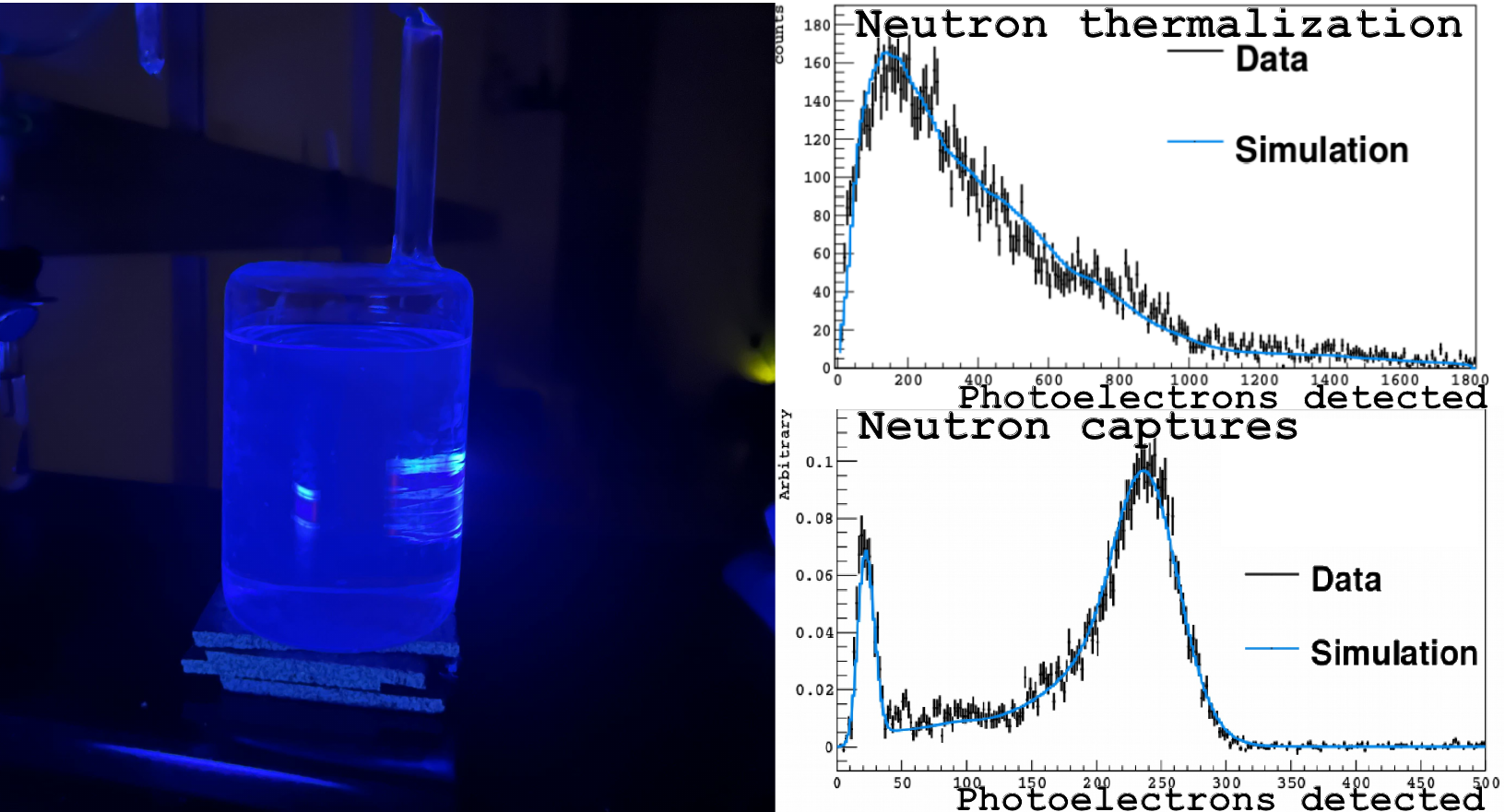}
    \caption{{\it Left} -- A neutron detector produced for this project, exposed to UV light, to be wrapped in a reflector and coupled to a SiPM array (not shown). {\it Right} -- Prompt thermalization and delayed capture signals of neutrons from a calibration source in DarkSide-50's neutron veto, from~\cite{westerdaleStudyNuclearRecoil2016}.}
    \label{fig:neutron-detectors}
\end{figure}
DAMSA will develop techniques for modeling and mitigating neutron backgrounds, using techniques the University of California, Riverside (UCR) developed for dark matter direct detection~\cite{westerdalePrototypeNeutronVeto2016,westerdaleQuenchingMeasurementsModeling2017,agnesVetoSystemDarkSide502016}. 
We aim to (1) design techniques for characterizing and modeling environmental neutrons and BRNs; (2) define shielding and coincidence timing requirements to mitigate neutrons.
Doing so requires measuring the rate and energies of thermal, fast, and high-energy neutrons.

The neutron detectors for this project, like those in Fig.~\ref{fig:neutron-detectors}, are based on the borated liquid scintillator neutron veto used for the neutron veto of the DarkSide-50 dark matter detector.
Commercial detectors typically choose between energy resolution and efficiency~\cite{mukhopadhyayReviewDirectNeutron2022}.
Liquid scintillator detectors produce signals proportional to the energy lost by neutrons but typically have low efficiency, especially for low-energy neutrons. 
On the other hand, \ce{^3He} detectors efficiently count thermal neutron captures, but are insensitive to neutron energies.
Borated liquid scintillator detectors capture the best of both techniques: the hydrogenous scintillator quickly and efficiently thermalizes neutrons, producing a prompt signal proportional to energy, and thermal neutrons capture on \ce{^10B}, via,
\begin{gather}
    \begin{aligned}
        n+\ce{^10B} \rightarrow \begin{cases}
            \ce{^7Li}\text{ (\SI{1015}{\keV})} + \alpha\text{ (\SI{1775}{\keV})} & (\SI{6.4}{\percent}) \\
            \ce{^7Li^*} + \alpha\text{ (\SI{1471}{keV})}, & (\SI{93.6}{\percent}) \\
            \qquad \ce{^7Li^*} \rightarrow \ce{^7Li}\text{ (\SI{839}{\keV})} + \gamma\text{ (\SI{478}{\keV})}
        \end{cases}
    \end{aligned}
\end{gather}
It was found that the $\alpha$ and \ce{^7Li} are quenched to \SI{30}{~\keV_{ee}} (\si{\keV} electron equivalent)~\cite{westerdaleQuenchingMeasurementsModeling2017}. 
Since they cannot escape the detectors, sensitivity to them allows for high tagging efficiency, regardless of the energy that a neutron deposits.
While, nuclear recoils in the thermalization signal are quenched to \SI{10}{\percent}, a quenching model was able to reconstruct their energies~\cite{westerdaleQuenchingMeasurementsModeling2017}, as demonstrated by Fig.~\ref{fig:neutron-detectors} (top right), which shows accurate reproduction of the neutron energy spectrum in DarkSide-50's neutron veto, using an \ce{^241Am}-\ce{^13C} \alphan\ calibration source.
As such, these detectors can achieve both high efficiency and energy resolution, with $\sim1$\,ns timing resolution for fast and high-energy neutron thermalization signals.
Figure~\ref{fig:neutron-detectors} shows these signals in the DarkSide-50 neutron veto, which achieved $>99.5\%$ tagging efficiency~\cite{westerdaleStudyNuclearRecoil2016}.

These detectors will use equal volumes of pseudocumene (PC: the primary scintillator) and trimethyl borate (TMB: the boron-loading agent), with \SI{3}{\gram\per\liter} of 2,5-Diphenyloxazole (PPO) and \SI{25}{\milli\gram\per\liter} 1,4-Bis(2-methylstyryl)benzene (bis-MSB) as primary and secondary wavelength shifters, to enhance the light yield and convert UV PC scintillation light to the visible spectrum, where photons will be efficiently detected by arrays of silicon photomultipliers (SiPM) coupled to the detectors' face.
Based on prior experience, we expect them to have a light yield of \SI{2}{\pe\per\keV}, high enough to efficiently measure neutron captures.

Thermal neutrons will produce isolated capture signals that will be detected with \SI{>99}{\percent} efficiency.
About \SI{50}{\percent} of ambient fast neutrons will thermalize in the detector, creating a signal proportional to the neutron's energy, while the capture signal, which we will detect down to low energies by triggering on the capture signal that follows with a time constant of \SI{2.2}{~\micro\second}.
High-energy neutrons are less likely to thermalize in the detector and will instead produce nuclear recoils without subsequent capture signals; their spectrum can be unfolded from the nuclear recoil energy spectrum.
Since PC scintillates with nanosecond timing, fast and high-energy BRNs can be measured in coincidence with the beam to infer the timing structure of neutrons produced.

We plan to build an array of \num{8} detectors, which will be placed around the lab and near the target detector and beam, to measure neutrons.
These measurements will be benchmarked against simulations, using the \alphan~\cite{westerdaleRadiogenicNeutronYield2017,gromovCalculationNeutronGamma2023a} and \ngamma~\cite{weimerG4CASCADEDatadrivenImplementation2024} modeling tools developed by UCR~\cite{cano-ottWhitePaperAlpha2024}.
These comparisons will validate our technique for characterizing background neutrons and define shielding and timing requirements needed to reduce neutron backgrounds.

\subsection{300 MeV Electron Beam \textsc{GEANT4} Simulation}
\label{subsec:300-MeV-geant4-simulation}
DAMSA collaboration has a plan to use 300~MeV electron beams at Fermilab's Facility for Accelerator Science and Technology (FAST) for initial studies of detector development and background Monte Carlo validation efforts, as phase~0 of the experiment.
As part of this effort, the interaction of the 300 MeV electron beam with the 10 cm tungsten target was simulated using \textsc{GEANT4} to estimate the production of background and signal sources at the target. Electron-nuclear reactions assume the photon approximation, where real $\gamma$'s are generated from virtual ones at the electromagnetic vertex. This cross section is valid for incident electrons and positrons across all energy ranges. The background estimation considers all sources of electromagnetic interactions within the electromagnetic calorimeter detector volume, with particular emphasis on the neutron-induced background.

\begin{figure}[t]
\centering
\includegraphics[width=0.49\linewidth]{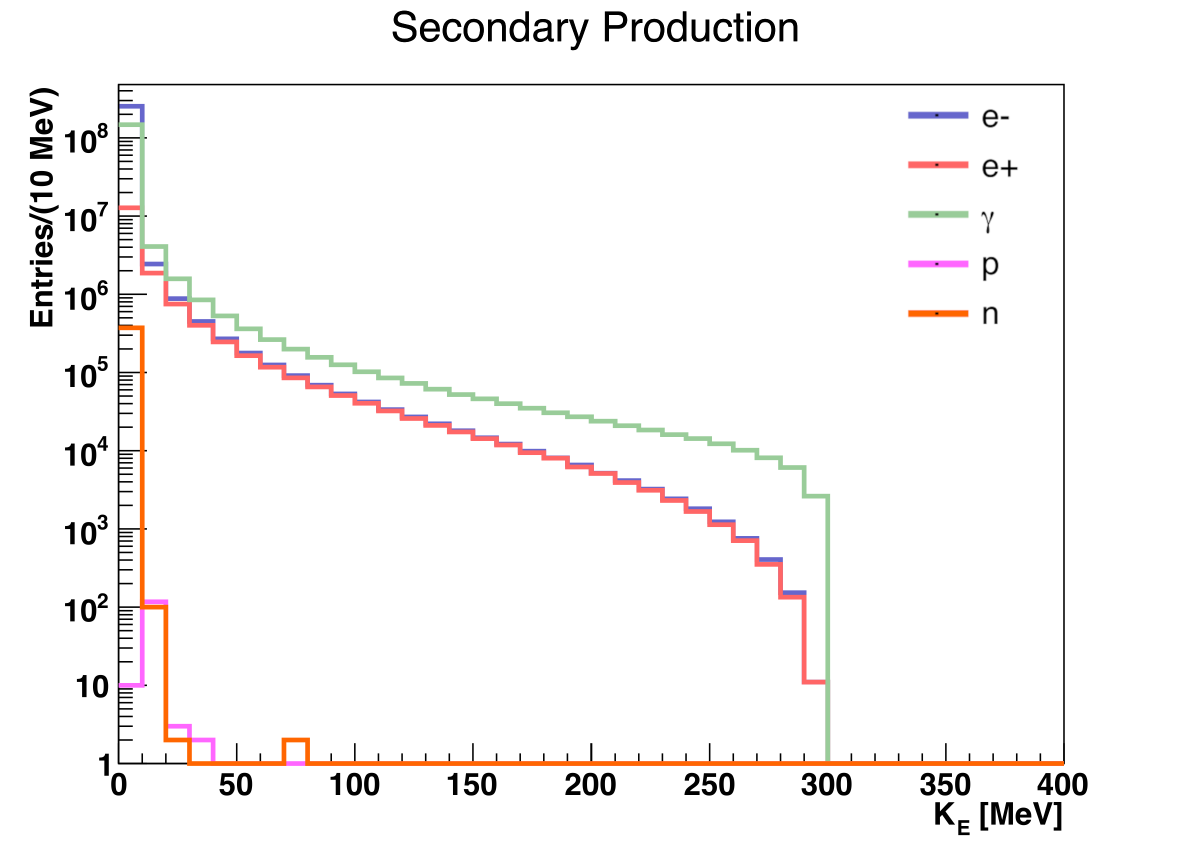}
\includegraphics[width=0.49\linewidth]{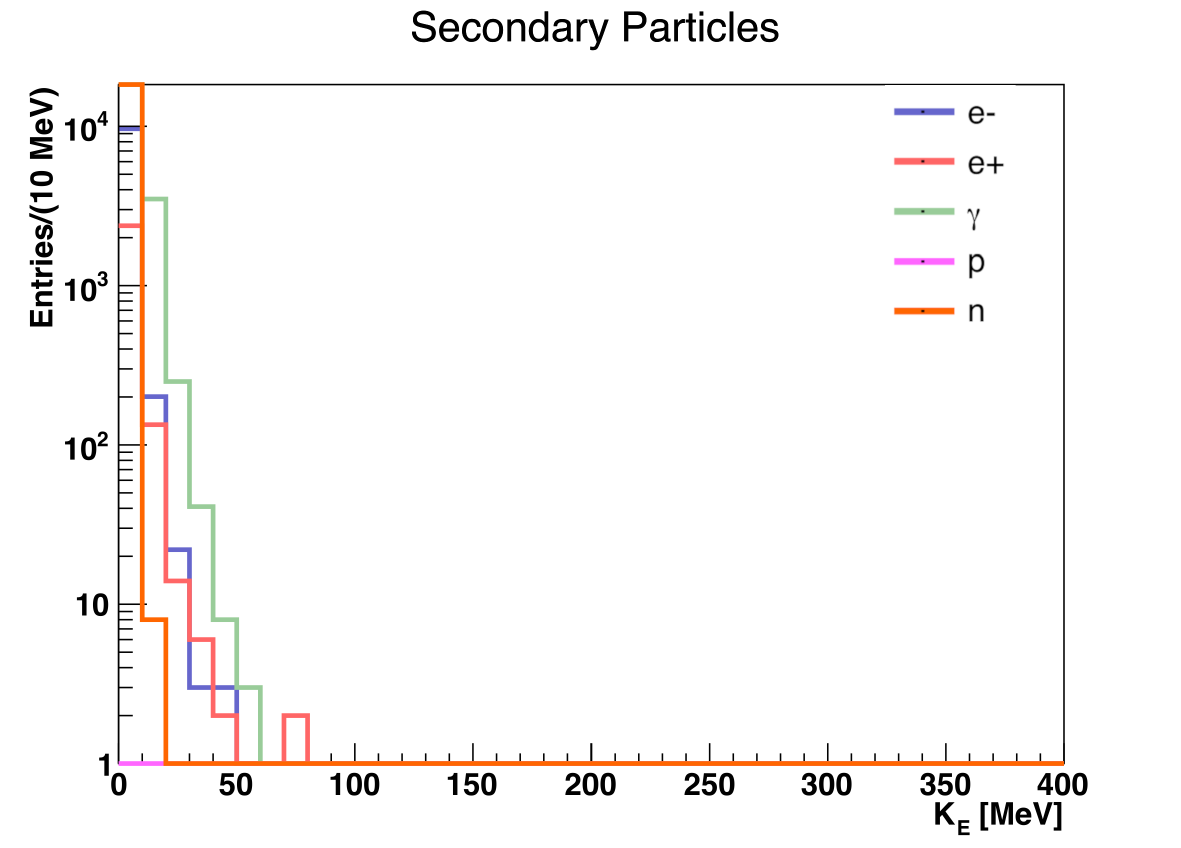}
\caption{Kinetic energy spectra of secondary particles, including electrons, positrons, photons, protons, and neutrons, within the target volume (left) and those escaping the target volume and potentially contributing to backgrounds (right), based on one million generated events.
}
\label{fig:prod}
\end{figure}

The \textsc{GEANT4} simulation uses a simplified geometry but considers all important features of background estimation, properly calculating all secondary particles and their interactions with the experimental structures. The left panel of Fig.~\ref{fig:prod} displays all particles produced inside the target volume, except for the tungsten isotopes. 

The background estimation is based on the energy deposited in the detector volume, including the secondary productions between a vacuum chamber and concrete walls. Neutrons below \SI{50}{\MeV} may not create signal-like photons, but their background energy deposition in the detector volume degrades the detector's performance. The right panel of Fig.~\ref{fig:prod} shows the energy distributions of secondary particles that escaped the target volume.

\begin{figure}[t]
\centering
\includegraphics[width=0.325\linewidth]{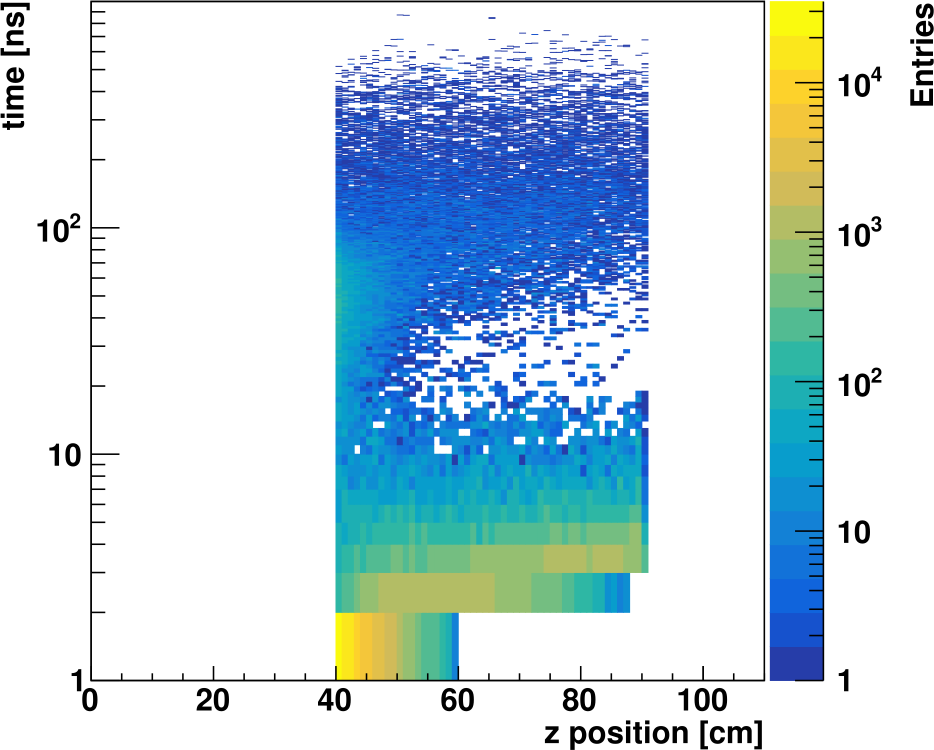}
\hspace{0.01cm}
\includegraphics[width=0.325\linewidth]{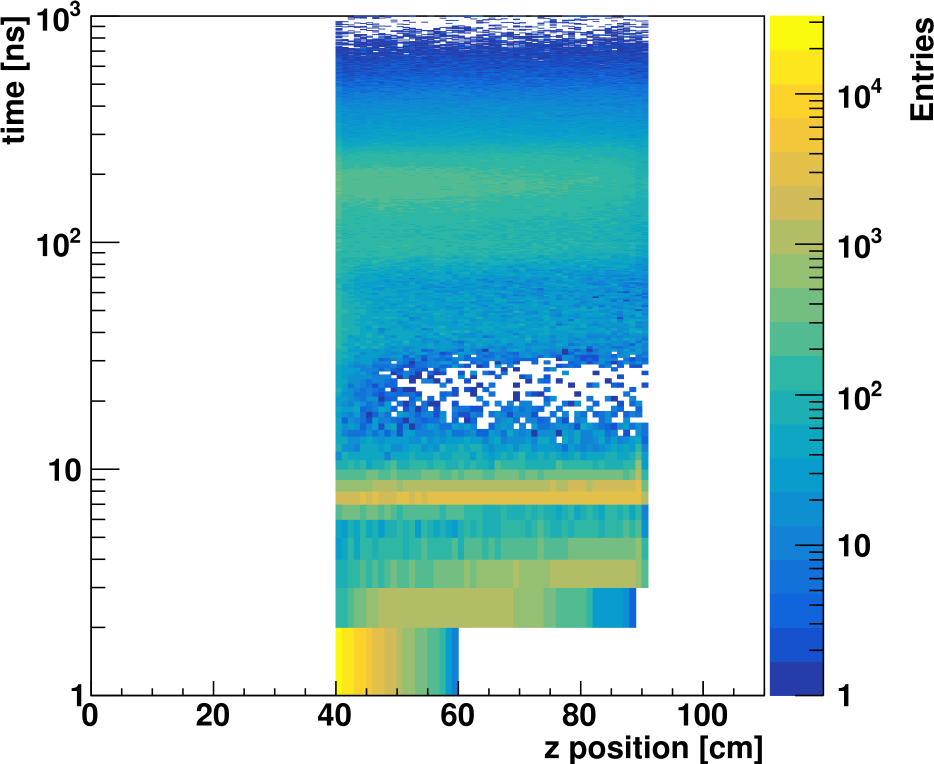}
\hspace{0.01cm}
\includegraphics[width=0.325\linewidth]{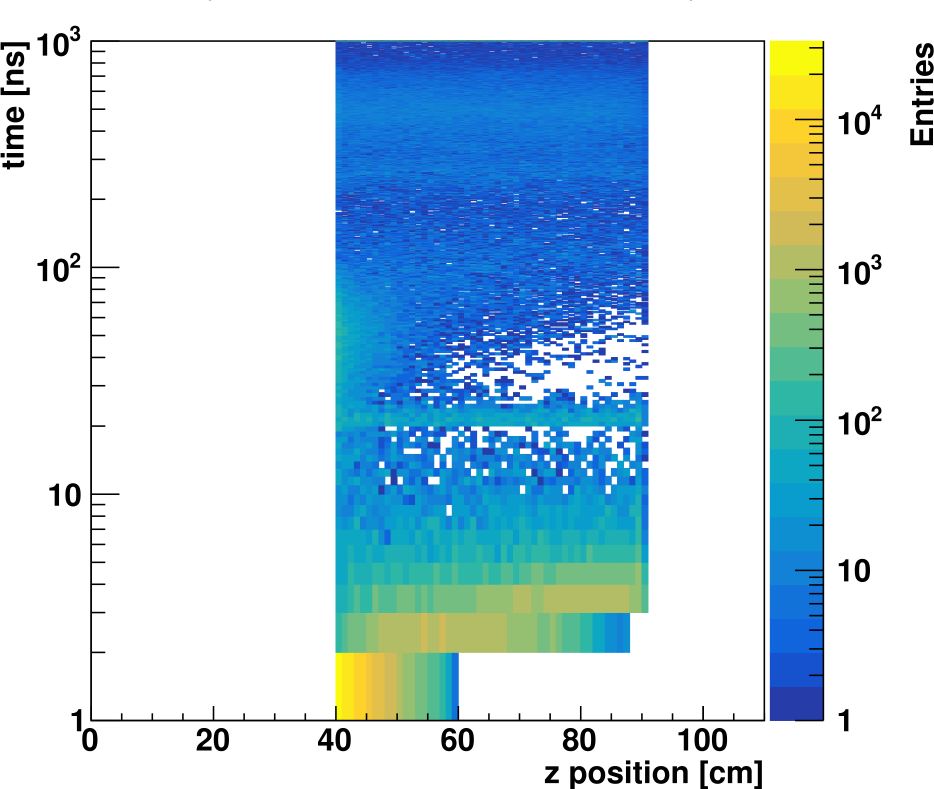}
\includegraphics[width=0.325\linewidth]{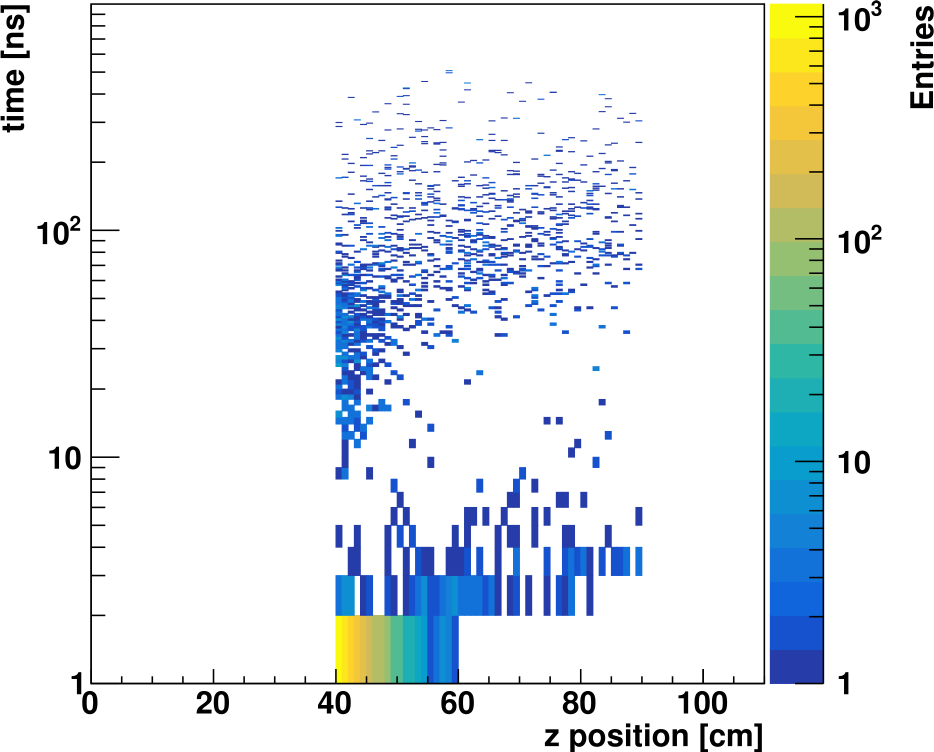}
\hspace{0.01cm}
\includegraphics[width=0.325\linewidth]{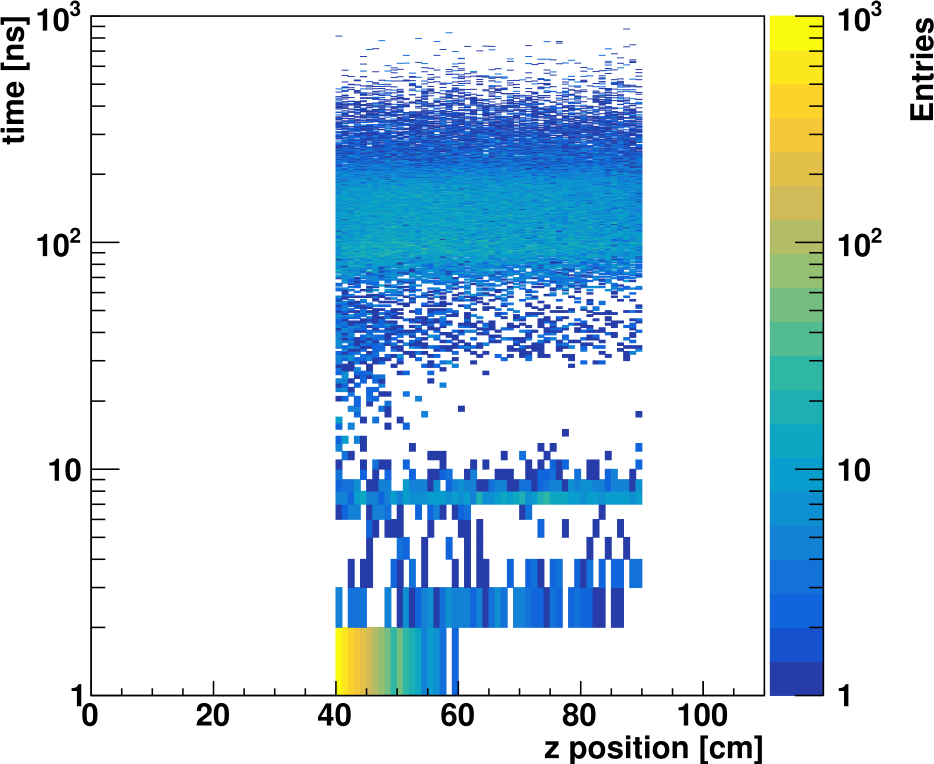}
\hspace{0.01cm}
\includegraphics[width=0.325\linewidth]{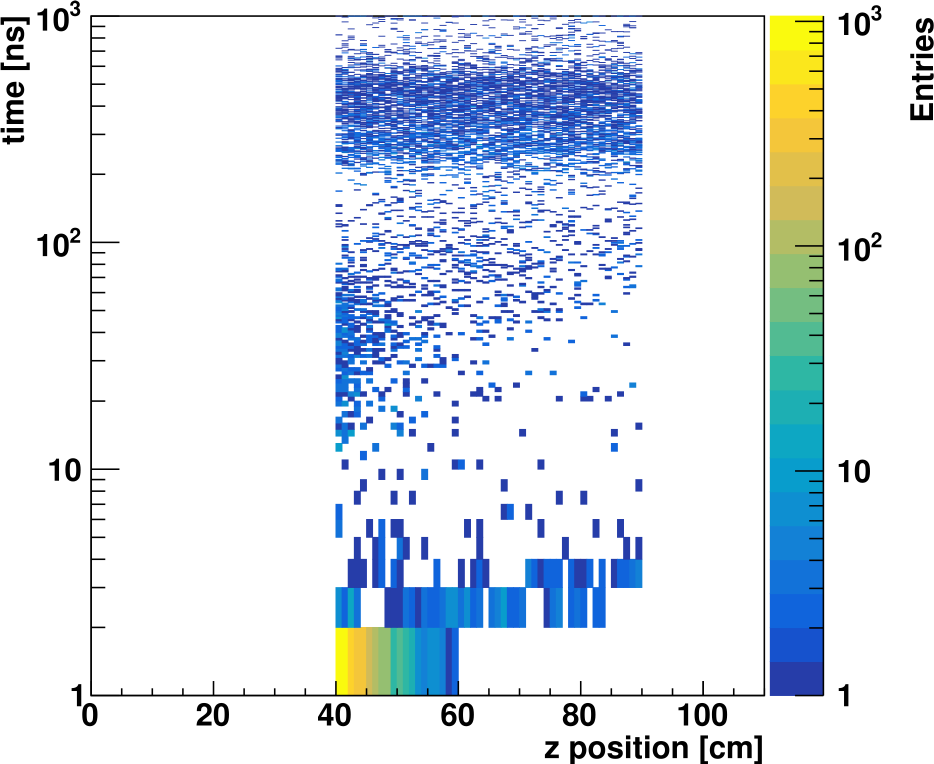}
\caption{All interactions of particles (top) and photons (bottom) escaping from the target and potentially contributing to the backgrounds at the detector are shown in the time(ns) vs $z$-position(cm) plane. Without a concrete wall (left panels), neutron interactions with the wall are not included. However, the 1 m (middle panels) and 3 m (right panels) thick walls show neutron interactions with the concrete wall, clearly illustrating the time difference between the two wall thicknesses.}
\label{fig:allWall}
\end{figure}

\textsc{GEANT4} simulations were used to accurately estimate background rates, particularly focusing on the time delay of neutron background interactions with the concrete walls. As a result, the particles (possibly behaving like prompt backgrounds) escaping from the target arrive at the detector within \SI{10}{\ns}, while neutron-induced backgrounds arrive several hundred nanoseconds later. However, most backgrounds reach the detector within \SI{1}{\ms}. Therefore, if the beam's pulse separation exceeds \SI{1}{\ms}, there is no significant overlap of backgrounds between beam bunches. The top panel of Fig.~\ref{fig:allWall} shows the timing of interactions between backgrounds and the detector for configurations without a concrete wall, with a \SI{1}{\m} wall, and a \SI{3}{\m} wall. The bottom panel of Fig.~\ref{fig:allWall} highlights the interactions of prompt and neutron-induced photon backgrounds. The \textsc{GEANT4} simulation does not include thermal neutrons, and the simulation threshold is set to the default value in \textsc{GEANT4}.

\subsection{Measurement at FTBF with 2 GeV Electron Beam and \textsc{GEANT4} Simulation}
\label{sec:2-GeV-electron-beam-geant4-simulation}
\begin{figure}[h]
\centering
\includegraphics[width=0.45\linewidth]{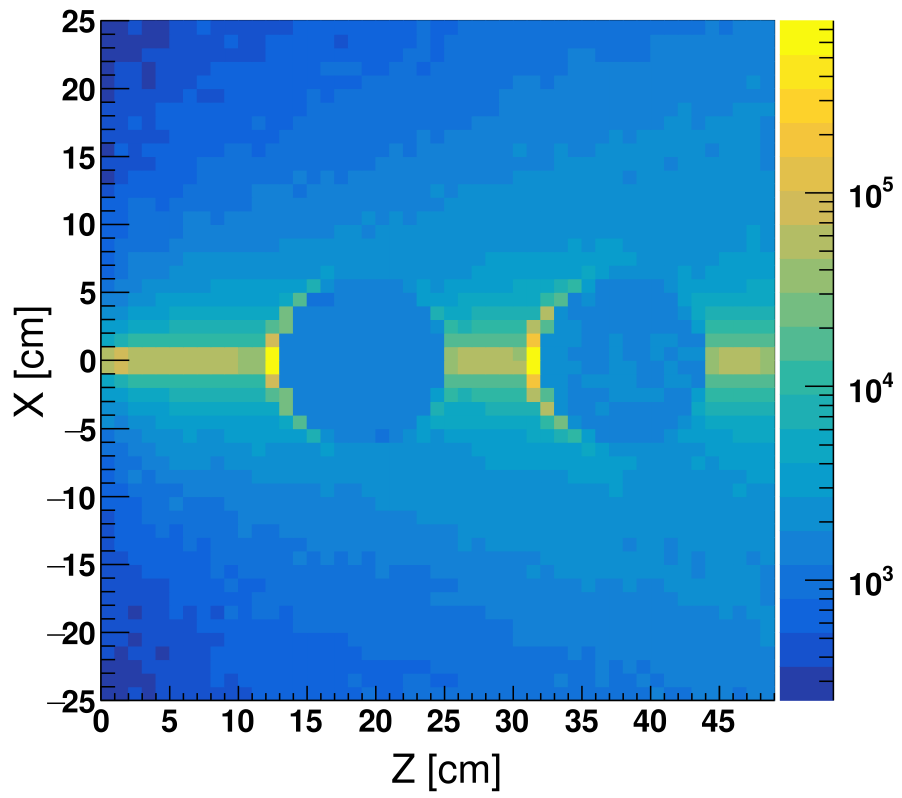}
\includegraphics[width=0.45\linewidth]{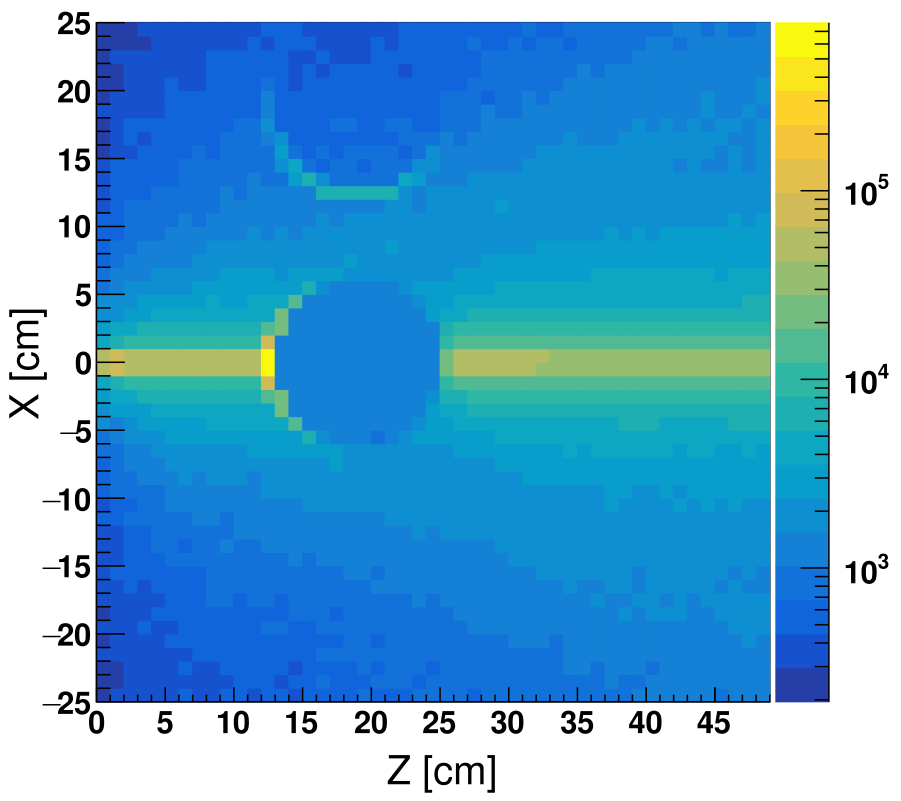}
\caption{The \textsc{GEANT4} simulation results of geometry configurations 1 (left) and 2 (right).} 
\label{fig:geoConf}
\end{figure}

To study the BRN background, we have conducted an experiment at a 2 GeV electron beam facility, irradiating a tungsten target. The experiment involved varying the target length and using two different geometric configurations for the liquid scintillator detectors. \textsc{GEANT4} simulations were performed to evaluate the efficiency of the liquid scintillator trigger, as well as the energy deposition and distribution of particles within the scintillator. Figure~\ref{fig:geoConf} shows the geometry configurations adopted in our \textsc{GEANT4} simulations.

\begin{figure}[t]
\centering
\includegraphics[width=0.49\linewidth]{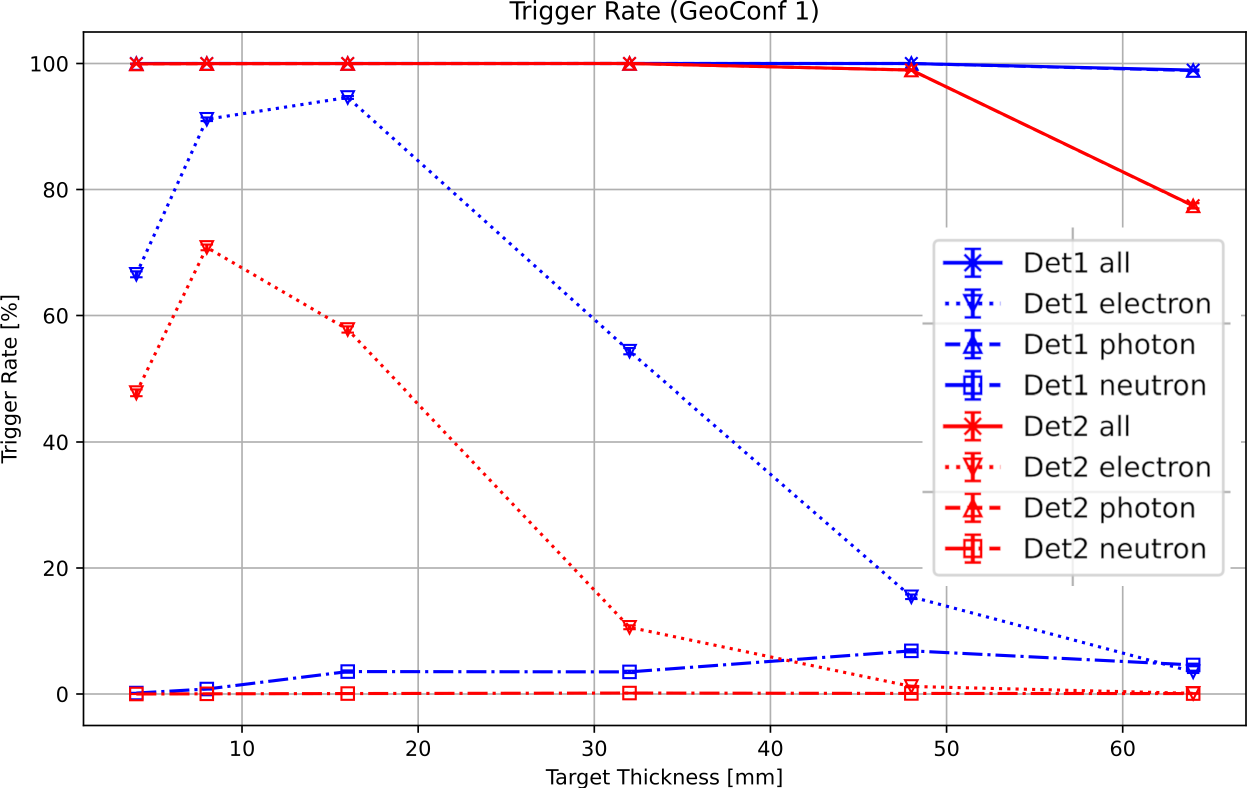}
\includegraphics[width=0.49\linewidth]{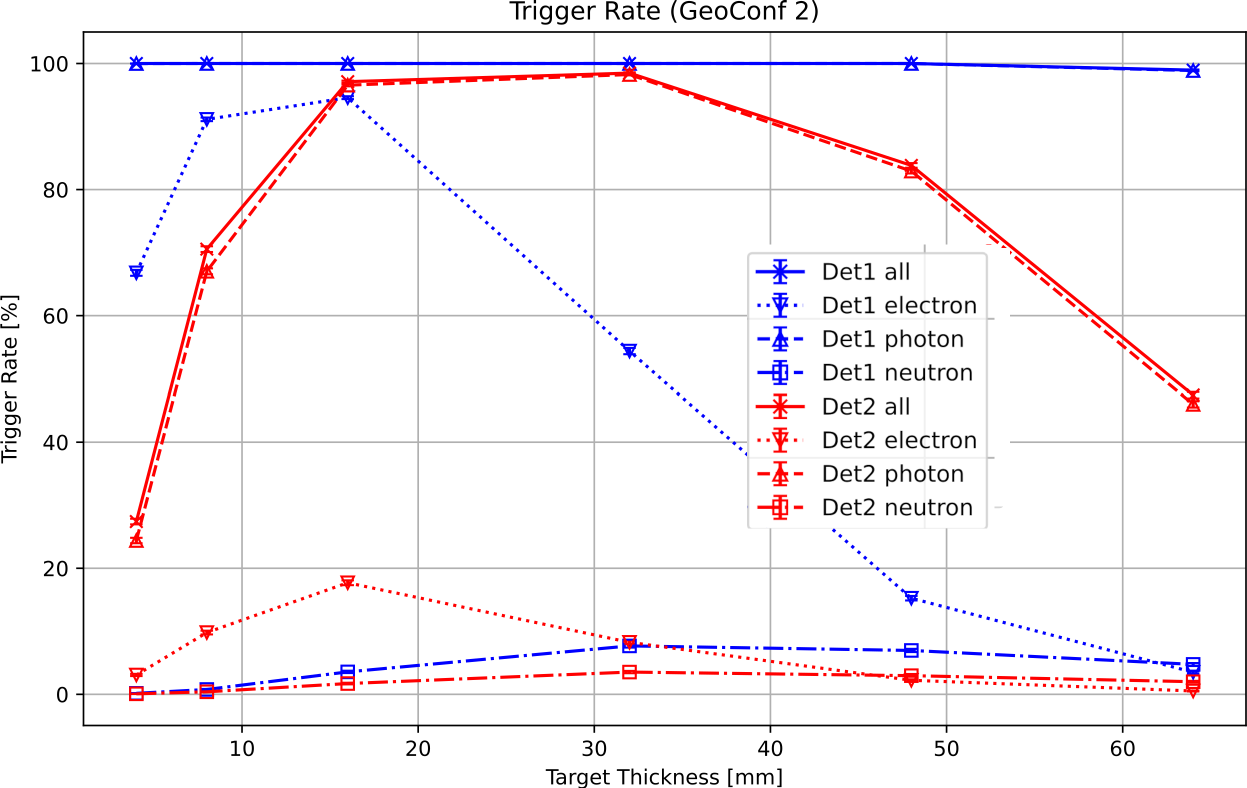}
\caption{The trigger rates of geometry configurations 1 (left) 2 (right). Detector 1 is near the beam target, while detector 2 is next to detector 1. The $x$-axis represents the target length(mm), and the $y$-axis represents the trigger rate(\%).} 
\label{fig:trigger}
\end{figure}

The trigger rate is determined by counting how many tracks from the target are produced within the detector. %
The threshold for tracking is set to the default value in \textsc{GEANT4}. Our findings indicate that most of the trigger rate contribution comes from photons. The trigger rates for the two geometric configurations are shown in Fig.~\ref{fig:trigger}.

\begin{figure}[t]
\centering
\includegraphics[width=0.45\linewidth]{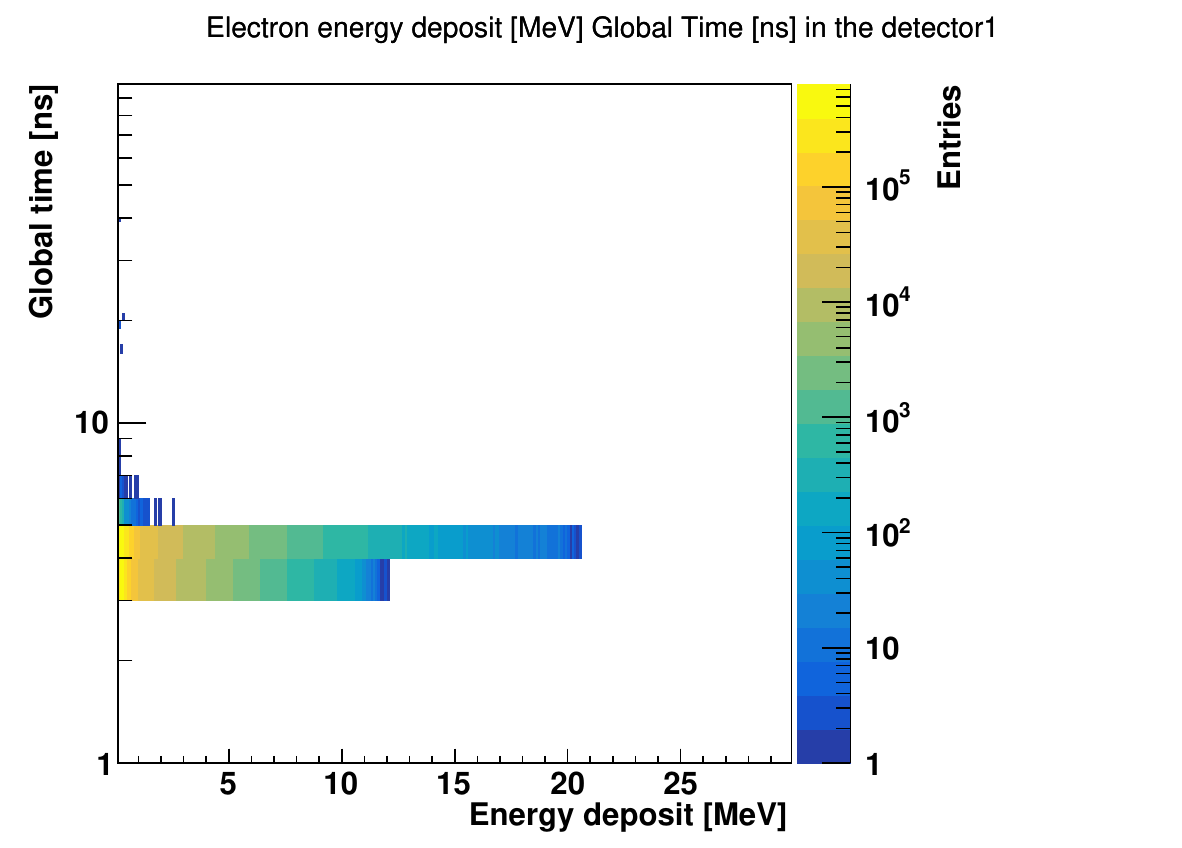}
\includegraphics[width=0.45\linewidth]{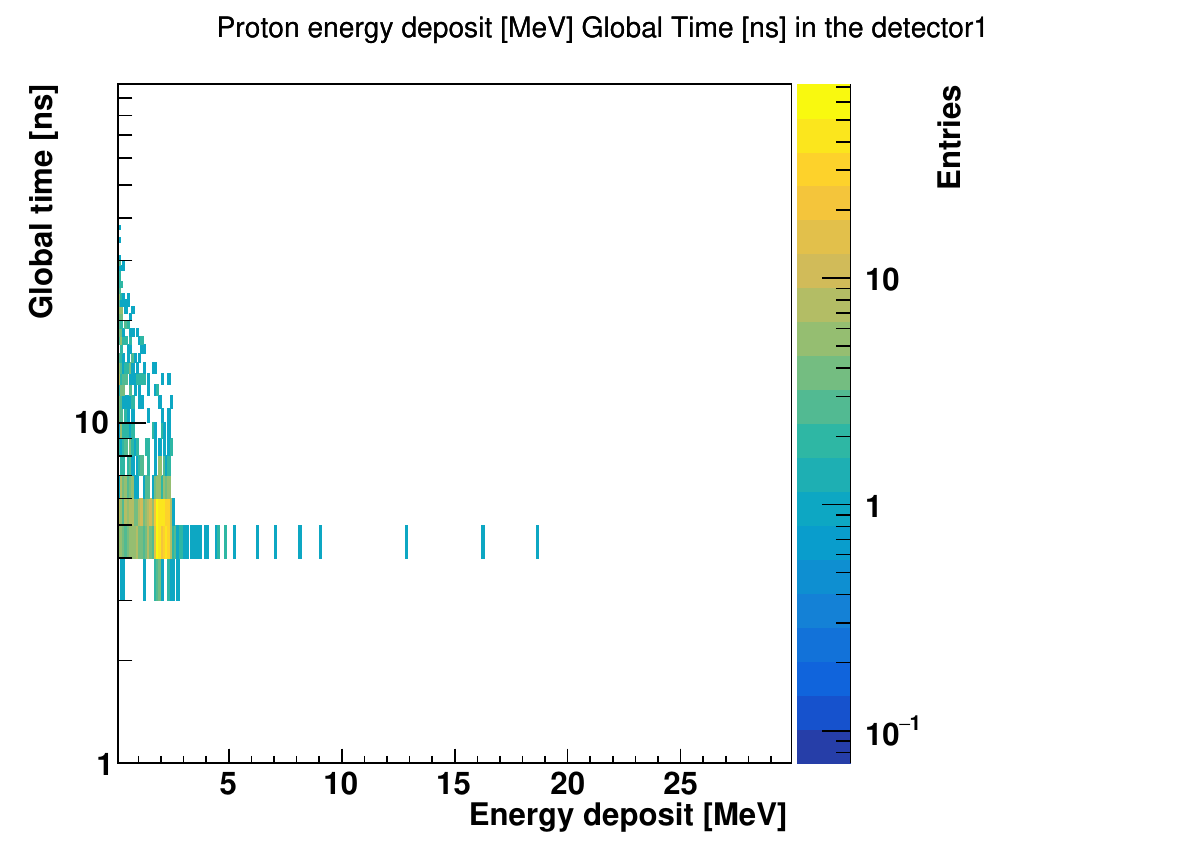}
\caption{Simulated energy depositions in the EJ-301 liquid scintillator, shown in the energy-deposit versus global-time plane, for a
2~GeV electron beam on a 0.4~cm tungsten target with $10^{5}$~EOT. EJ-301
scintillates only in response to charged particles, so incident photons and
neutrons escaping the target are registered indirectly, through the
secondary electrons and recoil protons they generate within the liquid.
(Left)~Energy deposited by secondary electrons---produced mainly by
Compton scattering and pair production of incident photons in the
scintillator. (Right) Energy deposited by recoil protons---produced
by elastic scattering of incident neutrons off hydrogen in the scintillator.
The plotted energy is therefore the deposit from these secondary charged
particles, not from the primary photon or neutron itself, and no separate
photon or neutron energy-deposit panel exists by construction. Two temporal
populations are visible: prompt particles depositing energy within
$\sim$6~ns of the beam pulse, and neutron-induced recoil protons depositing
energy over the following several tens of nanoseconds. See Fig.~\ref{fig:lseall}
for the corresponding incident-particle kinetic-energy spectra.} 
\label{fig:lsedep}
\end{figure}

\begin{figure}[tbh]
\centering
\includegraphics[width=0.65\linewidth]{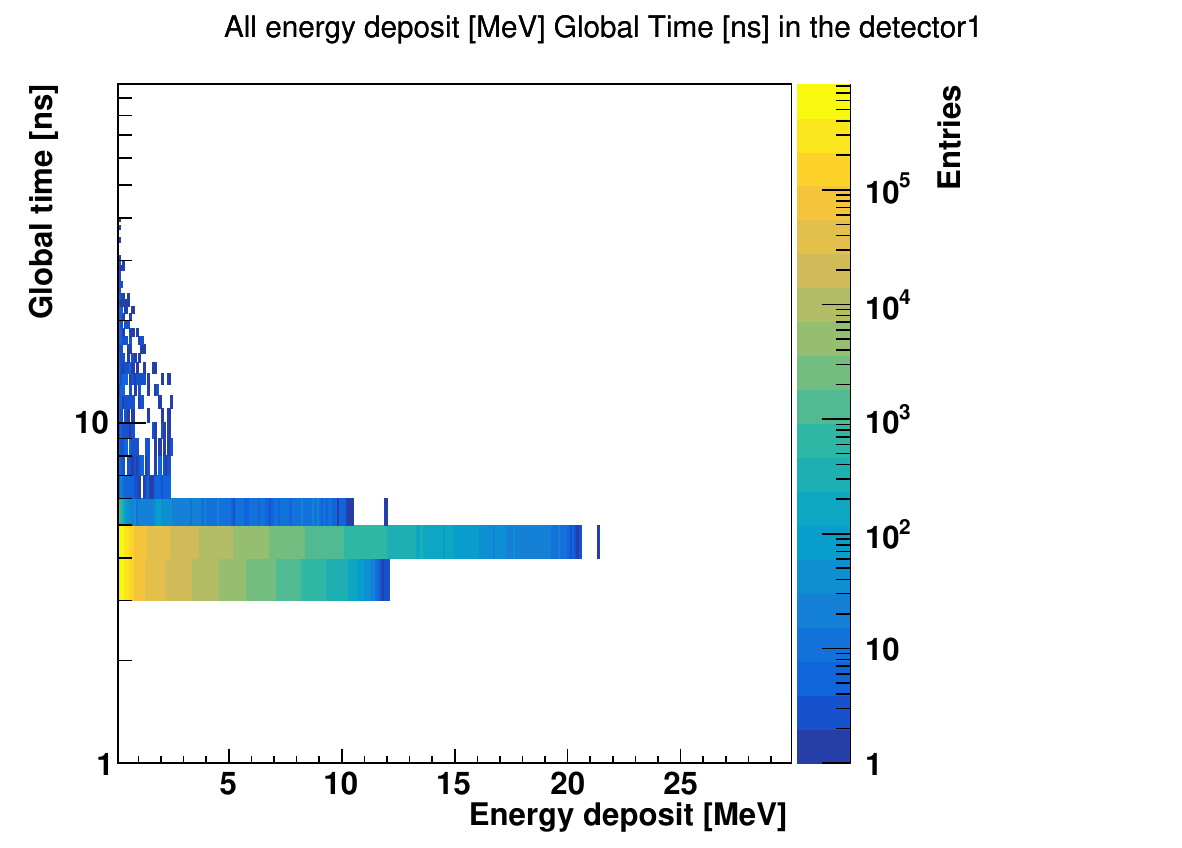}
\caption{Energy depositions by all particles in the EJ-301 liquid scintillator, shown in the energy-deposit-vs-time plane. The results clearly show two distinct groups: prompt particles deposit energy before approximately 6 ns, while neutron-induced particles deposit energy after that, continuing for several tens of nanoseconds. This result is based on the same configuration as in Fig.~\ref{fig:lsedep}.
} 
\label{fig:lsedall}
\end{figure}

The \textsc{GEANT4} simulation models the EJ-301 liquid scintillator and its aluminum housing to calculate the energy distribution and deposition of particles. To compare with experimental results, the energy deposition can be equated to the scintillator light counts observed in the experiment. However, the simulation does not attempt to model the liquid scintillator's response to electrons and protons. Instead, the energy deposition is correlated with signal counts. The \textsc{GEANT4} simulation results specifically depict the energy deposited by electrons and protons in the EJ-301 liquid scintillator (top panel of Fig.~\ref{fig:lsedep}), consistent with the scintillator's specifications. Neutrons and photons, classified as important background particles, 
leave no direct energy deposits as the EJ-301 liquid scintillator responds to electrons and protons (bottom panel of Fig.~\ref{fig:lsedep}). However, they indirectly deposit their energies via induced electrons and protons for which the energy distributions will be displayed later.
Figure~\ref{fig:lsedall} shows the energy deposition from all particles in detector 1 with a 0.4~cm thick tungsten target. The results clearly show two distinct groups: prompt particles depositing energy before approximately 6~ns, and neutron-induced particles depositing energy for several tens of nanoseconds afterward.

\begin{figure}[tbh]
\centering
\includegraphics[width=0.45\linewidth]{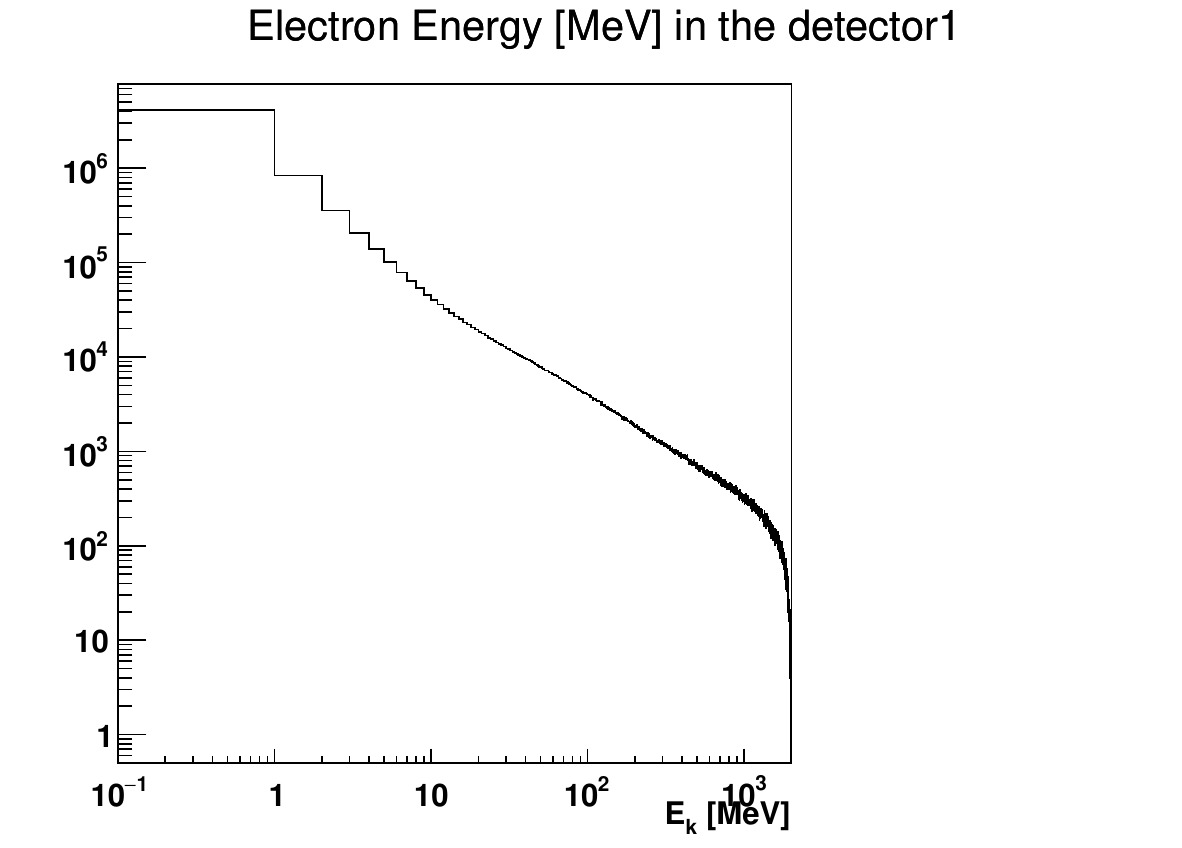}
\includegraphics[width=0.45\linewidth]{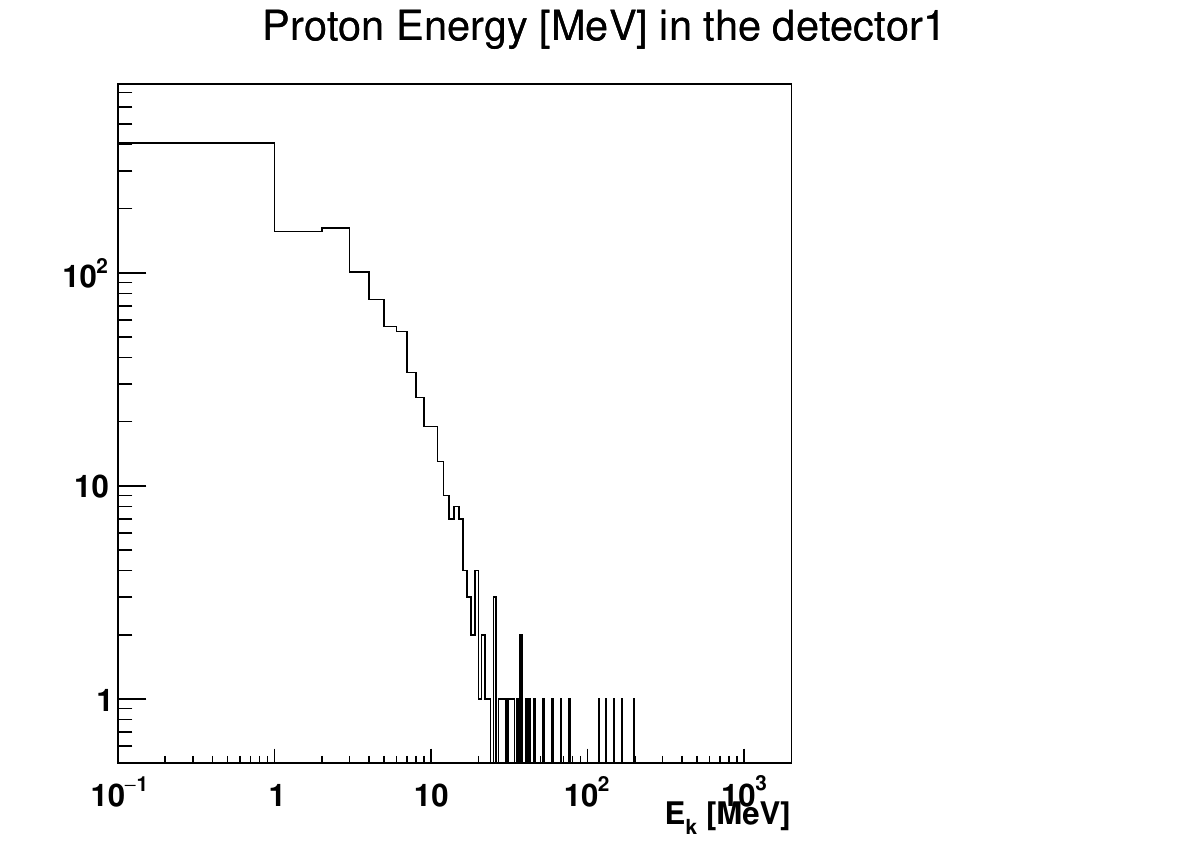}\\
\includegraphics[width=0.45\linewidth]{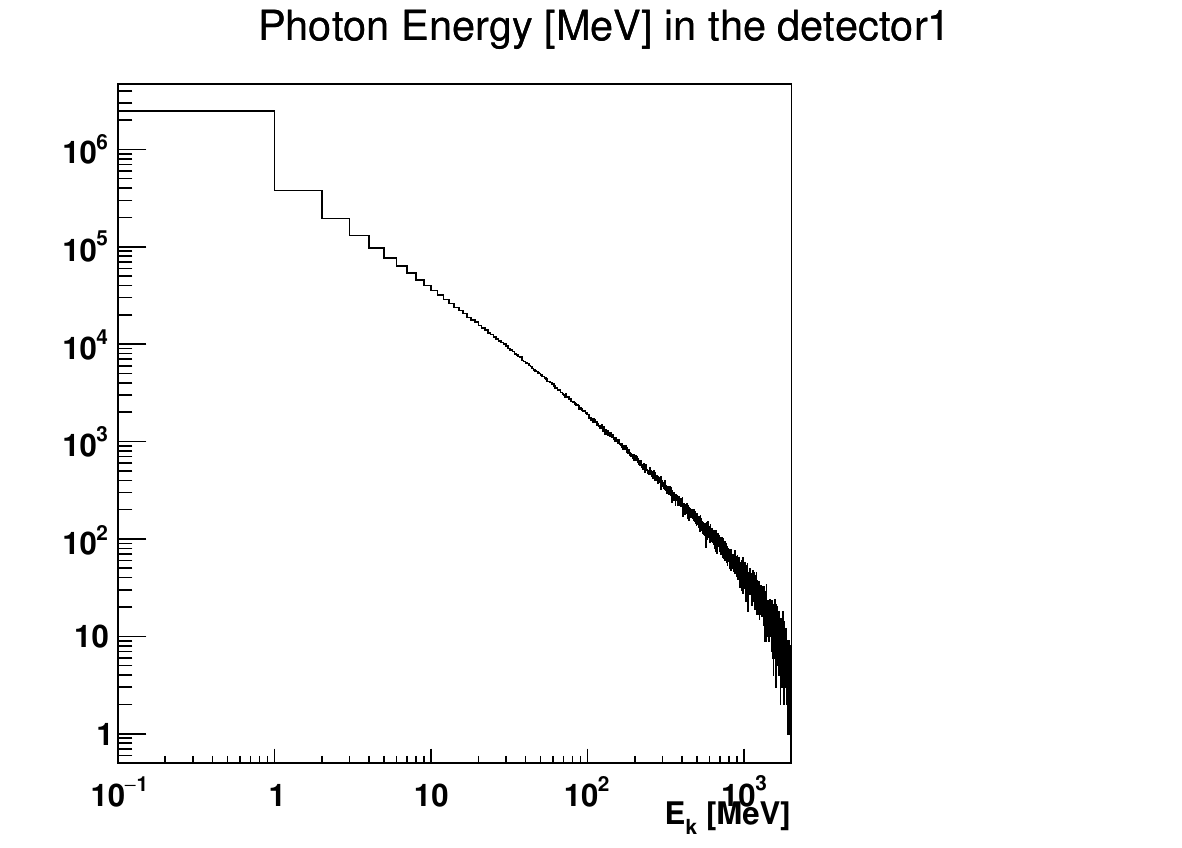}
\includegraphics[width=0.45\linewidth]{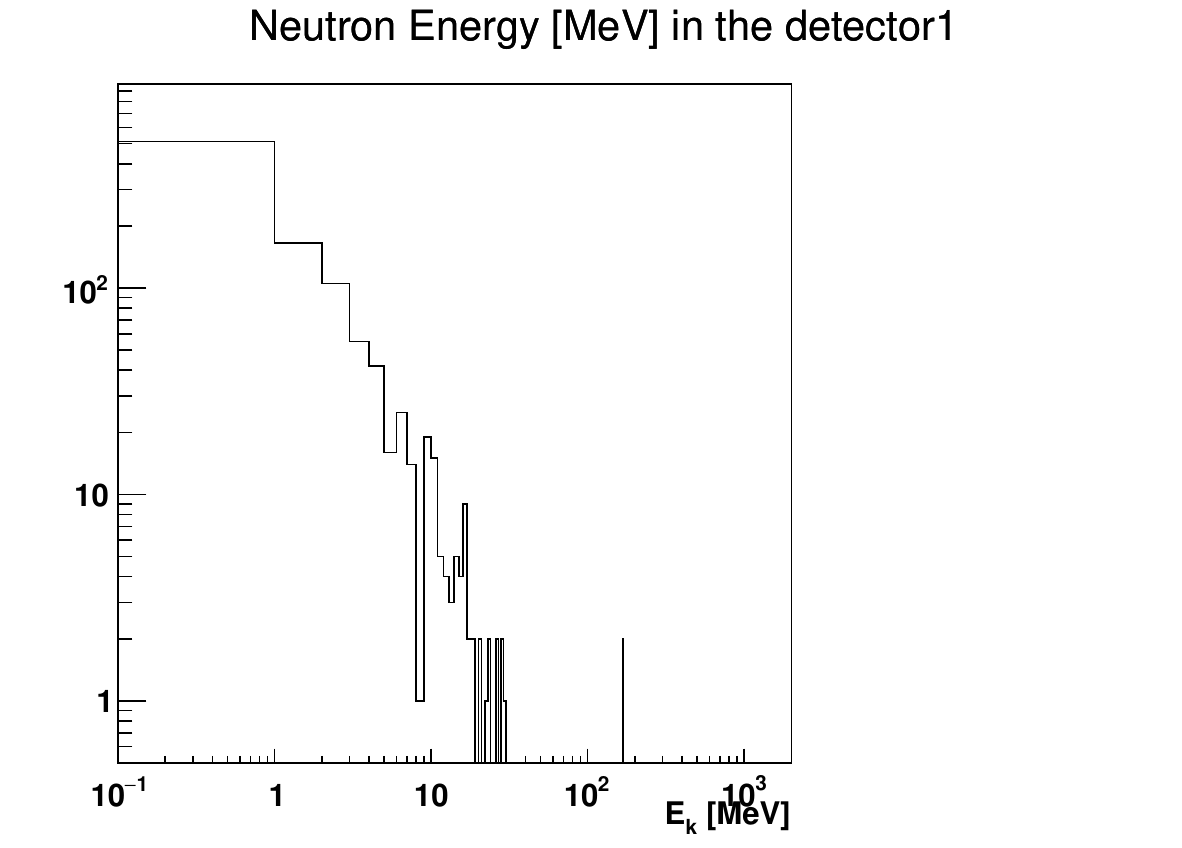}
\caption{Energy distributions of electrons, protons, photons, and neutrons in the EJ-301 liquid scintillator. 
The distributions of photons and neutrons are based on the energy deposited by electrons and protons induced by photon and neutron interactions.
This result is based on the same configuration as in Fig.~\ref{fig:lsedep}. } 
\label{fig:lseall}
\end{figure}

The \textsc{GEANT4} simulation calculated the energy distribution of particles in the detector, accounting for all physics process steps, meaning one particle can create multiple entries. These results will aid in interpreting the experimental data. The EJ-301 liquid scintillator responds only to electrons and protons, although other particles can generate a response indirectly by producing electrons and protons. Furthermore, identifying particles based on the signal shape is crucial for accurate experimental measurements. Figure~\ref{fig:lseall} shows the energy distributions of electrons, protons, photons, and neutrons from the \textsc{GEANT4} simulation. 
As mentioned above, the distributions of photons and neutrons are based on the energy deposited by electrons and protons induced by photon and neutron interactions.
This information helps interpret the measurement result by estimating and distinguishing the particle type and its energy.

\subsection{Background estimation at the 8 GeV electron beam}

The dominant backgrounds in the DAMSA 8~GeV electron beam configuration
originate from particles escaping the tungsten target and reaching the detector
volume. In an electron-beam facility such as SLAC, the 8~GeV beam loses its
energy in the target almost entirely through electromagnetic (EM) interactions,
producing an EM shower dominated by photons, electrons, and positrons. Hadronic
processes---and therefore BRN and charged hadrons---are
strongly suppressed relative to a proton beam of comparable power, in which
hadronic spallation is the primary energy-loss mechanism. This makes the
electron beam an intrinsically cleaner and more controlled environment in which
to establish the ultra-short-baseline technique, and it is one of the principal
reasons for staging DAMSA with an electron beam before moving to a high-energy
proton facility.

\begin{figure}[tbh]
    \centering
    \includegraphics[width=0.6\linewidth]{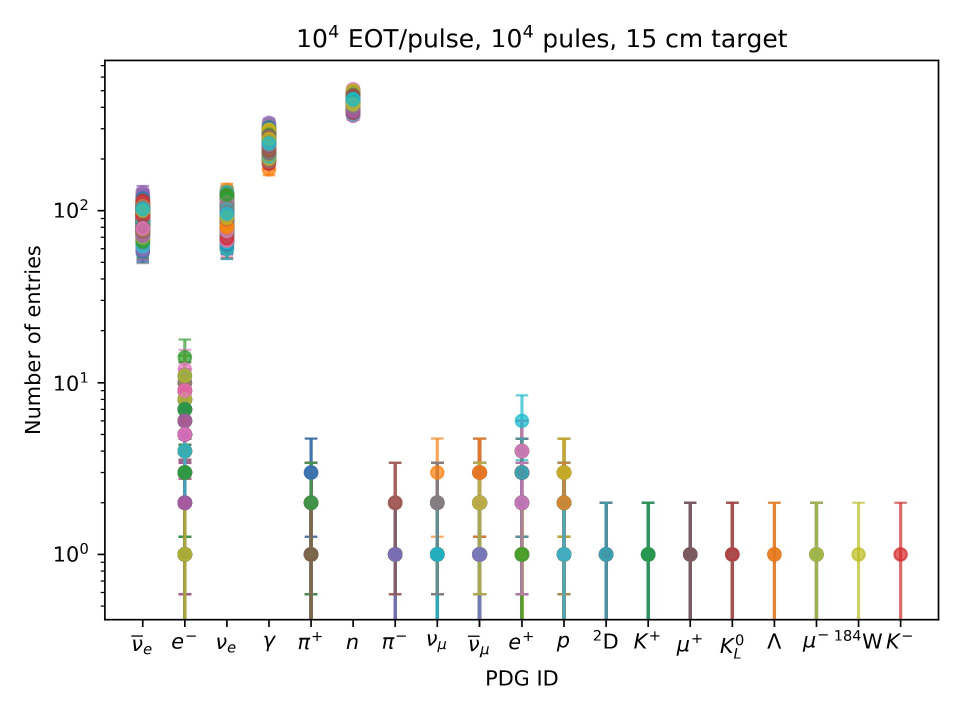}
    \caption{Leakage particles entering the detector per pulse ($10^4$~EOT
    (Electrons on Target)); neutrons and photons dominate.}
    \label{fig:leakagePU}
\end{figure}

To quantify these target-escaping backgrounds, a dedicated \textsc{Geant4}
simulation was performed using the full geometry. The simulation corresponds to
$10^{8}$~EOT, assuming $10^{4}$~EOT per pulse, for a total of $10^{4}$ simulated
pulses. The dominant leakage particles entering the detector are neutrons and
photons (Fig.~\ref{fig:leakagePU}), consistent with the EM-dominated shower
composition and with the trends observed at 2~GeV. The \textsc{Geant4}
simulation includes high-precision neutron transport but does not track thermal
neutrons explicitly; since the objective is to estimate the incident particle
flux and energy deposition at the detector, and thermal neutrons carry
negligible kinetic energy, this omission has a negligible effect---in both flux
and energy deposition---on the present estimate.

Table~\ref{tab:pu_rates} summarizes the target-escaping particle multiplicity,
broken down by species and energy threshold. The three thresholds illustrate
how the escaping flux hardens with energy: while neutrons and photons dominate
the total leakage, their yields fall steeply toward the 500~MeV signal
threshold, above which the photon rate in particular becomes extremely small.

\begin{table}[ht]
    \centering
    \caption{The target-escaping particle multiplicity per readout window at the
    nominal operating point of $10^{4}$~EOT per pulse and $10~\mathrm{\mu s}$
    pulse separation, evaluated from a simulation of $10^{4}$ consecutive pulses
    corresponding to a total exposure of $10^{8}$~EOT. Values are reported as
    mean~$\pm$~standard deviation, where the standard deviation denotes the
    pulse-to-pulse fluctuation.}
    \label{tab:pu_rates}
    \begin{tabular}{lccc}
        \hline\hline
        Threshold & Charged & Photon & Neutron \\
        \hline
        $E > 10~\mathrm{MeV}$  & $0.415 \pm 0.65$ & $2.39 \pm 1.6$ & $2.72 \pm 1.7$ \\
        $E > 100~\mathrm{MeV}$ & $0.335 \pm 0.58$ & $(4.00 \pm 74)\times10^{-3}$ & $1.03 \pm 1.0$ \\
        $E > 500~\mathrm{MeV}$ & $0.174 \pm 0.42$ & $(4.00 \pm 200)\times10^{-4}$ & $0.156 \pm 0.40$ \\
        \hline\hline
    \end{tabular}
\end{table}

\begin{figure}[tbh]
    \centering
    \includegraphics[width=0.49\linewidth]{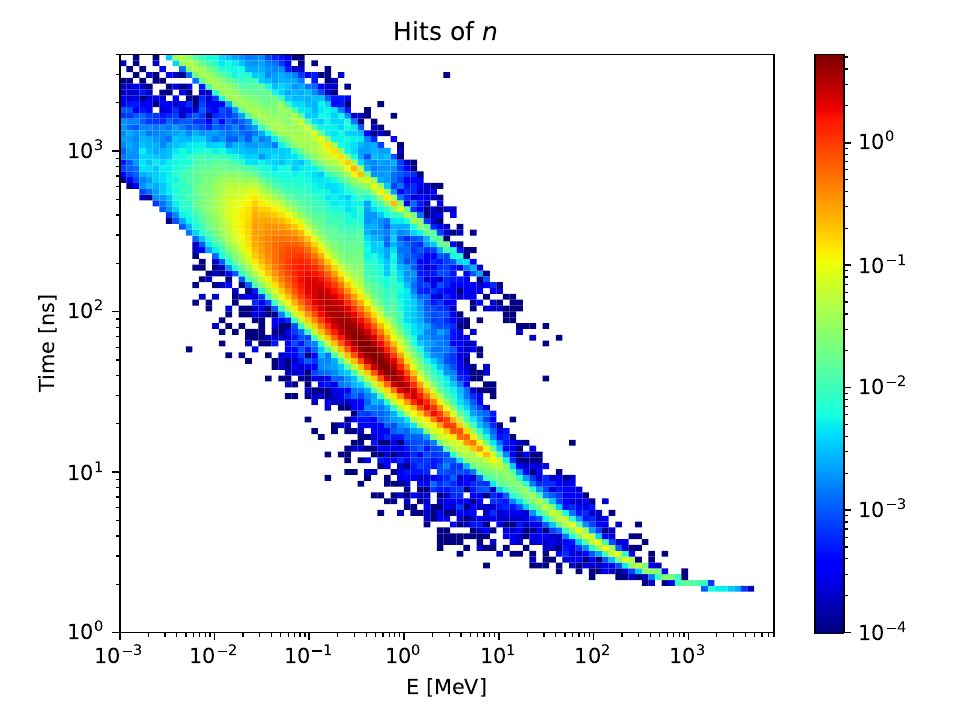}
    \includegraphics[width=0.49\linewidth]{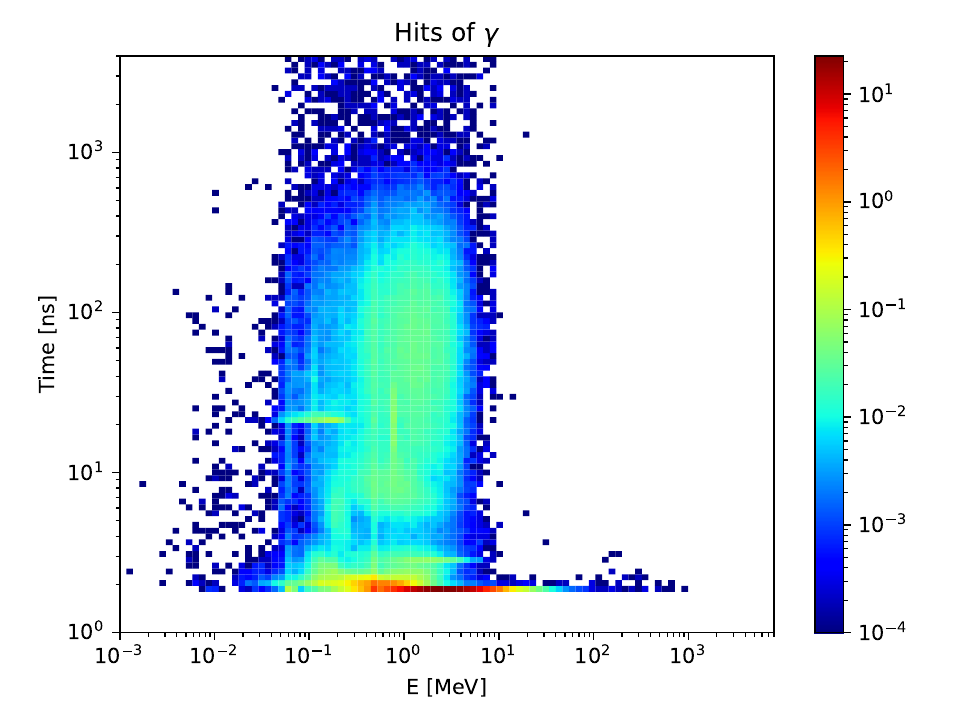}
    \caption{Energy--time distributions of neutrons (left) and photons (right),
    normalized to $10^4$~EOT. The x-axis shows kinetic energy (MeV), and the
    y-axis shows global time (ns). A secondary band at 1~$\mu$s in the neutron
    distribution arises from reflections off a 3~m radius concrete wall.}
    \label{fig:major_leakage}
\end{figure}

The temporal structure of these backgrounds (Fig.~\ref{fig:major_leakage})
shows that the leakage flux is dominated by slow neutrons and neutron-induced
photons with kinetic energies below 10~MeV, which largely vanish within
$\sim$1~$\mu$s. This delayed component is well separated in time from the prompt
EM leakage, so a bunch spacing of $\gtrsim 10~\mu$s at $10^{4}$~EOT per pulse is
sufficient to avoid pulse-to-pulse pile-up of delayed backgrounds and
constitutes a manageable operating condition for the detector.

\begin{figure}[tbh]
    \centering
    \includegraphics[width=0.7\linewidth]{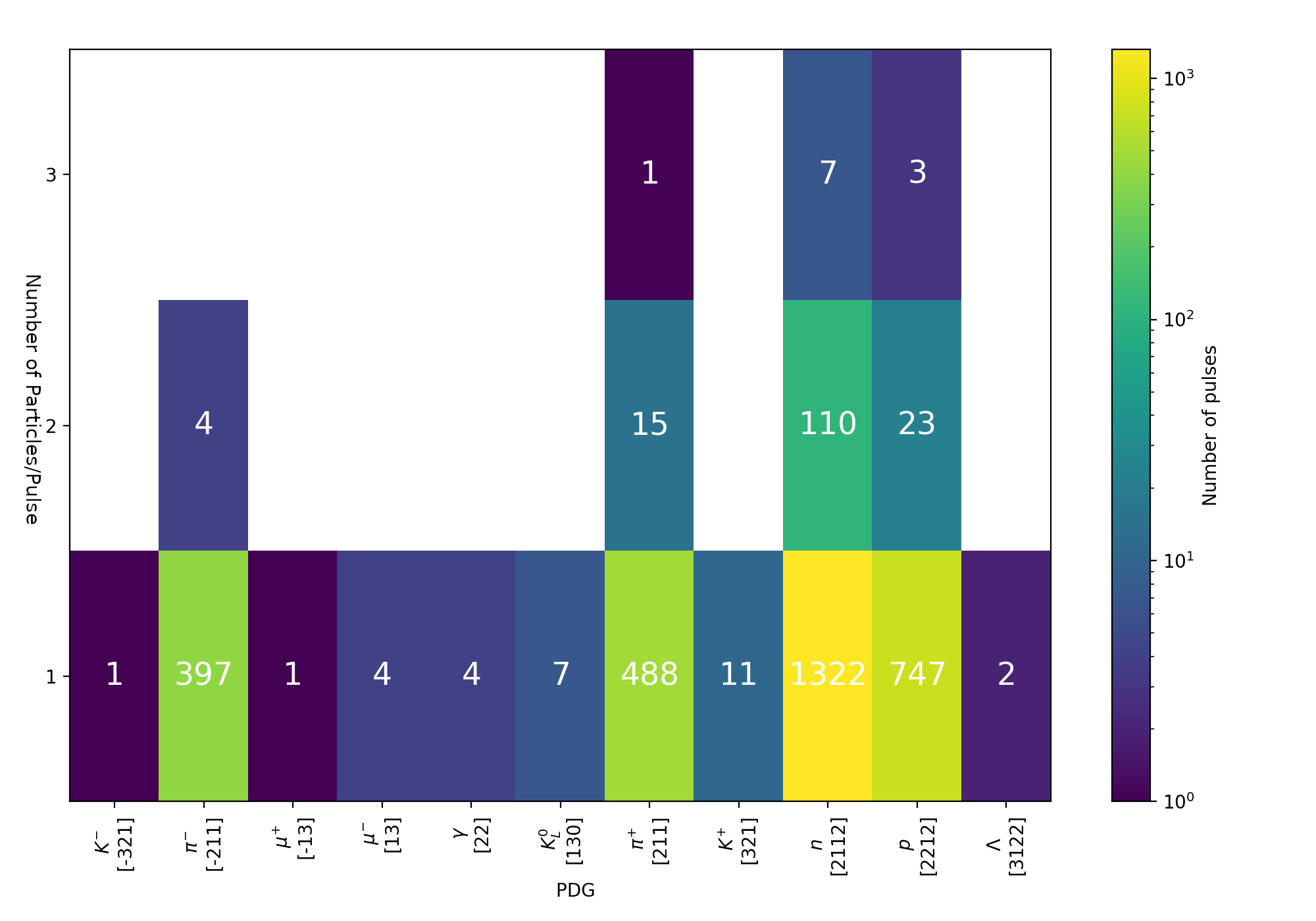}
    \caption{Leakage particles above 500~MeV. Particle yields are shown as
    entries per pulse ($10^4$~EOT) for $10^4$ simulated pulses. The x-axis
    denotes particle type, and the y-axis shows the number of particles of
    labeled PDG type per pulse.}
    \label{fig:totPU500}
\end{figure}

Applying the 500~MeV signal photon threshold, only single photons above
500~MeV are observed to enter the detector, appearing in four of the $10^{4}$
simulated pulses (Fig.~\ref{fig:totPU500})---a single-photon rate above
threshold of $\sim 4\times10^{-4}$ per pulse. Since the $a\to\gamma\gamma$
signal requires two photons above threshold, the relevant accidental
background is two uncorrelated above-threshold photons in the same pulse, which
is bounded combinatorial at
$\mathcal{O}\!\left[(4\times10^{-4})^{2}\right]\approx 2\times10^{-7}$ per pulse,
negligible relative to the projected signal exposure even before angular and
timing correlations are exploited. This supports the assumption of a nearly
background-free diphoton search. While higher statistics are needed, these
results indicate that signal-mimicking diphoton backgrounds are extremely rare,
and rare-event sampling techniques are being pursued to further refine this
estimate.

Two features of the DAMSA 8~GeV electron beam design suppress these
target-escaping backgrounds at their source. First, the choice of an electron
rather than a proton beam ensures that target energy loss proceeds through EM
showering, so that photon and $e^{\pm}$ production dominates while neutron and
hadron yields are minimized---in contrast to the proton-beam configuration,
where hadronic spallation produces by proton. Second,
because DAMSA is designed to probe the beam-dump ``ceiling'' region of parameter
space---where sensitivity is limited by the early decay of short-lived particles
rather than by accumulated statistics---it does not require the very high beam
intensities used by conventional long-baseline beam-dump experiments. Operating
at reduced intensity directly lowers the target-escaping particle flux, and
hence the background rate, without sacrificing sensitivity in the ceiling
region.
\vspace{0.5em}
\section{Conclusions}      
\label{sec:Conclusions}
DAMSA is a tabletop scale, ultra-short baseline dark-portal particle search and discovery experiment at an accelerator with excellent physics potential. 
DAMSA Phase I focuses on gaining a deeper understanding of beam-related neutron backgrounds in a well-controlled electron beam environment while searching for MeV-scale new physics, including ALPs and dark photons. By leveraging the ultra-short baseline between the target and detector and the 8~GeV electron beams at SLAC's LESA with highly flexable beam configuration, Phase I DAMSA can probe short-lived particles, testing the experimental concept and serving as the first step toward building a full-capability detector in the subsequent Phase II enables the experiment to be fully ready for a proton beam facility, such as CERN's beam dump facility. 
The proposed DAMSA Phase I and II experiments have significant potential to explore uncharted regions of ALP and other dark messenger parameter space, particularly for short-lived ones that have eluded previous experiments. 
With unprecedented sensitivity at SLAC's LESA facility, taking advantage of its flexibility, this innovative experimental setup could lead to the discovery of ALPs, dark photons, and other unseen dark sector states, or impose stringent new constraints on their interactions with ordinary matter, thereby advancing our understanding of light, weakly interacting particles.

    \section*{Acknowledgments}
    This material is based upon work supported by the U.S. Department of Energy (DOE) Office of Science and the National Research Foundation of Korea (NRF) [RS-2024-00356960, RS-2025-25442707]. 
    We are grateful to the Center for Theoretical Underground Physics and Related Areas (CETUP*), the Institute for Underground Science at Sanford Underground Research Facility (SURF), and the South Dakota Science and Technology Authority for their hospitality and financial support.

    \appendix
    
    \bibliography{main,detector}
    
    \newpage

\end{document}